\documentclass[aps,pre,floats,superscriptaddress,nofootinbib,floatfix,twocolumn,longbibliography,showkeys]{revtex4-1}
\usepackage[T1]{fontenc}
\usepackage{amssymb,amsmath}
\usepackage{braket}
\usepackage{graphicx}
\usepackage{dcolumn}
\usepackage{subfigure}
\usepackage{appendix}
\usepackage[colorlinks=true,linkcolor=blue,allcolors=blue]{hyperref}
\usepackage[noabbrev]{cleveref}
\usepackage[english]{babel}

\usepackage{bm}
\usepackage{comment}
\usepackage{hyperref}
\usepackage{makecell}
\usepackage{changepage}
\usepackage{multirow}
\usepackage[table]{xcolor}
\usepackage[mathscr]{euscript}
\usepackage{csquotes}
\usepackage{soul,xcolor}
\usepackage{ulem}

\begin{document}
\title{Stability of flocking in the reciprocal two-species Vicsek model: Effects of relative population, motility, and noise}

\author{Aditya Kumar Dutta}
\email{saisakd2137@iacs.res.in}
\affiliation{School of Mathematical \& Computational Sciences, Indian Association for the Cultivation of Science, Kolkata -- 700032, India.}

\author{Matthieu Mangeat}
\email{mangeat@lusi.uni-sb.de}
\affiliation{Center for Biophysics \& Department for Theoretical Physics,
Saarland University, 66123 Saarbr$\ddot{u}$cken, Germany}

\author{Heiko Rieger}
\email{heiko.rieger@uni-saarland.de}
\affiliation{Center for Biophysics \& Department for Theoretical Physics,
Saarland University, 66123 Saarbr$\ddot{u}$cken, Germany}

\author{Raja Paul}
\email{raja.paul@iacs.res.in}
\affiliation{School of Mathematical \& Computational Sciences, Indian Association for the Cultivation of Science, Kolkata -- 700032, India.}

\author{Swarnajit Chatterjee}
\email{swarnajit.chatterjee@cyu.fr}
\affiliation{Center for Biophysics \& Department for Theoretical Physics,
Saarland University, 66123 Saarbr$\ddot{u}$cken, Germany}
\affiliation{Laboratoire de Physique Th{\'e}orique et Mod{\'e}lisation, UMR 8089, CY Cergy Paris Universit{\'e}, 95302 Cergy-Pontoise, France}

\begin{abstract}
Natural flocks need to cope with various forms of heterogeneities, for instance, their composition, motility, interaction, or environmental factors. Here, we study the effects of such heterogeneities on the flocking dynamics of the reciprocal two-species Vicsek model [Phys. Rev. E {\bf 107}, 024607 (2023)], which comprises two groups of self-propelled agents with anti-aligning inter-species interactions and exhibits either parallel or anti-parallel flocking states. The parallel and anti-parallel flocking states vanish upon reducing the size of one group, and the system transitions to a single-species flock of the majority species. At sufficiently low noise (or high density), the minority species can exhibit collective behavior, anti-aligning with the liquid state of the majority species. Unequal self-propulsion speeds of the two species strongly encourage anti-parallel flocking over parallel flocking. However, when activity landscapes with region-dependent motilities are introduced, parallel flocking is retained if the faster region is given more space, highlighting the role of environmental constraints. Under noise heterogeneity, the colder species (subjected to lower noise) attain higher band velocity compared to the hotter one, temporarily disrupting any parallel flocking, which is subsequently restored. These findings collectively reveal how different forms of heterogeneity, both intrinsic and environmental, can qualitatively reshape flocking behavior in this class of reciprocal two-species models.
\end{abstract}

\maketitle

\section{Introduction}
\label{intro}
Flocking is ubiquitous in nature~\cite{vicsek2012collective} and denotes the transition of self-propelled, mutually aligning agents to coherent motion in one common direction. This collective behavior of active matter emerges in human gatherings~\cite{bottinelli2016emergent}, mammalian herds~\cite{gomez2022intermittent}, bird flocks~\cite{ballerini2008interaction}, and fish schools~\cite{becco2006experimental}, to microscopic systems including unicellular organisms like bacteria~\cite{peruani2012collective}, collective cell migration in dense tissues~\cite{giavazzi2018flocking}, and cytoskeletal filaments driven
by molecular motors~\cite{schaller2010polar}. Beyond living systems, flocking has also been experimentally realized in synthetic active colloids~\cite{bricard2013emergence,kaiser2017flocking} and in vibrated polar disks~\cite{deseigne2010collective}.

The Vicsek model (VM)~\cite{Vicsek} is a paradigmatic framework for studying flocking in active matter systems. It describes the dynamics of self-propelled particles moving in two dimensions with constant speed and aligning their velocities with those of their neighbors within a specified interaction radius, subject to random noise. Despite its simplicity, the model captures the essence of flocking behavior, where individual particles transition from disordered motion to coherent, collective movement as the noise level decreases or the density increases. The VM exhibits long-range order (LRO)~\cite{Toner1995LROXY,Toner1998flocks}, and this emergence of order in two-dimensional systems is particularly striking because it seemingly violates the Mermin-Wagner theorem, which prohibits the spontaneous breaking of continuous symmetries in two-dimensional equilibrium systems with short-range interactions at finite temperature. The apparent violation arises from the non-equilibrium active nature of the VM, where the constant input of energy at the particle level drives the system far from equilibrium. The presence of LRO in the VM is further supported by giant number fluctuations~\cite{Solon2015phase}, which are a hallmark of active matter systems and emphasize their fundamental departure from equilibrium statistical mechanics. 

Recently, there has been a growing interest in understanding active systems composed of multiple particle species with inter-species reciprocal and non-reciprocal interactions~\cite{menzel2012collective,NRVM,kreienkamp2022clustering,SwarnajitTSVM,NRASM,mangeat2024emergent,grauer2020swarm,tucci2024nonreciprocal,tucci2025hydrodynamic}. In Ref.~\cite{SwarnajitTSVM}, the flocking dynamics of two unfriendly species has been investigated in the framework of the two-species Vicsek model (TSVM), which is a two-species generalization of the VM with reciprocal anti-ferromagnetic inter-species interactions. The reciprocal TSVM exhibits two primary steady states of collective motion: the anti-parallel flocking (APF) state, where the two species form bands moving in opposite directions, and the parallel flocking (PF) state, where the bands travel in the same direction. In the low-density and high-noise region of the coexistence phase, PF and APF states undergo fluctuation-induced stochastic transitions, with the transition frequency decreasing as the system size increases. At higher densities and lower noise levels, the PF state disappears, leaving the APF state as the sole ordered liquid phase. In contrast, when the inter-species interaction is non-reciprocal, instead of parallel and anti-parallel flocking, the system exhibits chiral motion~\cite{NRVM}.

Natural environments are inherently heterogeneous, where multi-species swarms resemble moving ecosystems~\cite{ben2016multispecies}, and this heterogeneity in mixed populations influences dynamics across scales. Heterogeneity is a natural feature of collective behavior and exists even within a single species due to individual behavioral differences. Examples range from individual fish adjusting their behavior in groups~\cite{Herbert2013colmove} to the effect of cell aspect ratio on swarming bacteria~\cite{ilkanaiv2017effect,Peled2021bacteria}. The influence of such individual-level heterogeneity on collective behavior gets further amplified in multi-species systems, where inter-species differences introduce additional complexity. For instance, the ratio of two swarming bacterial species populations has been found to influence dynamics at all scales, from the microscopic speed distribution to mesoscopic vortex sizes and macroscopic colony structure~\cite{Jose2022mixedspecies}. In mixed‑species bacterial swarms, the population ratio can also dictate local segregation~\cite{Natan2022bacteria}. Further examples of heterogeneous systems of self-propelled agents include agents with varying motility~\cite{zuo2020dynamic,kolb2020active,pattanayak2020speed,Forget2022hetmol,maity2023spontaneous}, diffusivity~\cite{pigolotti2014selective}, responsiveness to external cues~\cite{book2017modeling}, interparticle and inter-species  interactions~\cite{khodygo2019homogeneous,lardet2025flocking}, temporal characteristics of the heterogeneity~\cite{khelfa2022heterogeneity}, and sensitivities to external noise~\cite{Ariel2014HetSPP,Netzer20191hetpop,ilker2020phase}. Environmental heterogeneity also plays a pivotal role in shaping collective motion across diverse systems. Examples range from bacteria adapting to light cues~\cite{Wilde2017prokaryotes, FrangipaneEcoli} and active Brownian particles in spatially varying activity landscapes~\cite{andreas2021microswimmer,auschra2021polarization,Wysocki_2022} to binary chiral particles under complex environmental noise~\cite{BCVM_Chirality_Huang2024}, run-and-tumble disks driven through a random obstacle array~\cite{reichhardt2018avalanche} and self-trapping of active particles in disordered media~\cite{saavedra2024self}. Remarkably, topological flocking models maintain long-range order even in spatially heterogeneous environments~\cite{rahmani2021topological}, in contrast to their metric counterparts. These insights highlight the profound role of heterogeneity in governing mesoscale dynamics across natural and synthetic active matter systems.

Motivated by the importance of heterogeneities in multi-species flocks, we consider in this paper the reciprocal TSVM~\cite{SwarnajitTSVM} with (a) {\it population heterogeneity}, where the two species have different densities; (b) {\it motility heterogeneity}, where particles of the two species differ in velocity; (c) {\it spatial heterogeneity or activity landscape}, involving two spatially segregated regions with counteracting motility heterogeneities, one species has a higher velocity in one region and a lower velocity in the other; and (d) {\it noise heterogeneity} where the two species experience different external noise. We aim to investigate how these heterogeneities influence collective motion and pattern formation in the system, particularly their influence on the emergence and stability of the PF and APF dynamical states. Note that, varying the interaction strengths instead of density, velocity, or noise, one reaches a whole new class of TSVM with completely different behavior~\cite{NRVM}, which is not within the scope of the present study.

\section{Model}
\label{model}

\begin{figure}[t]
    \includegraphics[width=\columnwidth]{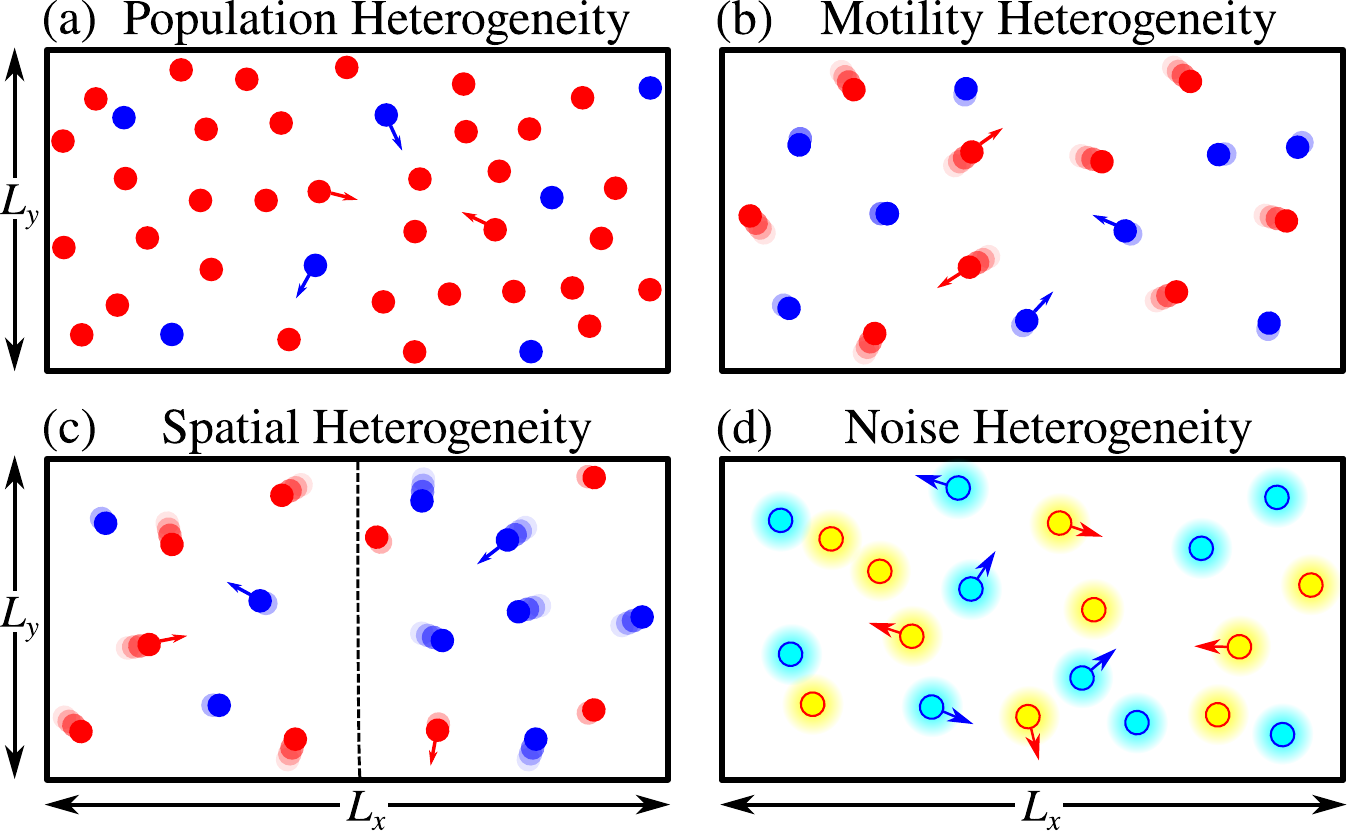}
    \caption{(color online) {\it Schematic of the different heterogeneities applied on the TSVM.} Particles of species A (B) are represented by red (blue) balls. (a) Population heterogeneity: $N_{\rm A} \neq N_{\rm B}$ with $v_{\rm A} = v_{\rm B}$ and $\eta_{\rm A} = \eta_{\rm B}$; (b) Motility heterogeneity: $v_{\rm A} \neq v_{\rm B}$ with $N_{\rm A} = N_{\rm B}$ and $\eta_{\rm A} = \eta_{\rm B}$; (c) Spatial heterogeneity: $v_{\rm A} > v_{\rm B}$ in left region and $v_{\rm A} < v_{\rm B}$ in right region, where the dotted line represents the inter-region interface, $N_{\rm A} = N_{\rm B}$ and $\eta_{\rm A} = \eta_{\rm B}$; (d) Noise heterogeneity: $\eta_{\rm A} \neq \eta_{\rm B}$ with $N_{\rm A} = N_{\rm B}$ and $v_{\rm A} = v_{\rm B}$, i.e. the red particles experience a higher noise amplitude (a ``hotter'' environment) compared to the blue particles (a ``colder'' environment). In all subsequent snapshots, $L_x$ and $L_y$ represent horizontal and vertical system sizes, respectively.}
    \label{fig:tsvm_het_scheme}
\end{figure}

Consider $N_{\rm{A}} (N_{\rm{B}})$ point-like self-propelled particles of species A (B) in a two-dimensional geometry of dimension $L_{x} \times L_{y}$ with periodic boundary conditions. The total number of particles is then $N = N_{\rm{A}}+N_{\rm{B}}$. The position and orientation vectors associated with each particle are $\bm{r}_i^t = \left(x_{i}^t, y_{i}^t\right)$ and $\bm{\sigma}_i^t = \left(\cos \theta_{i}^t, \sin \theta_{i}^t\right)$ respectively, where $\theta_{i}^t$ $\in \left[-\pi,\pi\right]$ is the orientation angle representing the self-propulsion direction of the particle. A static Ising-like spin variable, $s_i = \pm 1$ is used to define the species of the particle, $s_{i} = +1$ for A particles and $s_{i} = -1$ for B particles. Particles belonging to species A and B move with a constant speed $v_{\rm{A}}$ and $v_{\rm{B}}$, respectively, in the direction of $\bm{\sigma}_i$. 

At each discrete time step $\Delta t$, the $i^{\rm th}$ particle interacts with neighboring particles within a circular neighborhood of radius $R$, denoted by $\mathcal{N}_i$. Following these interactions, we obtain the orientation vector of the $i^{\rm th}$ particle as a spin-weighted sum of orientation vectors of neighboring particles after time $t$:

\begin{equation}
\bm{\bar{\sigma}}_i^t \;=\;
\frac{\displaystyle\sum_{j\in N_i}J_{ij}\,\sigma_j^t}
     {\bigl\lvert\sum_{j\in N_i}J_{ij}\,\sigma_j^t\bigr\rvert}\,
\label{eq:3}
\end{equation}

where $J_{ij}=s_{i} s_{j}$ is the exchange coupling between the particles $i$ and $j$. $J_{ij}=1$ signifies an intra-species ferromagnetic interaction whereas $J_{ij}=-1$ signifies inter-species anti-ferromagnetic interaction. Thus, the orientation angle of the $i^{\rm th}$ particle after time $t$ gets updated in the following way:

\begin{equation}
\theta_i^{t+\Delta t}
=\arg\bigl(\bm{\bar{\sigma}}_i^t\bigr)+\eta_i\,\xi_i^t
\ ,
\label{eq:1}
\end{equation}

where $\xi_{i}^{t}$ is a scalar noise uniformly distributed in $[-\pi, \pi]$ and uncorrelated for all sites and times: $\langle  \xi_{i}^{t}\rangle\ =\ 0$, and $\langle \xi_{i}^{t} \xi_{j}^{s}\rangle \sim \delta_{ts} \delta_{ij}$, and $\eta_i$ is the parameter controlling the noise strength. $\eta_i=\eta_{\rm A}$ for A particles and $\eta_i=\eta_{\rm B}$ for B particles.

On the other hand, considering that orientation vector at time $t+\Delta t$ will be $\bm{\sigma}_{i}^{t+\Delta t}$, the position update of the $i^{\rm th}$ particle after time $t$ is given by:
\begin{align}
\label{eq:2}
\bm{r}_{i}^{t+\Delta t}\ =\ \bm{r}_{i}^{t}\  +\ v_i \bm{\sigma}_{i}^{t+\Delta t} \Delta t \, ,
\end{align}
with $v_i=v_{\rm A}$ for A particles and $v_i=v_{\rm B}$ for B particles.

The model parameters include the species densities $\rho_s = N_s / L_x L_y$ (for $s \in {\rm \{A, B\}}$), the noise strengths $\eta_s$, and the velocity moduli $v_s$. For simplicity, we consider the following when these parameters are uniform across species: if all particles share the same speed, we set $v_{\rm A} = v_{\rm B} = v_0$, if the noise strengths are identical, we take $\eta_{\rm A} = \eta_{\rm B} = \eta$, and if the species densities are equal, we define $\rho_{\rm A} = \rho_{\rm B} = \rho / 2$, where $\rho = N / L_x L_y$ is the total particle number density of the system. We mostly consider a rectangular simulation box of high aspect ratio $L_{x} / L_{y}\ =\ 8$, $R = 1$, and $\Delta t =1$, unless stated otherwise. 

\section{Simulation details}
\label{simul_details}
Assigning random initial positions and orientations to the particles, numerical simulations of the stochastic process are performed with parallel updates of orientations and positions of the $N$ particles. The system evolves under three control parameters: average particle density, external noise, and particle velocity. After initialization, we equilibrate the system for $t_{\rm{eq}} = 10^5$ and then measure various quantities until the maximum simulation time, $t_{\rm{max}} = 10^6$.

The TSVM~\cite{SwarnajitTSVM} typically exhibits three phases: a low-density, high-noise gas phase, a low-noise, high-density liquid phase, and an intermediate liquid-gas coexistence region which can further be classified into two categories: (i) PF or \enquote{parallel flocking} state where bands of two species move in the same direction and (ii) APF or \enquote{anti-parallel flocking} state where A and B bands move in the opposite direction. To characterize the collective motion of the A and B species, the following order parameters are introduced~\cite{SwarnajitTSVM}:
\begin{equation}
\bm{v}_{+}^{t}\ =\ \frac{1}{N_{\rm A}} \sum_{i \in \rm {A}} \bm{\sigma}_{i}^{t},\; \; \; \; \; \bm{v}_{-}^{t}\ =\ \frac{1}{N_{\rm B}} \sum_{i \in \rm {B}} \bm{\sigma}_{i}^{t} \, .
\label{eq:5}
\end{equation}
Let $v_{\pm}=|\bm{v}_{\pm}^{t}|$, then $\langle v_{\pm} \rangle$ are the flocking order parameters, where $\langle (...) \rangle$ denotes the steady state time average and the ensemble average over independent runs. The PF and APF states are distinguished by:
\begin{subequations}
\label{eq:6}
\begin{equation}
\begin{aligned}[b]
\bm{v}_{s}^{t}\ =\ \frac{1}{N} \sum_{i=1}^{N} \bm{\sigma}_{i}^{t}\ &=\ \frac{1}{N} \left[N_{\rm A} \bm{v}_{+}^{t}\ +\ N_{\rm B} \bm{v}_{-}^{t}\right] \ ,
\end{aligned}
\label{eq:6a}
\end{equation}    
\begin{equation}
\begin{aligned}[b]
\bm{v}_{a}^{t}\ =\ \frac{1}{N} \sum_{i=1}^{N} s_{i}^{t} \bm{\sigma}_{i}^{t}\ &=\ \frac{1}{N} \left[N_{\rm A} \bm{v}_{+}^{t}\ -\ N_{\rm B} \bm{v}_{-}^{t}\right] \ .
\label{eq:6b}
\end{aligned}
\end{equation}
\end{subequations}
Using $v_{s(a)}=|\bm{v}_{s(a)}^{t}|$ from Eqs.~\eqref{eq:6}, $\langle v_{s} \rangle$ and $\langle v_{a} \rangle$ are defined as the order parameters of the PF and APF states, respectively. In the thermodynamic limit, $\langle v_{s} \rangle > 0$ and $\langle v_{a} \rangle = 0$ in the PF state and $\langle v_{s} \rangle = 0$ and $\langle v_{a} \rangle > 0$ in the APF state.

\section{Results}
\label{num_results}

In this section, we present numerical results of the TSVM under the following heterogeneities: 

{\bf A.} {\it Population heterogeneity} where $N_{\rm A} \neq N_{\rm B}$ but all particles move with the same velocity $v_0$ and experience the same noise $\eta$. 

{\bf B.} {\it Motility heterogeneity} where equal population $(N_{\rm A} = N_{\rm B})$ of A and B species respectively move with velocities $v_{\rm A}$ and $v_{\rm B}$ $(v_{\rm A} \neq v_{\rm B})$ under the same external noise $\eta$.

{\bf C.} {\it Spatial heterogeneity} where particle velocities are space-dependent. In one region of the simulation box, A moves faster than B ($v_{\rm A} > v_{\rm B}$), but in the other region, B has the greater velocity ($v_{\rm A} < v_{\rm B}$). Here also $N_{\rm A} = N_{\rm B}$ and $\eta_{\rm A} = \eta_{\rm B} = \eta$.

{\bf D.} {\it Noise heterogeneity} where one species is subjected to higher noise, analogous to a hotter environment while the other experiences a markedly reduced noise level, mimicking a colder regime, $\eta_{\rm A} \neq \eta_{\rm B} \, \text{with} \, N_{\rm A} = N_{\rm B}$ and $v_{\rm A} = v_{\rm B} = v_0$.

\subsection{Population heterogeneity}
\label{PH-TSVM}

\begin{figure}[t]
    \centering
    \includegraphics[width=\columnwidth]{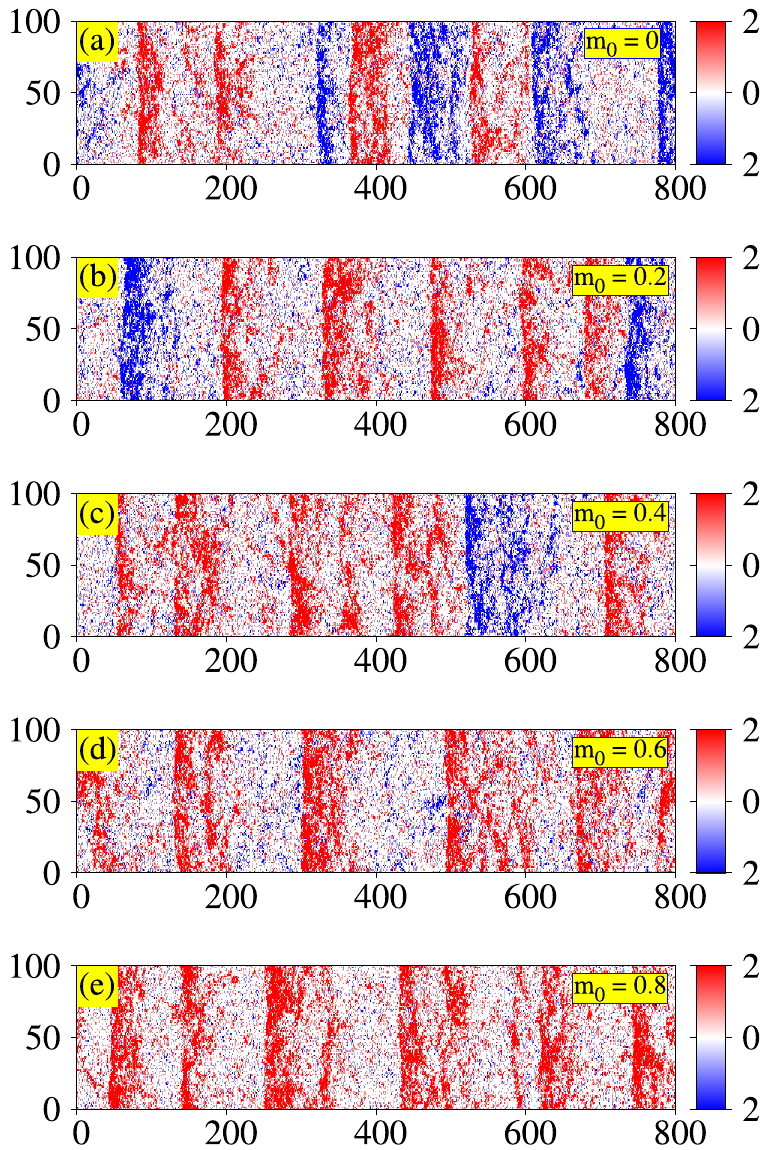}
    \caption{(color online) {\it Steady-state snapshots for varying population heterogeneity.} Particles of species A (B) are represented with red (blue) dots, and a local particle density is color-coded according to the color bar. (a) The homogeneous TSVM features an equal number of bands for A and B species. (b--c) The band number of species B decreases with increasing $m_{0}$. (d--e) B-particles can not form bands due to scarcity in numbers. Parameters: $\rho = 1$, $\eta = 0.3$, $v_{0}=0.5$, $L_{x}=800$, and $L_{y}=100$. A movie (\texttt{movie1}) of the same can be found at Ref.~\cite{zenodo}.}
    \label{fig:phtsvm_snap}
\end{figure}

First, we consider the TSVM with different populations of the two species, i.e., $N_{\rm A} \neq N_{\rm B}$. The strength of the heterogeneity is characterized by $m_{0} = (N_{\rm A}-N_{\rm B})/N$, and without any loss of generality, we only consider $m_0>0$. Fig.~\ref{fig:phtsvm_snap} shows the steady-state snapshots of the TSVM at $t=10^6$ and for increasing $m_0$, where the species with a greater population (here, A) exhibit more traveling high-density liquid bands than those with a lesser population (here, B). As $m_0$ increases, species B eventually fails to form any bands due to an insufficient number of particles and forms a solo gaseous state of B-particles for $m_0 \geq 0.6$. For the sake of generality, from now on, we will refer to species A $\left(N_{\rm A} \geq N_{\rm B}\right)$ as the {\it majority} species and species B as the {\it minority} species. The system thus displays a transition from a PF state at $m_0=0$ to a majority-species dominated single-species flocking (SSF) state characteristic of the VM~\cite{Solon2015phase} at $m_0=0.8$.

It is important to highlight that the microphase-separated band configurations displayed in Fig.~\ref{fig:phtsvm_snap} correspond to stable steady states that persist over long time scales. For each value of $m_0$, once the system undergoes phase separation, both the number and the form of the bands remain essentially constant, with no observable change up to times of at least $t=10^7$ (see Appendix~\ref{appA}). This indicates that the system undergoes minimal, if any, coarsening after the initial formation of bands. Although the resulting structures may resemble smectic order due to the apparent regularity in band spacing, significant fluctuations in inter-band distances preclude any sustained long-range translational order. Similar to the standard Vicsek model (VM), the TSVM displays giant number fluctuations (GNF)~\cite{SwarnajitTSVM}, which are fundamentally incompatible with smectic order~\cite{chen2013universality,adhyapak2013live}. These fluctuations disrupt crystalline arrangements and inhibit translational symmetry breaking. Consequently, even if quasi-periodic band spacing may appear over certain time intervals or regions, the system retains a dynamic, fluctuating banded structure rather than forming a true smectic or crystalline phase (see Appendix~\ref{appA}).

\begin{figure}[t]
    \includegraphics[width=\columnwidth]{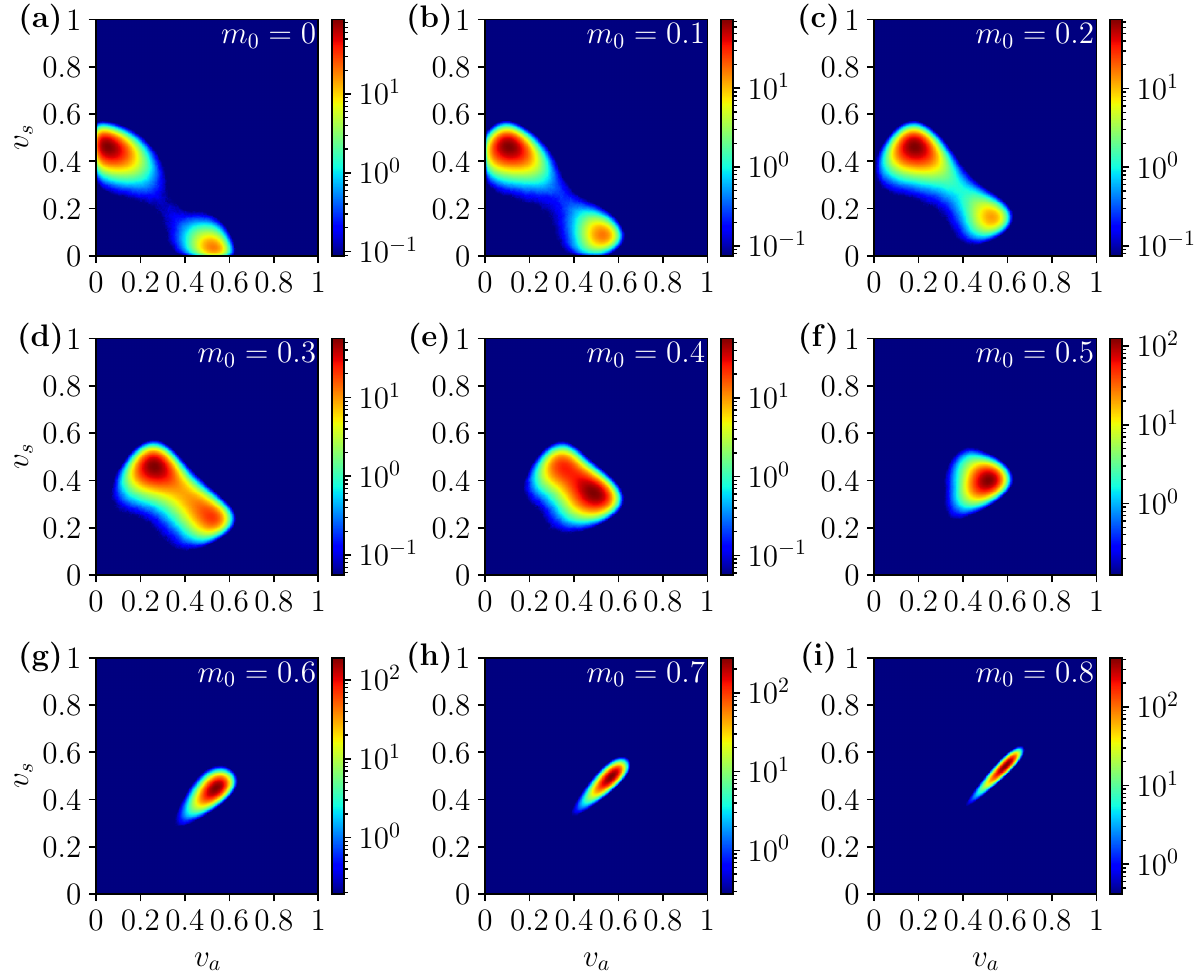}
    \caption{(color online) {\it Probability distribution $P(v_{s},v_{a})$ for varying population heterogeneity.} (a) Representation of the homogeneous TSVM ($N_{\rm A} = N_{\rm B}$) exhibiting stochastic switching between the PF and APF states. (b--i) The two peaks progressively converge as $m_0$ increases, signifying a collapse into a single state. Parameters: $\rho=0.5$, $\eta=0.24$, $v_{0}=0.5$, $L_{x} = 256$, and $L_{y} = 32$. A movie (Movie S1) of the same can be found at Ref.~\cite{SM}.}
    \label{fig:phtsvm_hist}
\end{figure}

In Fig.~\ref{fig:phtsvm_hist}, we present the probability distribution $P(v_{a},v_{s})$ for increasing $m_0$ constructed from the steady state time series of $v_{s(a)}=|\bm{v}_{s(a)}^{t}|$ across several independent realizations. For $m_0=0$ [Fig.~\ref{fig:phtsvm_hist}(a)], we observe a two-peak structure characteristic of the homogeneous TSVM, indicating the coexistence of PF $(v_s > v_a)$ and APF $(v_a>v_s)$ states, with fluctuation-induced stochastic switching between these dynamical states in the steady state~\cite{SwarnajitTSVM}. However, as $m_0$ increases, the peaks gradually converge [Fig.~\ref{fig:phtsvm_hist}(b--e)] until they merge into a single peak [Fig.~\ref{fig:phtsvm_hist}(f--i)], signaling the collapse of the PF and APF states into an SSF state with $v_s \sim v_a$.

\begin{figure}[t]
\includegraphics[width=\columnwidth]{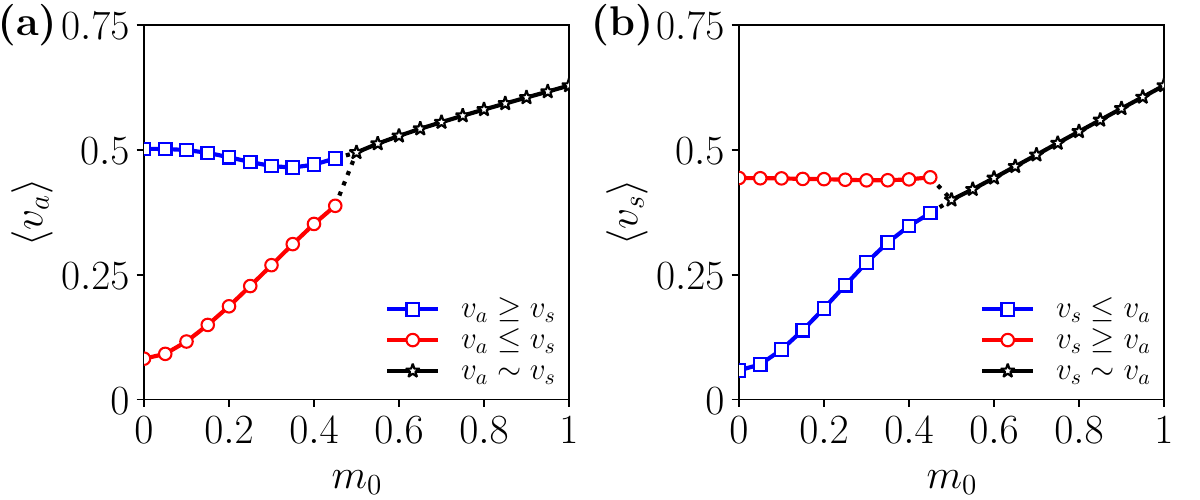}
\caption{(color online) {\it Order parameters for population heterogeneity.} $\langle v_{a} \rangle$ and $\langle v_{s} \rangle$ are shown in the restricted APF (blue squares), PF (red circles), and SSF (black stars) ensembles for varying $m_0$. (a) $\langle v_{a} \rangle$ remains relatively constant in the APF ensemble but increases monotonically in the PF ensemble. (b) $\langle v_{s} \rangle$ increases monotonically in the APF ensemble while remaining relatively constant in the PF ensemble. Parameters: $\rho = 0.5$, $\eta = 0.24$, $v_{0}=0.5$, $L_{x} = 256$, and $L_{y} = 32$.}
\label{fig:phtsvm_op}
\end{figure}

To characterize the behavior of the PF and APF states separately with $m_0$, in Fig.~\ref{fig:phtsvm_op}, we measure the time-averaged order parameters $\langle v_{s} \rangle$ and $\langle v_{a} \rangle$, averaging only over the ensemble defined by $v_a \leq v_s$ or only over the ensemble defined by $v_a \geq v_s$. Fig.~\ref{fig:phtsvm_op}(a) shows that $\langle v_{a} \rangle$ increases monotonically with $m_0$ in the former, and the system is in a PF state, while it remains nearly constant in the latter, where APF behavior dominates. Conversely, Fig.~\ref{fig:phtsvm_op}(b) shows the opposite trend for $\langle v_{s} \rangle$. The emerging general picture is that the order parameter (e.g., $\langle v_{s} \rangle$) associated with the less prevalent dynamical state (e.g., PF behavior in the $v_{a} \geq v_{s}$ ensemble) approaches that of the dominant state (e.g., $\langle v_{a} \rangle$ representing APF behavior in the same ensemble) as the population heterogeneity $m_0$ increases. This convergence signifies a collapse into a single state near $m_0 \sim 1$, corresponding to the VM limit. In the pure VM limit, $v_s$ and $v_a$ are equivalent $(v_s \simeq v_a)$, and the system can be described by a single Vicsek order parameter~\cite{Vicsek}. Note that the PF and APF dynamical states are only meaningful when both species form well-defined high-density liquid bands that move either parallel or anti-parallel to each other. When one species becomes significantly more abundant than the other, the concept breaks down, as the minority species can no longer form bands.

To understand the results shown in Fig.~\ref{fig:phtsvm_op}, let us consider the time-averaged order parameters presented in Eqs.~\eqref{eq:6}:
\begin{subequations}
\label{eq:7}
\begin{equation}
\begin{aligned}[b]
\langle \bm{v}_{s} \rangle  = \frac{1+m_0}{2} \langle \bm{v}_{+} \rangle + \frac{1-m_0}{2} \langle \bm{v}_{-} \rangle  \equiv {\bf m_A} + {\bf m_B} ,
\end{aligned}
\label{eq:7a}
\end{equation}    
\begin{equation}
\begin{aligned}[b]
\langle \bm{v}_{a} \rangle  = \frac{1+m_0}{2} \langle \bm{v}_{+} \rangle  - \frac{1-m_0}{2} \langle \bm{v}_{-} \rangle  \equiv {\bf m_A} - {\bf m_B} ,
\label{eq:7b}
\end{aligned}
\end{equation}
\end{subequations}
where ${\bf m_A}$ and ${\bf m_B}$ are the average magnetization vectors of species A and B, respectively. If species $s$ creates a band in the coexistence region then $\rho_{\rm gas} \leq \rho_s \leq \rho_{\rm liq}$, where $\rho_{\rm gas}$ and $\rho_{\rm liq}$ are respectively the gas and liquid binodal densities of a single species, with a liquid fraction (of species $s$) defined by
\begin{equation}
\phi_s\ =\ \frac{\rho_s\ -\ \rho_{\rm gas}}{\rho_{\rm liq} - \rho_{\rm gas}}\ .
\label{eq:10}
\end{equation}

Let $m_{\rm liq}$ be the magnetization of the liquid phase, related to $\rho_{\rm liq}$ and independent of the total density $\rho$. Hence, one can express the modulus of the individual species magnetizations as $m_s = m_{\rm liq} \phi_s$. However, the species $s$ stops flocking when $\rho_s<\rho_{\rm gas}$, meaning that one of the two species remains in the gas phase when $m_0>1-2\rho_{\rm gas}/\rho$. Moreover, we can deduce that the system is in the gas phase when $\rho < 2 \rho_{\rm gas}/(1+m_0)$, and in the SSF state when 
\begin{equation}
\label{eq:SSFstate}
\frac{2 \rho_{\rm gas}}{1+m_0} < \rho < \frac{2 \rho_{\rm gas}}{1-m_0}.
\end{equation}
 
For $m_0<1-2\rho_{\rm gas}/\rho$, species A and B are either in a PF or an APF state. For a PF state, the magnetization vectors ${\bf m_A}$ and ${\bf m_B}$ are parallel, then Eqs.~\eqref{eq:7} can be rewritten as
\begin{subequations}
\label{eq:13}
\begin{equation}
\langle v_{s} \rangle = m_{\rm A}+m_{\rm B} =  m_{\rm liq}\ \frac{\rho -2 \rho_{\rm gas}}{\rho_{\rm liq} - \rho_{\rm gas}} \, ,
\label{eq:13a}
\end{equation}
\begin{equation}
\langle v_{a} \rangle = |m_{\rm A}-m_{\rm B}| =  m_{\rm liq}\ \frac{\rho m_{0}}{\rho_{\rm liq} - \rho_{\rm gas}} \, ,
\label{eq:13b}
\end{equation}
\end{subequations}
using the expression of $\phi_s$ from Eq.~\eqref{eq:10}. This implies that $\langle v_s\rangle$ is independent of species fraction $m_0$ [see Fig.~\ref{fig:phtsvm_op}(b)], whereas $\langle v_a \rangle$ linearly increases with $m_0$ [see Fig.~\ref{fig:phtsvm_op}(a)].

Similarly, for the APF state, where the magnetization vectors ${\bf m_A}$ and ${\bf m_B}$ are anti-parallel, one can show that
\begin{subequations}
\label{eq:14}
\begin{equation}
\langle v_{s} \rangle = |m_{\rm A}-m_{\rm B}| =  m_{\rm liq}\ \frac{\rho m_{0}}{\rho_{\rm liq} - \rho_{\rm gas}} \, ,
\label{eq:14a}
\end{equation}
\begin{equation}
\langle v_{a} \rangle = m_{\rm A}+m_{\rm B} =  m_{\rm liq}\ \frac{\rho - 2 \rho_{\rm gas}}{\rho_{\rm liq} - \rho_{\rm gas}} \, .
\label{eq:14b}
\end{equation}
\end{subequations}
This implies that $\langle v_s\rangle$ linearly increases with $m_0$  [see Fig.~\ref{fig:phtsvm_op}(b)] and $\langle v_a\rangle$ remains constant [see Fig.~\ref{fig:phtsvm_op}(a)].

For $m_0>1-2\rho_{\rm gas}/\rho$, species A (majority species) form bands while species B (minority species) enters the gas phase ($\rho_{\rm B} < \rho_{\rm gas}$) implying ${\bf m_B} = 0$. Then, rewriting Eqs.~\eqref{eq:7} using the expression of $\phi_{\rm A}$ in Eq.~\eqref{eq:10} we obtain:
\begin{equation}
\langle v_{s} \rangle = \langle v_{a} \rangle = \frac{m_{\rm liq}}{2} \left( \frac{\rho m_0}{\rho_{\rm liq} - \rho_{\rm gas}} + \frac{\rho - 2 \rho_{\rm gas}}{\rho_{\rm liq} - \rho_{\rm gas}}\right) \, .
\label{eq:12}
\end{equation}
Thus, at large $m_0$ values, both $\langle v_s \rangle$ and $\langle v_a \rangle$ vary in an affine manner with $m_0$, as shown in Fig.~\ref{fig:phtsvm_op}.

\begin{figure}[t]
\centering
\includegraphics[width=\columnwidth]{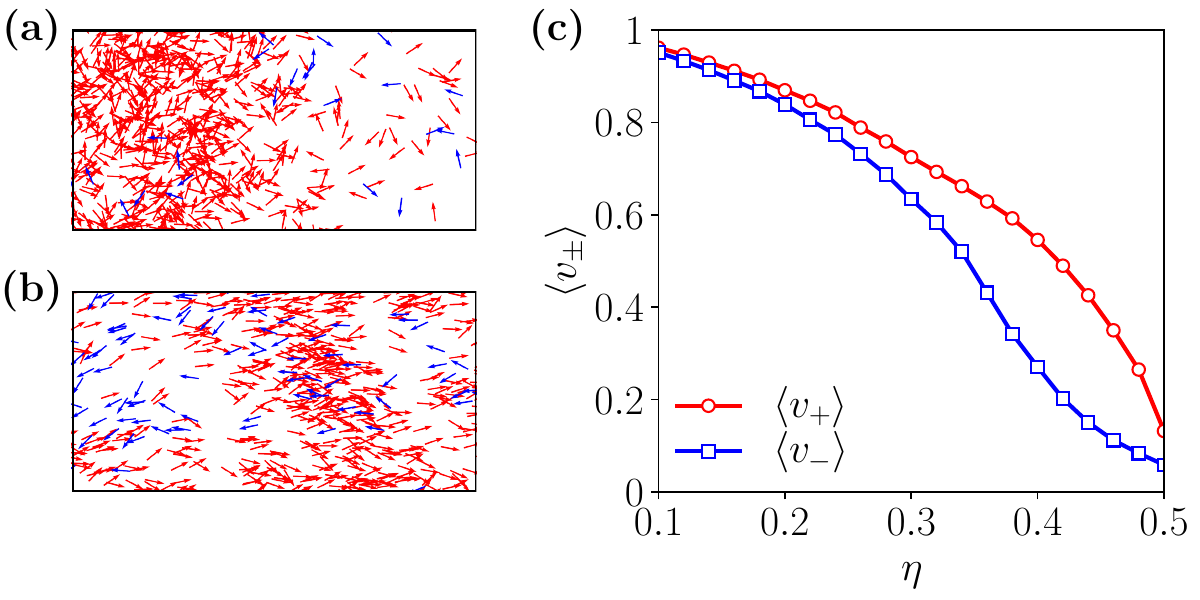}
\caption{(color online) {\it Single species and both species flocking at large population heterogeneity.} Snapshots of (a) SSF $(\eta=0.45)$ and (b) flocking of both species in an APF state $(\eta=0.2)$ are shown for a $20 \times 10$ section of a $800 \times 100$ simulation box. Red and blue arrows represent the orientation of A and B particles, respectively. (c) The time- and ensemble-averaged order parameters $\langle v_\pm \rangle$ as a function of $\eta$; $L_x=256$, $L_y=32$. Parameters: $\rho=2$, $v_0=0.5$, and $m_0=0.9$.} 
\label{fig:phtsvm_bsf_ssf}
\end{figure}

However, at high $m_0$, it is crucial to understand its effect on the collective dynamics of the minority species. At higher noise $(\eta=0.45)$, the minority species transitions into a disordered gaseous state due to its low density, while the majority species forms flocking bands [see Fig.~\ref{fig:phtsvm_bsf_ssf}(a)]. This defines the single-species flocking (SSF) state, analogous to the VM flocking behavior. At lower noise $(\eta=0.2)$, the minority species, although it cannot form bands as its density remains below  $\rho_{\rm gas}$, now exhibits a directed motion where both species flock in an APF state [see Fig.~\ref{fig:phtsvm_bsf_ssf}(b)]. In this state, the majority species generally remain in the liquid state due to low noise, while the minority species anti-aligns with the majority species due to the reciprocal anti-ferromagnetic interaction and forms an APF state. To further examine the impact of \(\eta\) on the collective dynamics of the majority (A) and minority (B) species under strong heterogeneity $(m_0 = 0.9)$, we plot the corresponding order parameters, $\langle v_+ \rangle$ and $\langle v_- \rangle$, against $\eta$ in Fig.~\ref{fig:phtsvm_bsf_ssf}(c). At higher noise $(\eta \gtrsim 0.3)$, $\langle v_- \rangle$ decays more sharply than $\langle v_+ \rangle$ due to the much lower density of species B, signifying the SSF state. For lower $\eta$, although $\langle v_- \rangle < \langle v_+ \rangle$, the magnitude of $\langle v_- \rangle$ indicates that B-particles also exhibit an ordered state.

\begin{figure}[t]
\centering
\includegraphics[width=\columnwidth]{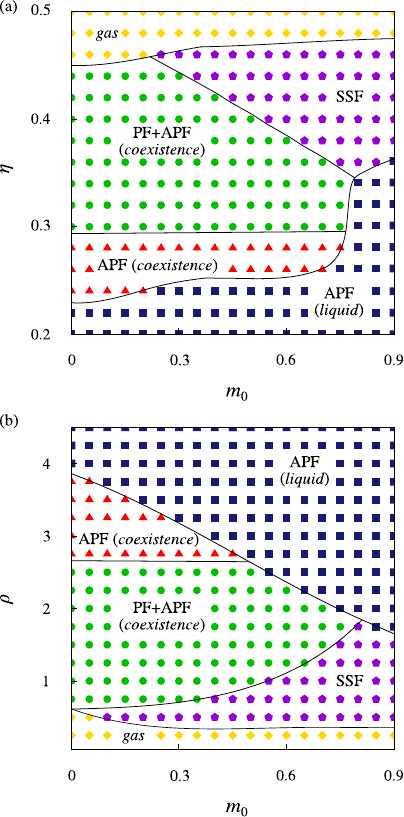}
\caption{(color online) {\it Phase diagrams for population heterogeneity.} (a) $\eta - m_0$ phase diagram for $\rho = 2$; (b) $\rho - m_0$ phase diagram for $\eta = 0.3$. For both cases, the velocity modulus is fixed ($v_0 = 0.5$). Alongside the homogeneous TSVM phases, a single-species flocking state emerges. The boundary lines are included as visual guides.} 
\label{fig:phtsvm_phasediag}
\end{figure}

Fig.~\ref{fig:phtsvm_phasediag} presents the $\eta-m_0$ and $\rho-m_0$ phase diagrams, constructed with the aid of snapshots, density profiles, and the order parameters defined in Eq.~\eqref{eq:5}. In Fig.~\ref{fig:phtsvm_phasediag}(a), for low $m_0$ values, we recover the phase behavior of the homogeneous TSVM~\cite{SwarnajitTSVM}. As $m_0$ increases, reflecting greater population heterogeneity, the majority species gathers enough particles to form the SSF state $(m_0 > 0.3)$, at high noise levels, as the minority species remains in a completely gaseous state. As the noise is reduced, the threshold of $m_0$ for the transition between the PF+APF state and the SSF state increases, since the minority species can now exhibit collective motion for a smaller species density. At low noise $(\eta \lesssim 0.34)$, the minority species exhibit directed motion even at high heterogeneity $(m_0 > 0.8)$ and form an APF liquid state.

In Fig.~\ref{fig:phtsvm_phasediag}(b), the SSF state is observed roughly within the range $0.5 \lesssim \rho \lesssim 1.5$, inside an interval given by Eq.~\eqref{eq:SSFstate}. At a fixed $m_0$, increasing $\rho$ increases the density of both species, allowing the minority species to form bands. Increasing $\rho$ further, we observe a transition from the PF+APF state to APF coexistence, and eventually to the APF liquid, similar to the behavior in the homogeneous TSVM. For strong heterogeneity, minority species band formation is less probable, and we observe a direct transition from the SSF state to the APF liquid state as $\rho$ increases, as mentioned in Fig.~\ref{fig:phtsvm_bsf_ssf}. Note that, depending upon the interplay of $\eta$ and $\rho$, for intermediate $m_0$, the minority species can also form bands while the majority species remain in a liquid state. Such a configuration is not possible in the homogeneous TSVM~\cite{SwarnajitTSVM}.

In summary, strong population heterogeneity ultimately eliminates the PF and APF states, leading to a single unified flocking state (SSF) at high noise (or low density), where the minority species remains in a disordered gas phase. However, at low noise (or high density), the minority species continues to exhibit directed motion, forming an APF-like liquid state.

\subsection{Motility heterogeneity, or unfriendly \enquote{fast} and \enquote{slow} particles}
\label{MH-TSVM}

We next investigate the motility heterogeneity in the TSVM by assigning different particle velocities to the two species $(v_{\rm A} \neq v_{\rm B})$. The key parameter of interest is the relative velocity, $\Delta v = v_{\rm A} - v_{\rm B}$. To maintain symmetry and reciprocity, the velocity modulus of species B is kept constant, $v_{\rm B} = v_0 = 0.5$, while that of species A is varied within the range $v_{\rm A} \in [0,1]$. 

\begin{figure}[t]
    \centering
    \includegraphics[width=\columnwidth]{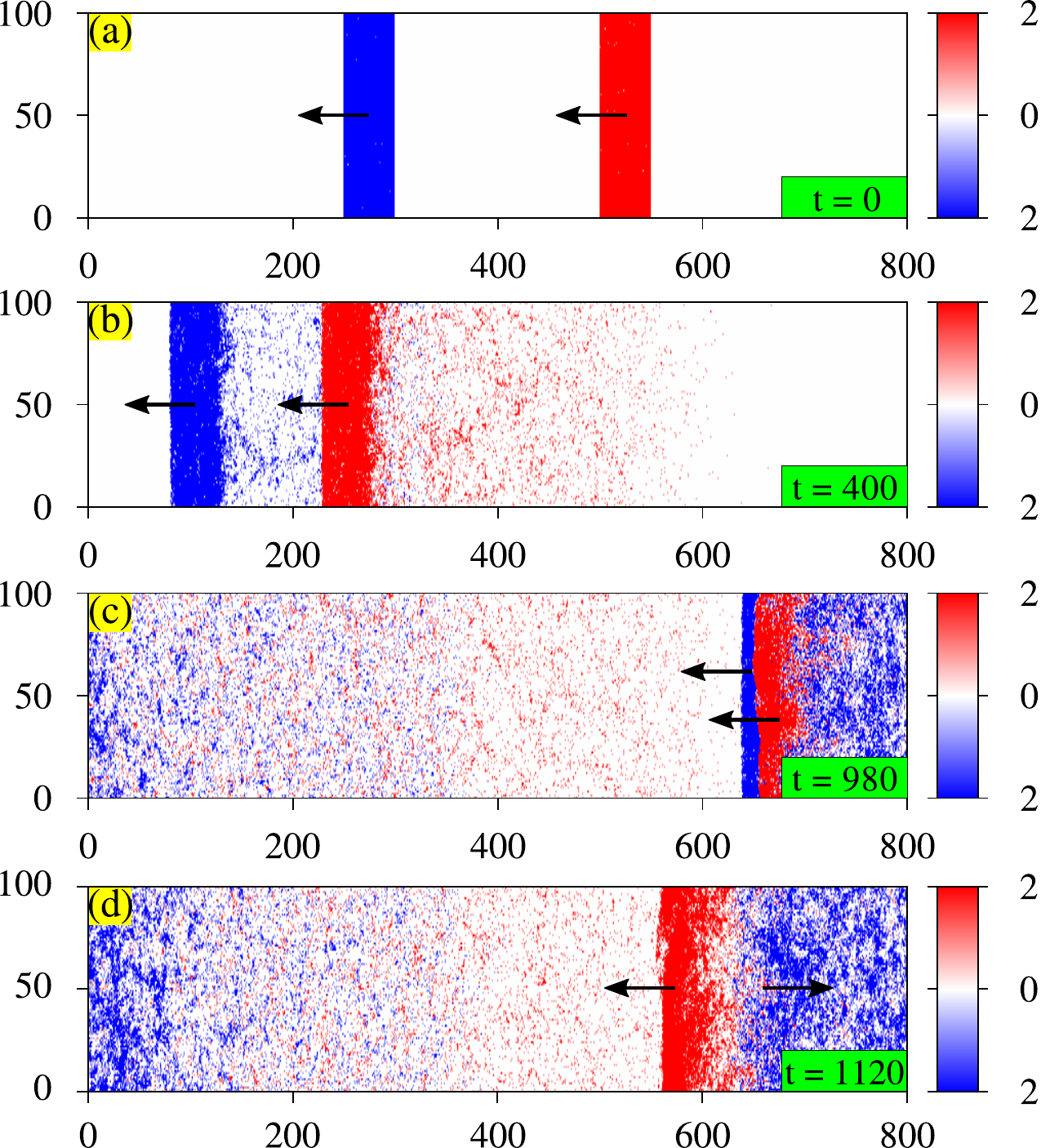}
    \caption{(color online) {\it Time-evolution with motility heterogeneity.} The flocking direction is indicated by black arrows. (a) Particles of species A (red dots) and species B (blue dots) are shown, with local particle density color-coded according to the color bar. (b--c) The faster-moving A-band catches the B-band and collides. (d) Species B reverses direction and forms an APF state. Parameters: $\rho = 1$, $\eta = 0.3$, $v_0=0.5$, $\Delta v = 0.3$, $L_{x}=800$, and  $L_{y}=100$. A movie (\texttt{movie2}) of the same can be found at Ref.~\cite{zenodo}.}
    \label{fig:mhtsvm_snap}
\end{figure}

For $\Delta v = 0.3$, the time evolution of such a system is presented in Fig.~\ref{fig:mhtsvm_snap}. We initialize the system in a PF state by placing a band of high-velocity particles behind a band of low-velocity particles [Fig.~\ref{fig:mhtsvm_snap}(a)] and then allow the system to evolve. Over time, the faster band (species A, red) closes the gap with the slower band (species B, blue) [Fig.~\ref{fig:mhtsvm_snap}(b)] and eventually collides [Fig.~\ref{fig:mhtsvm_snap}(c)]. Upon collision, due to the anti-alignment interaction between the species, the B-particles reverse direction, transitioning to an APF state [Fig.~\ref{fig:mhtsvm_snap}(d)]. As the A-band penetrates the B-band, it gradually reverses the orientation of the B-particles layer by layer. Consequently, after the A-band fully passes through, the previously dense B-band disperses. Notably, the A-band itself does not reverse, as the denser ``head'' of the band dominates the orientation update [Eq.~\eqref{eq:1}], impacting the minority B-particles more than the majority A-particles within the interaction area. If we express the alignment rule in Eq.~\eqref{eq:3} using the variable \(\bm{\alpha}_{i} \equiv s_{i} \bm{\sigma}_i\):
\begin{equation}
\bm{\bar{\alpha}}_{i}^{t} = \sum_{j \in \mathcal{N}_i} s_i^2 s_j \bm{\sigma}_j^t = \sum_{j \in \mathcal{N}_i} s_j \bm{\sigma}_j^t = \sum_{j \in \mathcal{N}_i} \bm{\alpha}_{j}^{t} \, ,
\label{eq:16}
\end{equation}
regardless of species type, each particle aligns its \(\bm{\alpha}\) variable with its neighbors. In the PF state, \(\bm{\alpha}\) vectors are anti-parallel between species, leading to stability only when spatially separated. Motility heterogeneity causes one band to overtake the other (see Fig.~\ref{fig:mhtsvm_snap}), eliminating inter-species segregation and transforming the PF state into the more stable APF state.

Starting from an APF state instead would result in the retention of APF behavior because the APF order is stronger than the PF order in TSVM~\cite{SwarnajitTSVM} due to inter-species anti-ferromagnetic interactions. In the APF state, $\bm{\alpha}$ vectors are parallel, allowing particles to perceive more ``correctly aligned'' neighbors, reducing fluctuations. This highlights the role of velocity asymmetry in driving the reorganization of the bands, which leads to a persistent APF state in the coexistence regime. 

\begin{figure}[t]
    \includegraphics[width=\columnwidth]{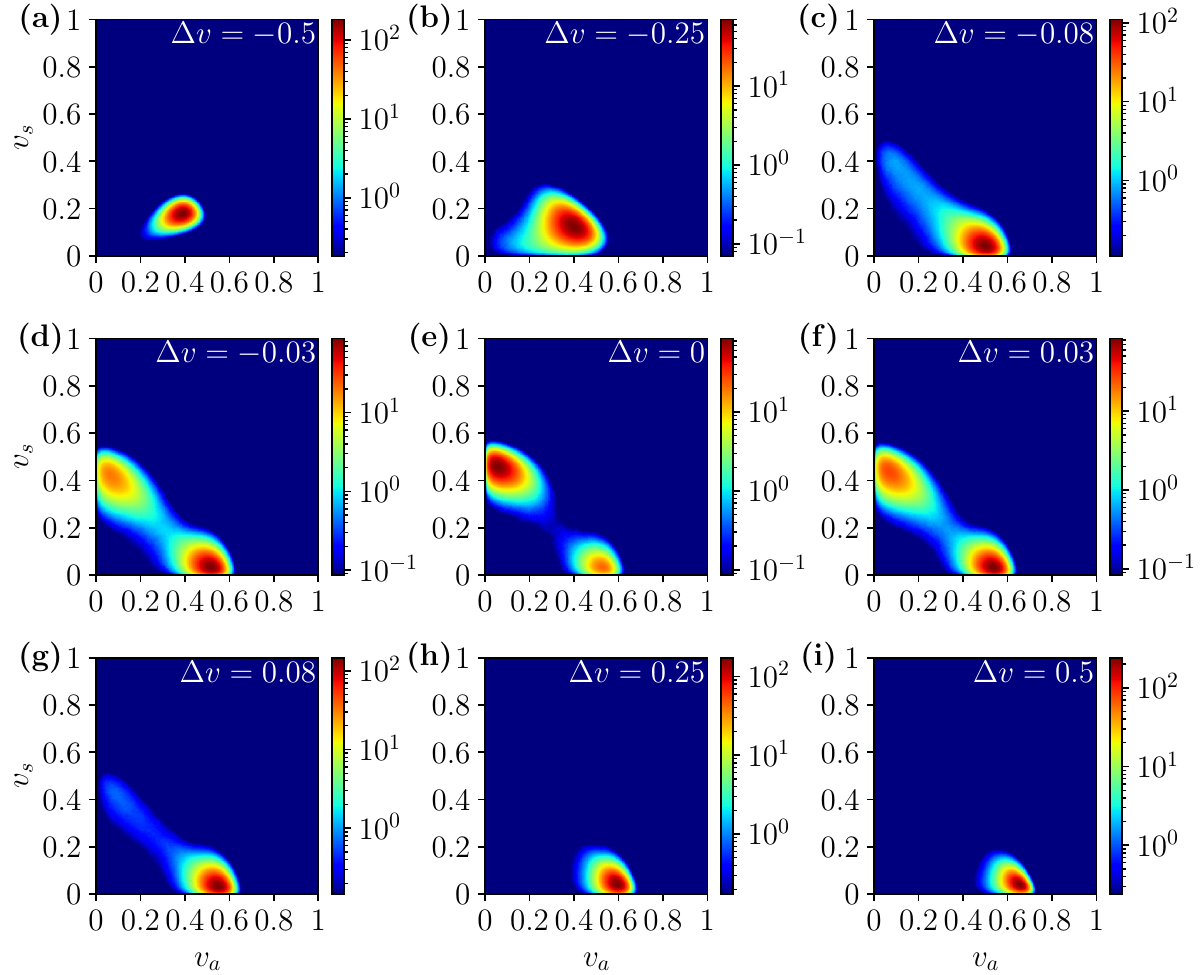}
    \caption{(color online) {\it Probability distribution $P(v_{s},v_{a})$ for varying motility heterogeneity.} (a-c, g-i) At high motility heterogeneity ($|\Delta v| > 0.1$), only APF state remains. (e) Homogeneous TSVM $(\Delta v = 0)$, characterized by the stochastic switching between the PF and APF states. (d, f) PF-APF stochastic switching with a stronger APF for moderate heterogeneity. Parameters: $\rho=0.5$, $\eta=0.24$, $v_0 = 0.5$, $L_{x} = 256$, and $L_{y} = 32$. A movie (Movie S2) of the same can be found at Ref.~\cite{SM}.}
    \label{fig:mhtsvm_hist}
\end{figure}

The probability distribution \(P(v_a, v_s)\) in Fig.~\ref{fig:mhtsvm_hist} clearly demonstrates the dominance of the APF state as \(|\Delta v|\) increases. Near the homogeneous TSVM limit $(\Delta v \sim 0)$, the typical two-peak structure [Fig.~\ref{fig:mhtsvm_hist}(d–f)] is observed, indicating stochastic PF-APF switching in the coexistence regime. For \(|\Delta v| > 0.1\), singular peaks emerge in the APF state region [Fig.~\ref{fig:mhtsvm_hist}(a–c) and Fig.~\ref{fig:mhtsvm_hist}(g–i)], with the mean value of the \(v_a\) order parameter increasing as \(v_{\rm A}\) increases [Fig.~\ref{fig:mhtsvm_hist}(g–i)]. The remaining PF traits $(v_s \neq 0)$ for $|\Delta v| \sim 0.1$ [Fig.~\ref{fig:mhtsvm_hist}(c, g)] suggest a transition from PF+APF to pure APF behavior as the relative velocity \(\Delta v\) increases.

\begin{figure}[t]
    \includegraphics[width=\columnwidth]{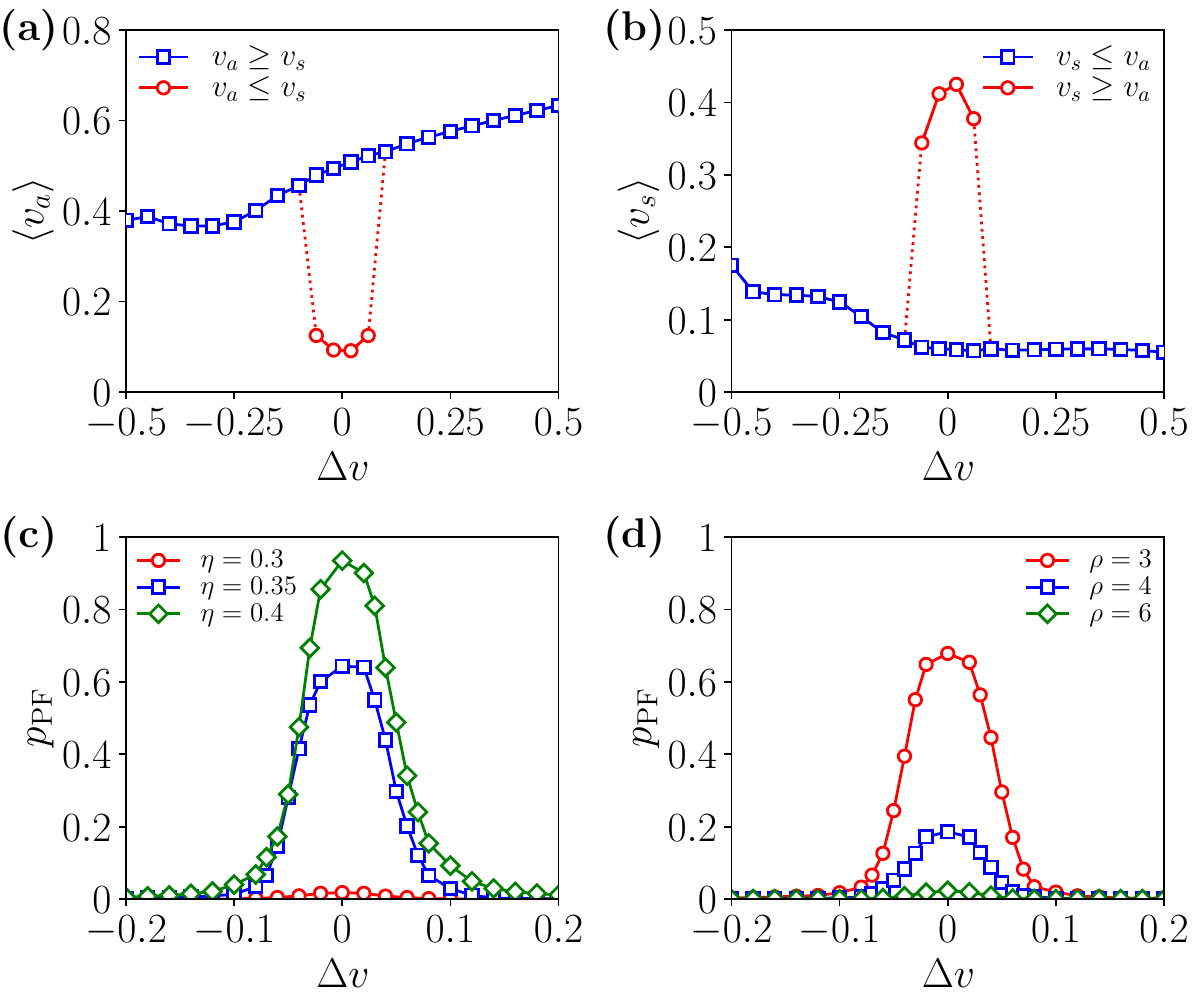}
    \caption{(color online) {\it Order parameters and PF state probability for motility heterogeneity.} (a) $\langle v_{a} \rangle$ and (b) $\langle v_{s} \rangle$ in the restricted APF (blue square) and PF (red circle) ensembles versus $\Delta v$ for $\rho = 0.5$ and $\eta = 0.24$. (c--d) Probability of the PF state ($p_{\rm PF}$) vs $\Delta v$ for (c) varying noise strength $\eta$ [$=0.30$ (circle), $0.35$ (square) and $0.40$ (diamond)] keeping $\rho = 2$ fixed and (d) varying particle density $\rho$ [$=3$ (circle), $4$ (square) and $6$ (diamond)] keeping $\eta = 0.4$ fixed. Parameters: $v_0 = 0.5$, $L_{x} = 256$, and $L_{y} = 32$.}
    \label{fig:mhtsvm_op}
\end{figure}

Fig.~\ref{fig:mhtsvm_op}(a--b) provides a quantitative analysis of the data presented in Fig.~\ref{fig:mhtsvm_hist}, illustrating the impact of motility heterogeneity on the stability of PF and APF states through the order parameters \(\langle v_{s} \rangle\) and \(\langle v_{a} \rangle\) as functions of \(\Delta v\). Near the homogeneous TSVM limit (\(\Delta v = 0\)), in the PF-dominant ensemble (\(v_{a} \leq v_{s}\)), unsurprisingly, the APF order parameter \(\langle v_{a} \rangle\) exhibits a local minimum [Fig.~\ref{fig:mhtsvm_op}(a)], while the PF order parameter \(\langle v_{s} \rangle\) shows a local maximum [Fig.~\ref{fig:mhtsvm_op}(b)]. Beyond this region, as depicted in Fig.~\ref{fig:mhtsvm_hist}, the APF state prevails. In the APF-dominant ensemble (\(v_{a} \geq v_{s}\)), conversely, \(\langle v_{a} \rangle\) increases with \(\Delta v\) as the APF order gets stronger, whereas \(\langle v_{s} \rangle\) shows a decreasing trend. The local extrema in Fig.~\ref{fig:mhtsvm_op}(a--b) directly correspond to the stochastic PF-APF switching observed in Fig.~\ref{fig:mhtsvm_hist}(d–f). 

For motility heterogeneity, PF behavior emerges predominantly near the homogeneous TSVM limit $(\Delta v \to 0)$. Fig.~\ref{fig:mhtsvm_op}(c--d) illustrates that this phenomenon is primarily driven by the interaction between system noise $(\eta)$ and particle density $(\rho)$. We compute the probability of the PF state $p_{\rm PF}$ $(p_{\rm APF}=1-p_{\rm PF})$, defined as the ratio of the time the system remains in the PF state $(t_{\rm PF})$ to the total time $(t)$ after reaching a steady state at time \(t_{\rm eq}\): \(p_{\rm PF} = t_{\rm PF}/t\) where \(t = t_{\rm max} - t_{\rm eq}\). The system is considered to be in the PF state when \(v_{s} > v_{a}\). In Fig~\ref{fig:mhtsvm_op}(c), \(p_{\rm PF}\) is plotted against $\Delta v \in [-0.2,0.2]$ for several values of $\eta$, keeping $\rho=2$ constant, and in Fig.~\ref{fig:mhtsvm_op}(d) for several values of $\rho$, keeping $\eta=0.4$ constant. The plots reveal two primary regimes: PF+APF (with a stochastic switching between these two states) near $\Delta v = 0$, and a weak PF behavior beyond this range. Near $\Delta v = 0$, PF behavior is the weakest for low noise $(\eta = 0.3)$ or high density $(\rho=6)$, as the system tends to be in a liquid phase, which is identified as APF in the TSVM~\cite{SwarnajitTSVM}. As noise increases or density decreases, PF behavior becomes more pronounced as the system transitions from the APF liquid state to a PF+APF coexistence regime.

\begin{figure}[t]
    \centering    
    \includegraphics[width=\columnwidth]{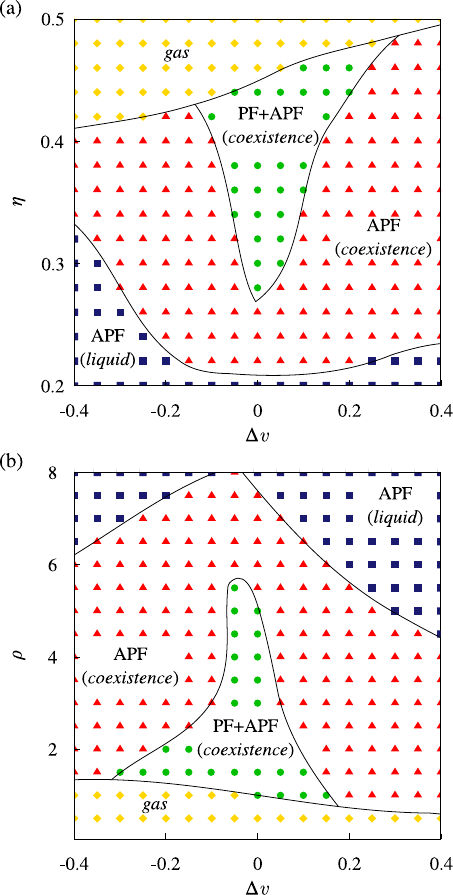}
    \caption{(color online) {\it Phase diagrams for motility heterogeneity.} (a) $\eta - \Delta v$ phase diagram for fixed $\rho = 1.5$; (b) $\rho - \Delta v$ phase diagram for fixed $\eta = 0.4$. For both cases, the velocity moduli of B species is fixed ($v_{\rm B} = 0.5$). The boundary lines act as a guide to the eyes.} 
    \label{fig:mhtsvm_phasediag}
\end{figure}

In Fig.~\ref{fig:mhtsvm_phasediag}, we present the \(\eta - \Delta v\) and \(\rho - \Delta v\) phase diagrams for motility heterogeneity, confirming the dominance of the APF state away from \(\Delta v = 0\). The system remains in a gaseous state at very high noise and low density for all \(\Delta v\). As noise decreases or density increases, the system transitions into a liquid-gas coexistence regime, showing PF+APF coexistence for intermediate noise and density values around \(\Delta v = 0\). Moving further from \(\Delta v = 0\) along with reducing noise or increasing density, the system first exhibits an APF coexistence state and then eventually enters the APF liquid state at very low noise or very high density. It is worth noting that, although reversing the sign of motility heterogeneity, $\Delta v \to -\Delta v$, simply swaps the roles of the fast and slow species, the phase diagrams in Fig.~\ref{fig:mhtsvm_phasediag} are not symmetric under this transformation. This asymmetry arises because $\Delta v$ is varied while keeping $v_{\rm B} = 0.5$ fixed, so that $v_{\rm A} < 0.5$ for negative $\Delta v$ and $v_{\rm A} > 0.5$ for positive $\Delta v$. For fixed noise $\eta$ and density $\rho$, the steady state for $\Delta v < 0$ is generally less ordered than that for $\Delta v > 0$.

In the homogeneous TSVM, inter-species anti-ferromagnetic interactions result in flocking either when the two species spatially separate and move in the same direction (PF), or when they move in opposite directions and satisfy the anti-alignment interaction (APF). Motility heterogeneity disrupts this arrangement, as differences in particle velocities prevent spatial segregation of the two species. Consequently, when heterogeneity is significant, APF remains the only viable state to satisfy the anti-alignment interaction. However, as observed, when heterogeneity is weak, the system can still exhibit a PF state.

\subsection{Spatial heterogeneity, or \enquote{activity landscape}}
\label{SH-TSVM}

\begin{figure}[t]
    \centering
    \includegraphics[width=\columnwidth]{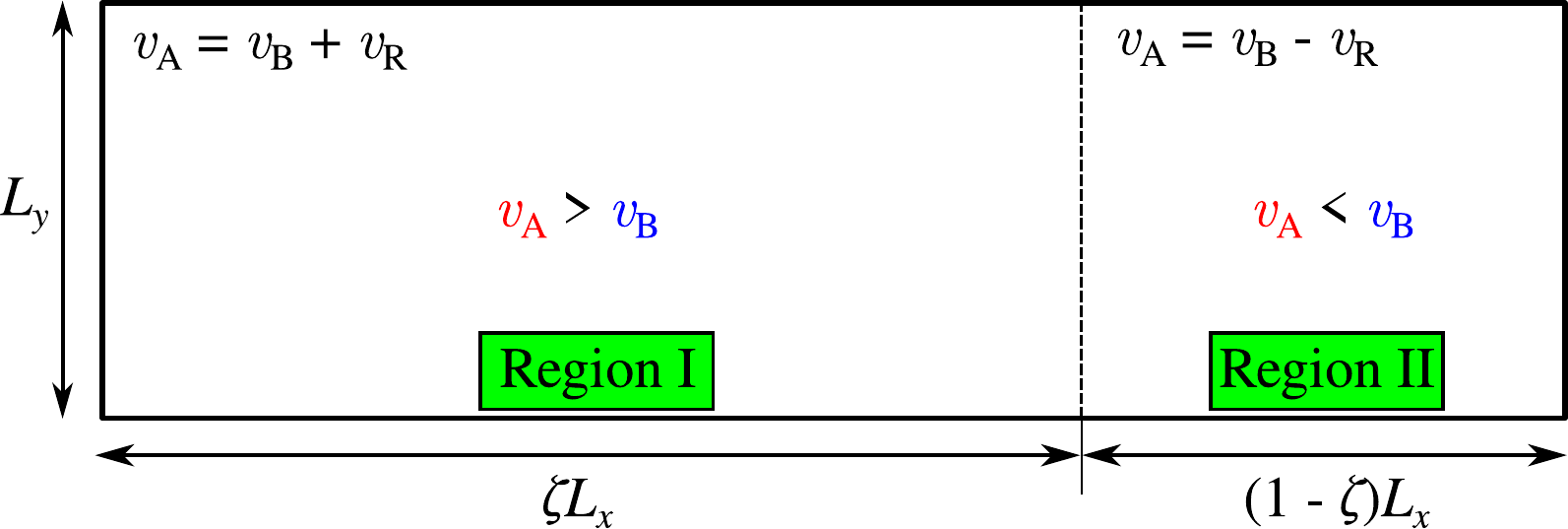}
    \caption{(color online) {\it Schematic representation of the activity landscape}. $v_{\rm B}$ is constant throughout the geometry. $\zeta$ is the fraction of space where $v_{\rm A} > v_{\rm B}$. This arrangement ensures that A-particles change their velocity twice: once at the interface boundary (vertical dotted line) and again due to the periodic boundary condition.} 
    \label{fig:shtsvm_scheme}
\end{figure}

We apply spatial heterogeneity on the TSVM by constructing an {\it activity landscape} which signifies space-dependent particle motility. In this model, we consider different velocities of A-particles in different regions while keeping the velocities of B-particles the same throughout the landscape. We construct the activity landscape by defining $\zeta \in [0,1]$ as the regional width fraction along the horizontal dimension ($L_x$) without affecting the vertical dimension (\enquote{height}) $L_y$.

In the region of width $\zeta L_x$ (fast region I), A-particles move faster than B-particles ($v_{\rm A} > v_{\rm B}$), while $v_{\rm A} < v_{\rm B}$ in the remaining region of width $(1-\zeta) L_x$ (slow region II). However, we keep the inter-species velocity moduli difference the same irrespective of the region, i.e, $v_{\rm R}=|v_{\rm A} - v_{\rm B}|$. This construction resembles two laterally attached regions with motility heterogeneity $\Delta v = + v_{\rm R}$ (in the left) and $\Delta v = - v_{\rm R}$ (in the right). A schematic of this arrangement is presented in Fig.~\ref{fig:shtsvm_scheme}. 

\begin{figure*}[t]
    \includegraphics[width=\textwidth]{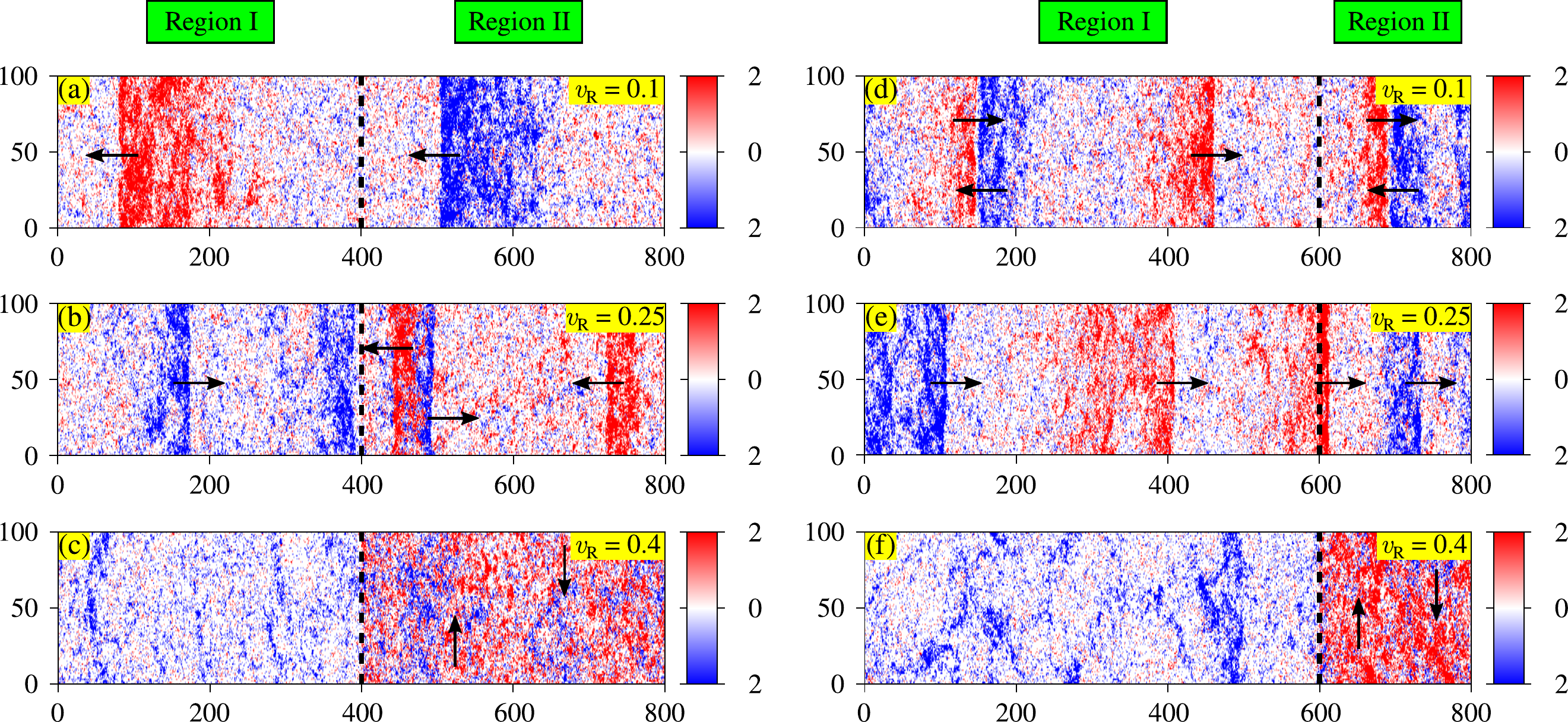}
    \caption{(color online) {\it Steady-state snapshots for spatial heterogeneity.} A (B) particles are represented by red (blue) dots, with local particle density color-coded according to the color bar. Black arrows indicate the direction of flock propagation. The dashed vertical line separates the $v_{\rm A} > v_{\rm B}$ region on the left from the $v_{\rm A} < v_{\rm B}$ region on the right. (a--c) Equal region sizes $(\zeta = 0.5)$. (d--f) Unequal region sizes $(\zeta = 0.75)$. Parameters: $\rho = 1$, $\eta = 0.3$, $v_{\rm B} = 0.5$, $L_{x} = 800$, and $L_{y} = 100$. A movie (\texttt{movie3}) of the same can be found at Ref.~\cite{zenodo}.} 
    \label{fig:shtsvm_snap}
\end{figure*}

In Fig.~\ref{fig:shtsvm_snap}, we demonstrate how $v_{\rm R}$ and $\zeta$ impact the behavior of the system. We first discuss the impact of $v_{\rm R}$ when the fast and slow regions have an equal size [$\zeta=0.5$, Fig.~\ref{fig:shtsvm_snap}(a--c)]. The system exhibits a PF state when $v_{\rm R}$ is small [$v_{\rm R}=0.1$, Fig.~\ref{fig:shtsvm_snap}(a)] but transitions to an APF state when $v_{\rm R}$ increases [$v_{\rm R}=0.25$, Fig.~\ref{fig:shtsvm_snap}(b)], since the enhanced velocity difference promotes APF behavior (see Sec.~\ref{MH-TSVM}). As A-particles traverse the region I much faster than region II, it leads to wider, more diffuse bands in the fast region and more condensed bands in the slow region. For sufficiently large $v_{\rm R}$ [$v_{\rm R}=0.4$, Fig.~\ref{fig:shtsvm_snap}(c)], A-particles move rapidly through region I, which limits their interaction time with B-particles, and upon entering region II, they slow down significantly and become almost trapped. The average number of A-particles in one region is proportional to the time spent in that region, which is $\zeta L_x/(v_{\rm B} + v_{\rm R})$ in region I and $(1-\zeta) L_x/(v_{\rm B} - v_{\rm R})$ in region II. The average density then reads
\begin{equation}
    \rho_{\rm A}^{\rm I/II} = \frac{v_{\rm B} \mp v_{\rm R}}{\zeta(v_{\rm B} - v_{\rm R}) + (1-\zeta)(v_{\rm B} + v_{\rm R})} \rho_{\rm A} \, ,
\end{equation}
in regions I and II, respectively. For fixed $\zeta$, when $v_{\rm R} \to v_{\rm B}$, $\rho_{\rm A}^{\rm I}/\rho_{\rm A}^{\rm II} \simeq (v_{\rm B} - v_{\rm R}) / 2 v_{\rm B} \ll 1$ shows a strong trapping of A-particles in region II. This trapping and large density of A-particles in region II favors the two species to organize themselves into a vertical APF liquid state, after one or several stochastic switching between horizontal APF and PF states, whereas the A-particles remain in the gas phase in region I.

Next, we discuss the case of fast and slow regions with unequal sizes [$\zeta=0.75$, Fig.~\ref{fig:shtsvm_snap}(d--f)]. For $v_{\rm R}=0.1$, the system exhibits an APF state [Fig.~\ref{fig:shtsvm_snap}(d)], in contrast to the corresponding $\zeta=0.5$ case, but recovers the PF state at $v_{\rm R}=0.25$ [Fig.~\ref{fig:shtsvm_snap}(e)]. This suggests that the emergence of parallel flocking in our activity landscape depends on the interplay between $v_{\rm R}$ and $\zeta$, with a larger $\zeta$ requiring a higher $v_{\rm R}$ to sustain the PF state. Similar to Fig.~\ref{fig:shtsvm_snap}(c), for large enough $v_{\rm R}$ [$v_{\rm R}=0.4$, Fig.~\ref{fig:shtsvm_snap}(f)], we observe the trapping of A-particles and the emergence of a vertical APF liquid state in the slow region. However, the trapping is more pronounced for $\zeta=0.75$, due to a narrower slow region, stabilizing even more the vertical APF state since the average density in region II is increased ($\rho_{\rm A}^{\rm II}=\rho_{\rm A}/(1-\zeta)$ when $v_{\rm R} \to v_{\rm B}$).

\begin{figure*}[t]
    \includegraphics[width=\textwidth]{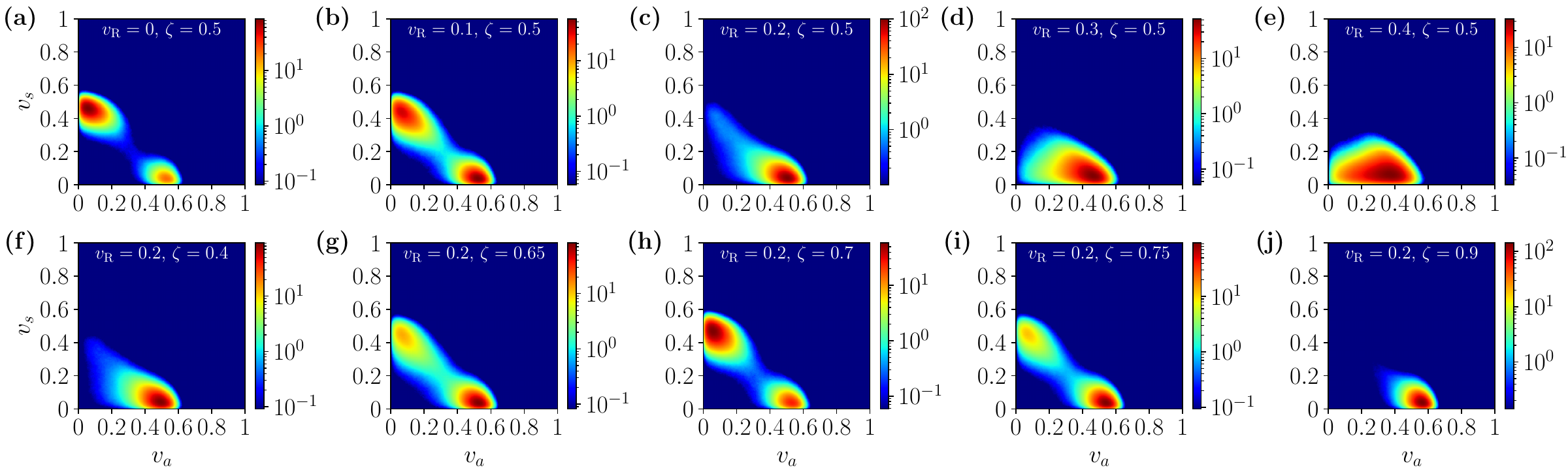}
    \caption{(color online) {\it Probability distribution $P(v_{s},v_{a})$ for spatial heterogeneity.} (a--e) For constant $\zeta=0.5$ and varying $v_{\rm R}$. (f--j) For constant $v_{\rm R}=0.2$ and varying $\zeta$. Parameters: $\rho=0.5$, $\eta=0.24$, $v_{\rm B}=0.5$, $L_{x} = 256$, and $L_{y} = 32$. Movies (Movie S3 and Movie S4) of the same can be found at Ref.~\cite{SM}.} 
    \label{fig:shtsvm_hist}
\end{figure*}

We next plot $P(v_{a}, v_{s})$ for a fixed $\zeta = 0.5$ with varying $v_{\rm R}$ in Fig.~\ref{fig:shtsvm_hist}(a--e) and for a fixed $v_{\rm R} = 0.2$ with varying $\zeta$ in Fig.~\ref{fig:shtsvm_hist}(f--h). As observed in Sec.~\ref{MH-TSVM}, motility heterogeneity causes the system to transition into an APF state with activity landscape, but only at sufficiently high \( v_{\rm R} \) [Fig.~\ref{fig:shtsvm_hist}(d)]. This behavior arises from the imposed spatial heterogeneity, which ensures that the average relative speed between the species is zero. In region I, species A particles attempt to catch up with species B particles but fall behind in region II, where B particles pursue A with an equal relative velocity. This results in a more balanced effect on spatial segregation compared to the simple motility heterogeneity, where the relative velocity difference is \(\Delta v = v_{\rm R}\), thereby promoting greater retention of the PF state [Fig.~\ref{fig:shtsvm_hist}(a--c)]. As \( v_{\rm R} \) increases, the probability of inter-species interactions within the finite widths of each region grows, diminishing spatial separation effects and ultimately destroying any remaining PF behavior [Fig.~\ref{fig:shtsvm_hist}(d, e)].

As we increase $\zeta$, keeping $v_{\rm R} = 0.2$ fixed [Fig.~\ref{fig:shtsvm_hist}(f--h)], we observe a non-monotonic behavior of the system concerning the APF state. While an increase in region I $(v_{\rm A} > v_{\rm B})$ encourages parallel flocking, the velocity difference $v_{\rm R}$ also plays a key role. In region I, due to comparatively higher particle velocities $(v_{\rm A}=0.7, v_{\rm B}=0.5)$, more horizontal space $(\zeta L_x)$ is required for A-particles to catch up, and eventually overtake B-particles. This can lead to A- and B-particles gaining spatial segregation in a PF state on increasing $\zeta$ up to a limit depending upon $v_{\rm R}$, exhibiting PF behavior [Fig.~\ref{fig:shtsvm_hist}(g,h)]. Such overtaking maneuvers can not occur for the simple motility heterogeneity (Sec.~\ref{MH-TSVM}) as one species is consistently faster than the other. Note that, with increasing $\zeta$, the segregation in region II is also decreasing concurrently, discouraging overtakes within the reducing horizontal space $[(1-\zeta) L_x]$ and the dominance of the APF behavior is gradually regained [Fig.~\ref{fig:shtsvm_hist}(i, j)]. 

\begin{figure}[t]
    \centering
    \includegraphics[width=\columnwidth]{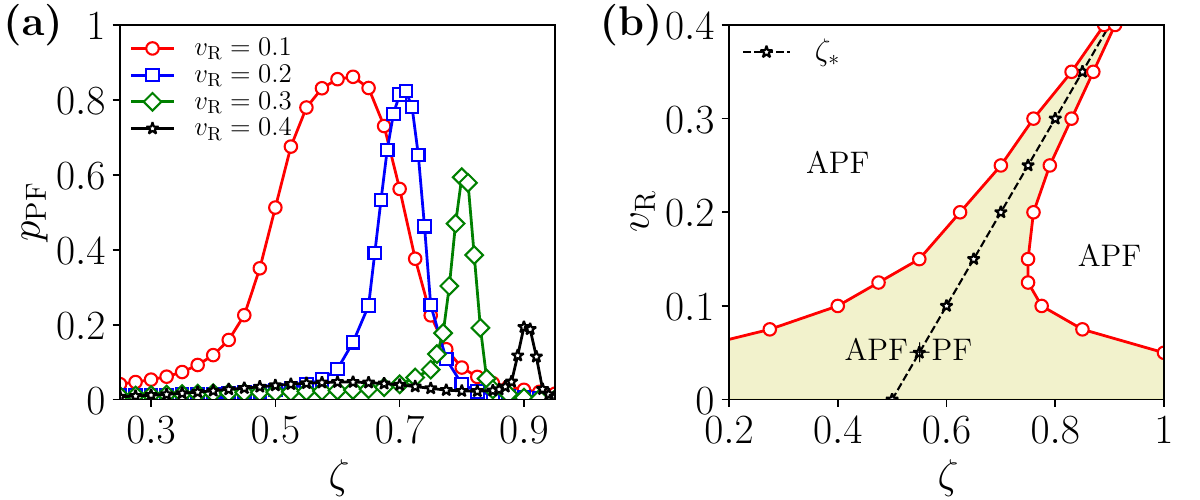}
    \caption{(color online) {\it PF state probability and phase diagram for spatial heterogeneity.} (a)  $p_{\rm PF}$ versus $\zeta$ for $L_{x} = 256$ and $L_{y} = 32$. On increasing $v_{\rm R}$, the peaks move towards higher $\zeta$ and become thinner. (b) $v_{\rm R}-\zeta$ phase diagram. The shaded region denotes the APF+PF regime, while the black dotted line represents $\zeta_*$. Parameters: $\rho=0.5$, $\eta=0.24$, and $v_{\rm B}=0.5$.} 
    \label{fig:shtsvm_ppf}
\end{figure}

However, as depicted in Fig.~\ref{fig:shtsvm_ppf}, the existence of the PF state depends on the combination of $v_{\rm R}$ and $\zeta$. Fig.~\ref{fig:shtsvm_ppf}(a) presents the probability of the PF state $(p_{\rm {PF}})$ against $\zeta$ for various values of $v_{\rm R}$ which exhibits $p_{\rm {PF}}$ attains its maximum at a certain width $\zeta_*(v_{\rm R})$ $(>0.5)$. As relative velocity $v_{\rm R}$ increases, the maximum value of $p_{\rm {PF}}$ decreases and shifts toward higher $\zeta_*$. This shifting signifies that enhanced motility heterogeneity needs a larger region I for the system to exhibit PF behavior. Additionally, the gradual lowering of the peak heights and narrower $p_{\rm {PF}}$ curves indicate that APF increasingly dominates the steady state as the velocity difference between the two species grows. With spatial heterogeneity, the total travelling times of species A and B across the system are, respectively, given by:
\begin{equation}
    t_{\rm A} = \frac{\zeta L_x}{v_{\rm B} + v_{\rm R}} + \frac{(1-\zeta)L_x}{v_{\rm B}-v_{\rm R}}, \quad t_{\rm B} = \frac{L_x}{v_{\rm B}} \, .
\end{equation}
The condition to ensure a stable PF state is $t_{\rm A} \simeq t_{\rm B}$ which gives:  
\begin{equation}
\label{eq:zeta}
    \zeta_* \simeq \frac{v_{\rm B} + v_{\rm R}}{2 v_{\rm B}} \, ,
\end{equation}
and shown in Fig.~\ref{fig:shtsvm_ppf}(b).

In Fig.~\ref{fig:shtsvm_ppf}(b), we present the $v_{\rm R}-\zeta$ phase diagram for $v_{\rm B}=0.5$. The phase diagram is primarily dominated by the APF state, while the shaded region, representing the APF+PF regime, shrinks and shifts to higher $\zeta$ values as the inter-species velocity difference  $v_{\rm R}$ increases. From the $p_{\rm {PF}}$ vs $\zeta$ plots, we extract $\zeta(v_{\rm R})$, where $p_{\rm {PF}}$ reaches its maximum, and plot it as the black dotted line in Fig.~\ref{fig:shtsvm_ppf}(b), which matches very well with Eq.~\eqref{eq:zeta}. Fig.~\ref{fig:shtsvm_ppf}(b) illustrates that for antagonistic species with differing velocities, the fast region I needs to be larger to maintain the A and B species separated and avoid the anti-alignment interaction. However, as the relative velocity increases further, the probability of maintaining this separation progressively decreases.

In summary, we show how spatial geometry acting as an activity landscape can affect the two-species flocking dynamics. Although APF behavior dominates the steady state, which could be horizontal or vertical depending on the relative velocity, parallel flocking behavior can be maintained within the overall geometry.

\subsection{Noise heterogeneity, or unfriendly \enquote{hot} and \enquote{cold} particles}
\label{NHTSVM}

So far, heterogeneity in species density $\rho_{\rm A(B)}$ and velocity $v_{\rm A(B)}$ is explored as a means of introducing variability in the otherwise homogeneous TSVM~\cite{SwarnajitTSVM}. However, heterogeneity can also be introduced through (athermal) noise~\cite{Ariel2014HetSPP,Netzer20191hetpop}, leading to two sub-populations with differing sensitivities to external noise. Species exposed to higher noise levels can be described as {\it hot} species, while those exposed to lower noise levels can be considered {\it cold} species. We will maintain a constant noise level $\eta_{\rm B} = \eta$ for species B, while varying the noise parameter $\eta_{\rm A}$ for species A with $\Delta \eta = \eta_{\rm A}-\eta_{\rm B}$.

\begin{figure}[t]
    \includegraphics[width=\columnwidth]{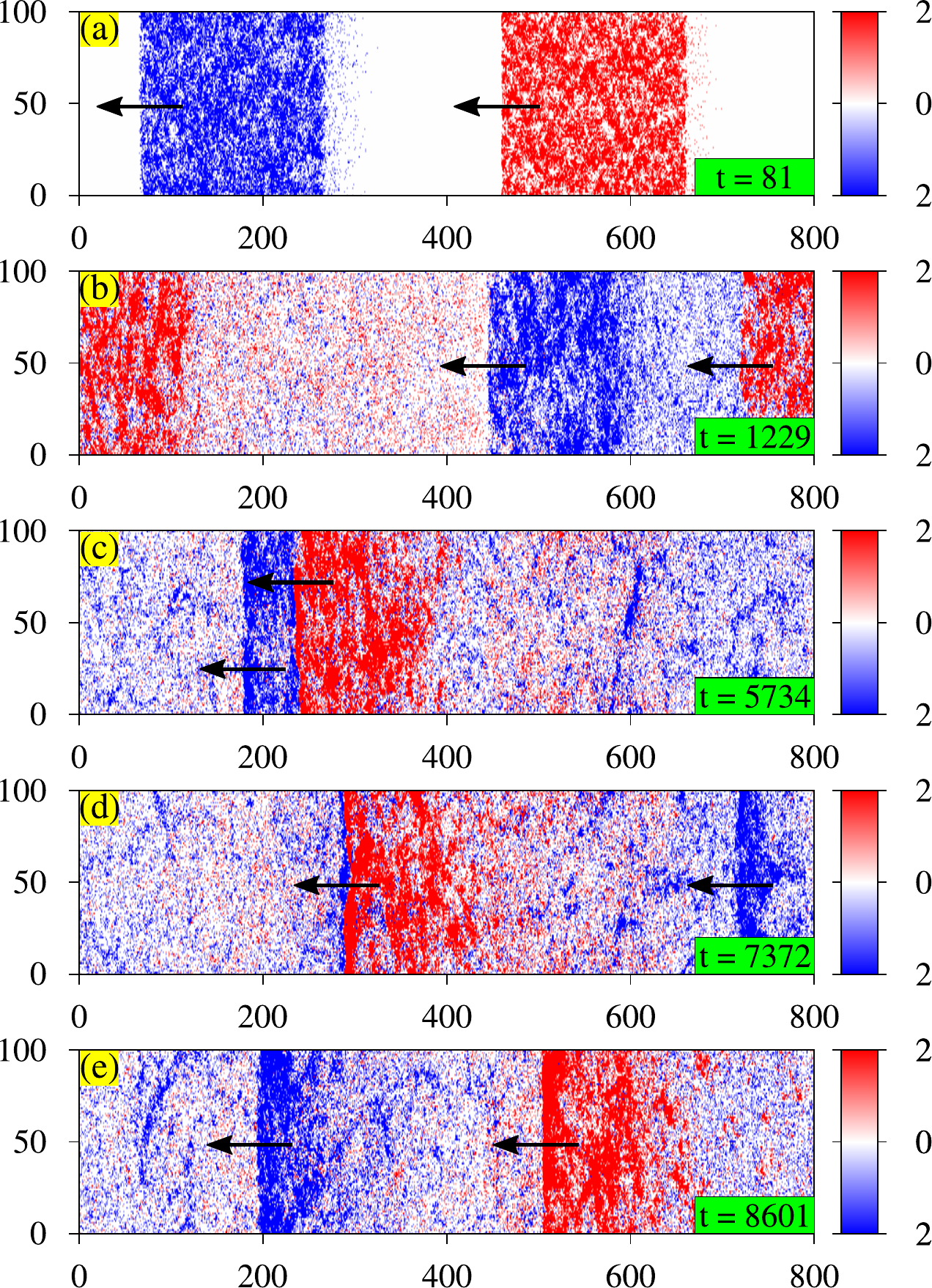}
    \caption{(color online) {\it Time evolution of the TSVM under noise heterogeneity.} (a) Initially, the bands of two species are in a PF state. (b--c) Over time, species A (red), having a higher group velocity, catches up to species B (blue), leading to a collision-mediated transition. (d--e) This interaction disrupts the flocking band of species B, causing fragmentation and subsequent reorganization into a new PF state. Parameters: $\rho = 1$, $\eta=0.3$, $v_0 = 0.5$, $\Delta \eta = -0.2$, $L_{x}=800$, and $L_{y}=100$. A movie (\texttt{movie4}) of the same can be found at Ref.~\cite{zenodo}.}
    \label{fig:nhtsvm_snap}
\end{figure}

Starting from an initial PF configuration, Fig.~\ref{fig:nhtsvm_snap} illustrates the time evolution of the TSVM under noise heterogeneity, with \(\eta = 0.3\) and \(\Delta \eta = -0.2\), at a fixed system density \(\rho = 1\) and self-propulsion speed \(v_0 = 0.5\). Initially, the two bands are organized in a PF state [Fig.~\ref{fig:nhtsvm_snap}(a)]. Over time, species A (red) advances faster than species B (blue), leading to an inter-species collision [Fig.~\ref{fig:nhtsvm_snap}(b--c)]. This collision triggers a transition where the anti-alignment interaction disrupts the order within species B, causing fragmentation and subsequent reorganization into a new PF state [Fig.~\ref{fig:nhtsvm_snap}(d--e)].  

This difference in flocking speeds arises despite both species having the same intrinsic self-propulsion speed $v_0$. The key factor governing their motion is the {\it band velocity}, a property shaped by noise heterogeneity. At lower noise, particles align more effectively, leading to a stronger collective motion and a larger band velocity. Conversely, higher noise reduces alignment, decreasing the band velocity. In Fig.~\ref{fig:nhtsvm_snap}, species A, with lower noise, achieves a higher band velocity than species B, with larger noise. As species A maintains a higher band velocity, this cycle of collision, PF state destruction, and reformation continues over time. Fig.~\ref{fig:nhtsvm_snap} thus highlights that in a two-species system, effective flocking velocity is not merely a direct consequence of individual propulsion speed but an emergent property governed by noise and inter-species interactions. It also indicates that the ``cold'' species (A) dominates the dynamics (as also observed in Ref.~\cite{Ariel2014HetSPP}) as its higher band velocity drives the recurring collisions and reorganizations of the flocking bands.

\begin{figure}[t]
    \includegraphics[width=\columnwidth]{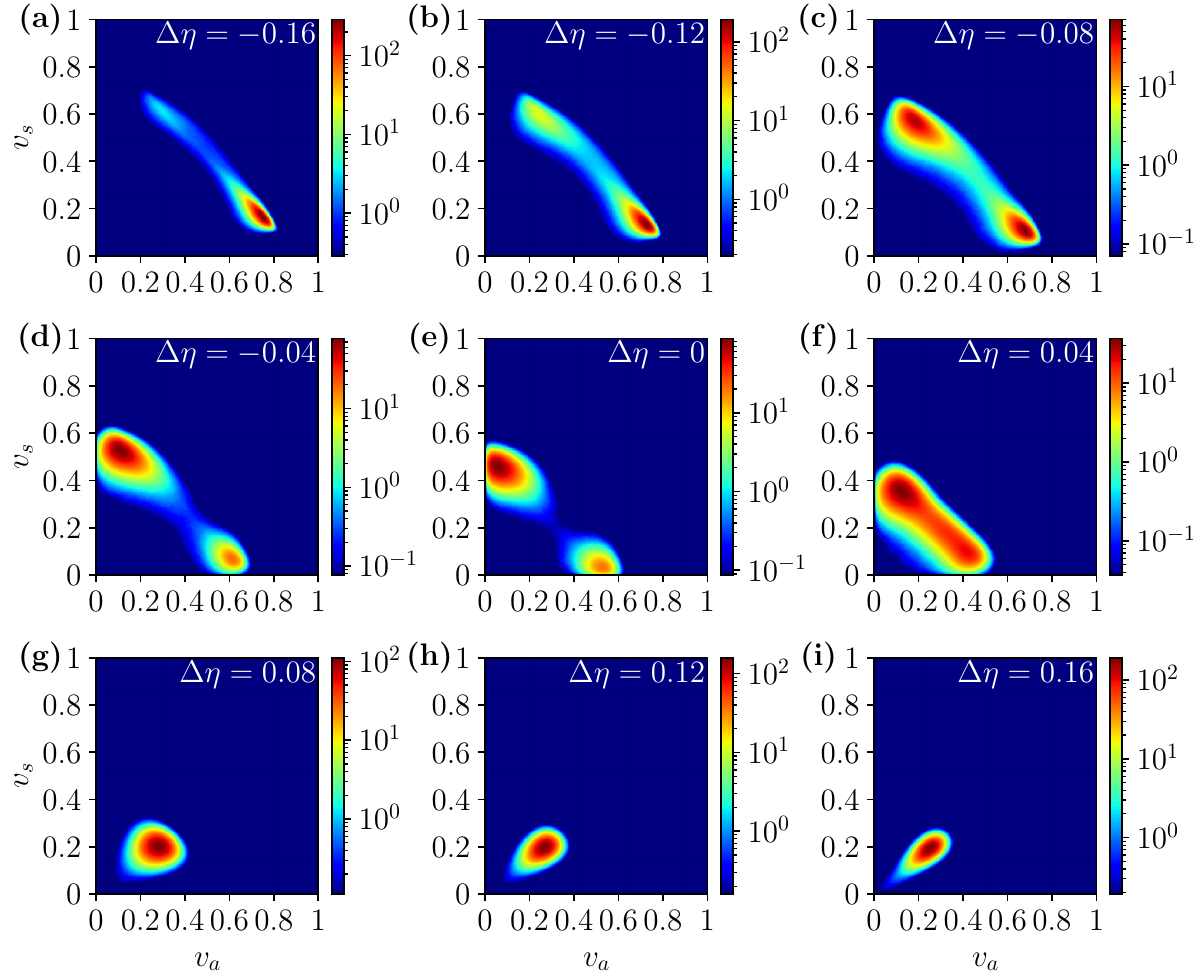}
    \caption{(color online) {\it Probability distribution $P(v_{s},v_{a})$ for varying noise heterogeneity.} (a--b) shows PF-APF stochastic switching with strong APF behavior and strong ordering. (c--f) denotes this switching with more prominence of PF behavior as $\Delta \eta$ increases. (g--i) demonstrates the dissolution of the dual flocking states as the system shows a SSF state at high $\Delta \eta$. Parameters: $\rho=0.5$, $\eta = 0.24$, $v_{\rm 0}=0.5$, $L_{x} = 256$, and $L_{y} = 32$. A movie (Movie S5) of the same can be found at Ref.~\cite{SM}.}
    \label{fig:tsvm_hc_pd}
\end{figure}

In Fig.~\ref{fig:tsvm_hc_pd}, the probability distribution $P(v_{a},v_{s})$ is presented for increasing values of $\Delta \eta$, with $\eta=0.24$ fixed. For $\Delta \eta<0$, the system transitions from a highly ordered APF state at very low $\eta_{\rm A}$ [Fig.~\ref{fig:tsvm_hc_pd}(a)] to PF+APF configurations, exhibiting stochastic switching between these two dynamic states as $\Delta \eta$ increases [Fig.~\ref{fig:tsvm_hc_pd}(b--c)]. When $\Delta \eta \sim 0$, the system oscillates between PF and APF states. As $\Delta \eta$ increases, the two states become gradually equiprobable, with small order parameter values due to the high noise, leading the system toward the SSF state discussed in Sec.~\ref{PH-TSVM} for $\Delta \eta \geq 0.08$.

\begin{figure}[t]
    \includegraphics[width=\columnwidth]{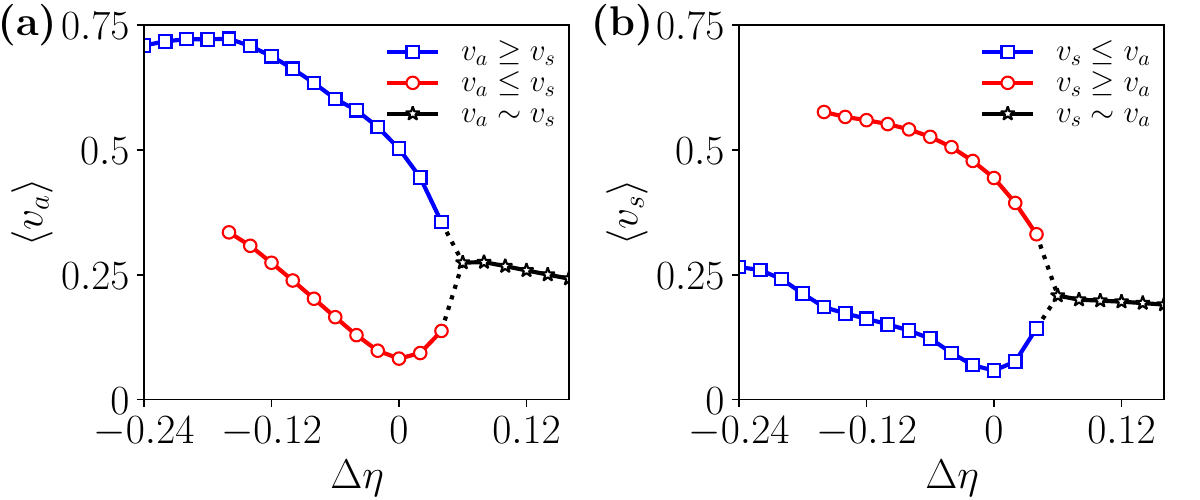}
    \caption{(color online) {\it Order parameters for noise heterogeneity.} $\langle v_{a} \rangle$ and $\langle v_{s} \rangle$ in the restricted APF (blue square), PF (red circle), and SSF (black star) ensembles for varying $\Delta \eta$. Parameters: $\rho = 0.5$, $\eta = 0.24$, $v_{0} = 0.5$, $L_{x} = 256$, and $L_{y} = 32$.}
    \label{fig:tsvm_hc_op}
\end{figure}

In Fig.~\ref{fig:tsvm_hc_op}, the order parameters of the system are plotted against $\Delta \eta$, showing a decline in their respective dominant ensembles (e.g., $\langle v_{a} \rangle$ in the $v_{a} \geq v_{s}$ ensemble or $\langle v_{s} \rangle$ in the $v_{s} \geq v_{a}$ ensemble) as $\Delta \eta$ increases, before stabilizing with $v_{a} \sim v_{s}$ where the SSF state is dominant (similar to Fig.~\ref{fig:phtsvm_op}). This reflects an increase in the overall disorder in the system as species A becomes ``hotter''. A more notable observation is the behavior of the order parameters in ensembles where they do not represent the dominant flocking behavior (e.g., $\langle v_{a} \rangle$ in the $v_{a} \leq v_{s}$ ensemble or $\langle v_{s} \rangle$ in the $v_{s} \leq v_{a}$ ensemble). There is a consistent decrease in the order parameter until the noise reception of the two species is equal, $\Delta \eta \simeq 0$, reflecting the trend of the dominant ensemble. However, beyond this point ($\Delta \eta>0$), up until the SSF regime ($\Delta \eta \geq 0.08$), a sharp increase in the order parameter is observed. In the PF+APF coexistence regime with $\Delta \eta \in [0, 0.08]$, the two ensembles converge as $\Delta \eta$ is increased and merge in the $v_{a} \sim v_{s}$ ensemble. This behavior can be attributed to the diminishing contribution of the order parameter from the ``hotter'' species, causing the system to resemble the SSF state discussed in Sec.~\ref{PH-TSVM}.

\begin{figure}[t]
    \includegraphics[width=\columnwidth]{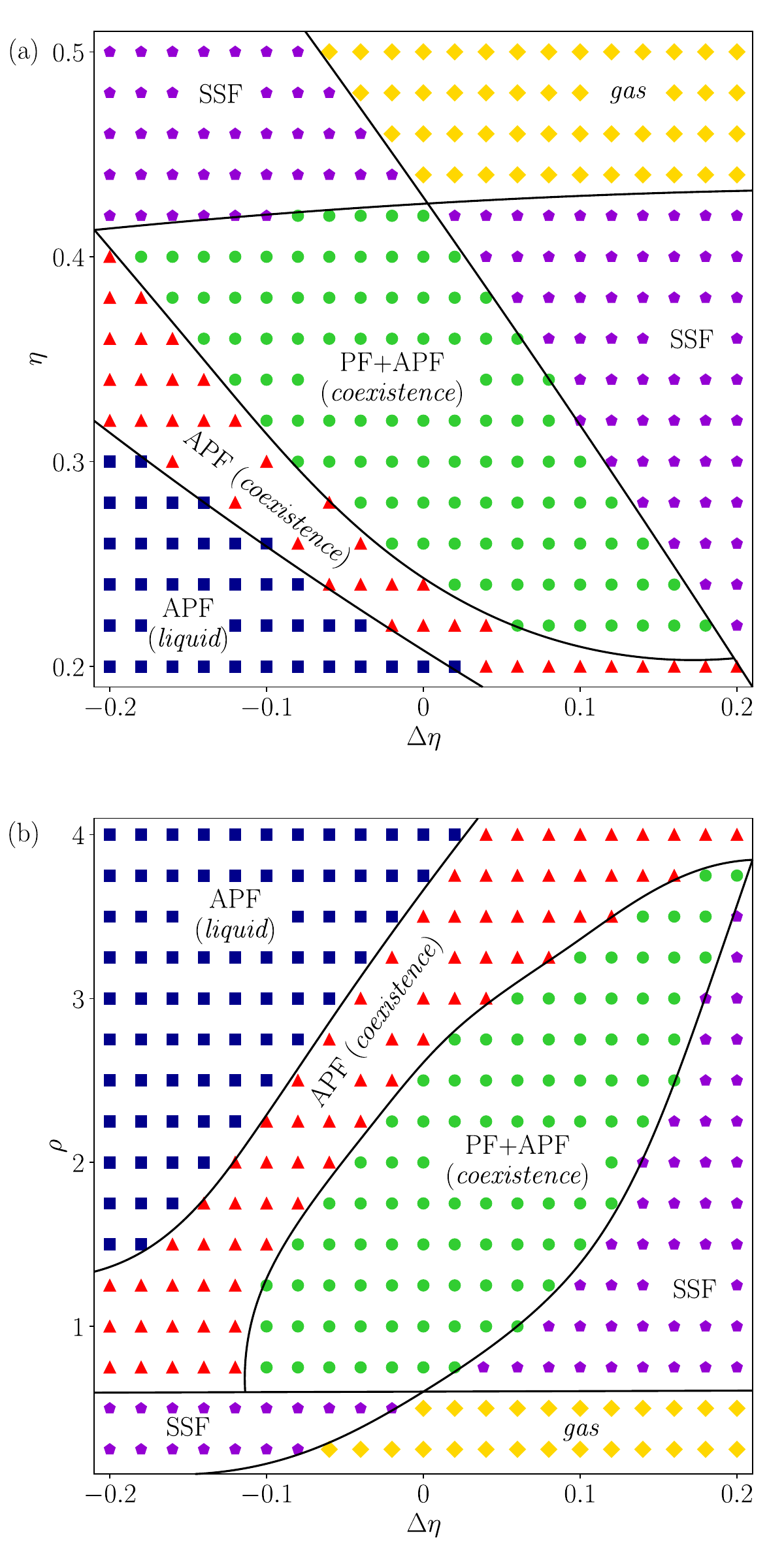}
    \caption{(color online) {\it Phase diagrams for noise heterogeneity.} (a) $\eta - \Delta \eta$ phase diagram for fixed $\rho = 1.5$; (b) $\rho - \Delta \eta$ phase diagram for fixed $\eta = 0.3$. For both cases, the velocity modulus is $v_0 = 0.5$. The boundary lines act as a guide to the eyes.}
    \label{fig:nhtsvm_phasediag}
\end{figure}

In Fig.~\ref{fig:nhtsvm_phasediag}, we present the \(\eta - \Delta \eta\) and \(\rho - \Delta \eta\) phase diagrams for noise heterogeneity, confirming the presence of the PF+APF coexistence state away from $\Delta \eta = 0$. The system remains in a gaseous state at high noise and low density for $\Delta \eta \ge 0$ (species A has higher noise), while the system exhibits an SSF state formed by A bands for $\Delta \eta < 0$ (species A has lower noise). As noise decreases or density increases, the system transitions into a liquid-gas coexistence regime, showing PF+APF coexistence, for $\Delta \eta < 0$, and into an SSF state formed by B bands for $\Delta \eta > 0$ (species A has higher noise). For even lower noise or larger density, the system transitions into APF coexistence state and then eventually enters the APF liquid state, analogously to the homogeneous TSVM.

In summary, noise heterogeneity, where species differ in their sensitivity to external noise, leads to distinct band velocities. The cold (lower-noise) species dominates flocking dynamics through higher band velocity, driving repeated collisions and reorganizations. This noise heterogeneity also gives rise to two distinct single-species flocking (SSF) states, depending on which species experiences the lower noise level.

\section{Discussion}
\label{summary&discussion}

We investigate the impact of various heterogeneities on multi-species flocking dynamics using the two-species Vicsek model (TSVM)~\cite{SwarnajitTSVM}. In the presence of strong population heterogeneity, at high noise (or low density), the PF and APF states vanish into a single flock dominated by the majority species, while the minority species remains in a disordered state, resembling the behavior of the single-species VM. However, at low noise (or high density), the minority group becomes polarized, leading to an APF liquid state where both species move in opposing directions.

For strong motility heterogeneity, we established the absence of the PF state in the coexistence regime as the spatial segregation between species is compromised. However, considering activity landscapes with region-dependent motilities, the dynamical behavior contrasts with the simple motility heterogeneity. We find high retention of the PF behavior for a given geometry, where the fast region is larger than the slow region, and the emergence of a vertical APF state for large relative velocity, due to particle trapping in the slow region. This shows that interruption by environmental constructs plays a big role in shaping the nature of flocking. In this regard, our current implementation is based on sharp spatial variations in motility. A natural extension would be to incorporate smooth motility gradients or time-evolving landscapes, which more accurately reflect realistic environments. Gradual changes in activity could soften inter-species collisions and mitigate sharp density mismatches, while temporally varying or fluctuating landscapes may act as continuous sources of disorder, disrupting segregation and potentially stabilizing new dynamic patterns. Exploring such scenarios presents an interesting direction for future work.

We also find that species motility is significantly affected by noise heterogeneity, where the colder species (subjected to lower noise) moves faster than the hotter one due to its higher band velocity, eventually catching up and transiently disrupting any PF structure. However, the pattern re-emerges, reflecting a dynamic yet robust response to noise asymmetry. This behavior contrasts with motility heterogeneity, where differing species velocities lead to the absence of the PF state in the coexistence regime.

Similar heterogeneities can be experimentally realized in vibrationally excited granular active matter (vibrobots)~\cite{deseigne2010collective, kumar2014flocking, koumakis2016mechanism} or programmable robotic swarms~\cite{chen2024emergent} through controlled adjustment of agent design (e.g. shape, size, surface properties etc.)~\cite{bera2020motile} and environmental parameters (e.g. population density, vibration frequency or robot speed, substrate characteristics, communication range, alignment strength, repulsion thresholds etc.)~\cite{chen2024scale}. In biological systems, such heterogeneities are especially relevant, both in multi-species interactions and within a single species exhibiting internal diversity. Examples may include flocks exhibiting a variety of collective escape patterns under predation~\cite{storms2019complex,papadopoulou2022emergence}, fish from high predation populations forming more cohesive groups~\cite{herbert2017predation}, strong ecological interactions leading to partner intermixing in microbial communities~\cite{momeni2013strong}, social behavior of mixed-species flocks emerging from species-specific interaction rules~\cite{farine2014collective}, consistent collective decision-making across heterogeneous taxonomic groups~\cite{papageorgiou2024compromise} and intermittent collective dynamics in sheep herds emerging from individual-level behavioral shifts~\cite{ginelli2015intermittent}. 

In confined environments, stability emerges more easily if components interact frequently, but it can also be disrupted by the surrounding \enquote{habitat}. For instance, if one species relies on a resource for growth while the other does not, this imbalance can induce indirect inter-species interactions. A classic case is antagonistic predator-prey dynamics: if a predator consumes prey that depends on an external resource (e.g., vegetation or water), fluctuations in that resource indirectly affect the predator, even though the predator interacts only with the prey. Our framework could be extended to capture such environmental feedback by introducing a localized, depletable resource field that influences only one species. Another natural extension would be to incorporate individual-level feedback mechanisms such as quorum sensing, where particles adjust their motility or alignment in response to local density or species composition. Within the framework of our model, this can be achieved by allowing each particle’s self-propulsion speed or alignment strength to depend dynamically on local crowding or the relative concentration of different species.

Furthermore, non-uniform system noise and obstacles~\cite{martinez2018collective} can be considered as other convincing candidates for imparting spatial heterogeneity. Another natural extension would be to let particle velocities depend on the local crowding by introducing a density-dependent self-propulsion speed similar to the softcore restriction considered in Ref.~\cite{karmakar2023jamming}. Finally, the TSVM~\cite{SwarnajitTSVM} does not have the factor of agent size. We can bring size (and later, even shape) into the picture by first working with finite-size hard discs~\cite{martinez2018collective} instead of point particles in the future. 

\appendix

\section{On the stability and translational order of the micro-phase separated state}
\label{appA}

\begin{figure*}[htbp]
  \centering
  \includegraphics[width=\textwidth]{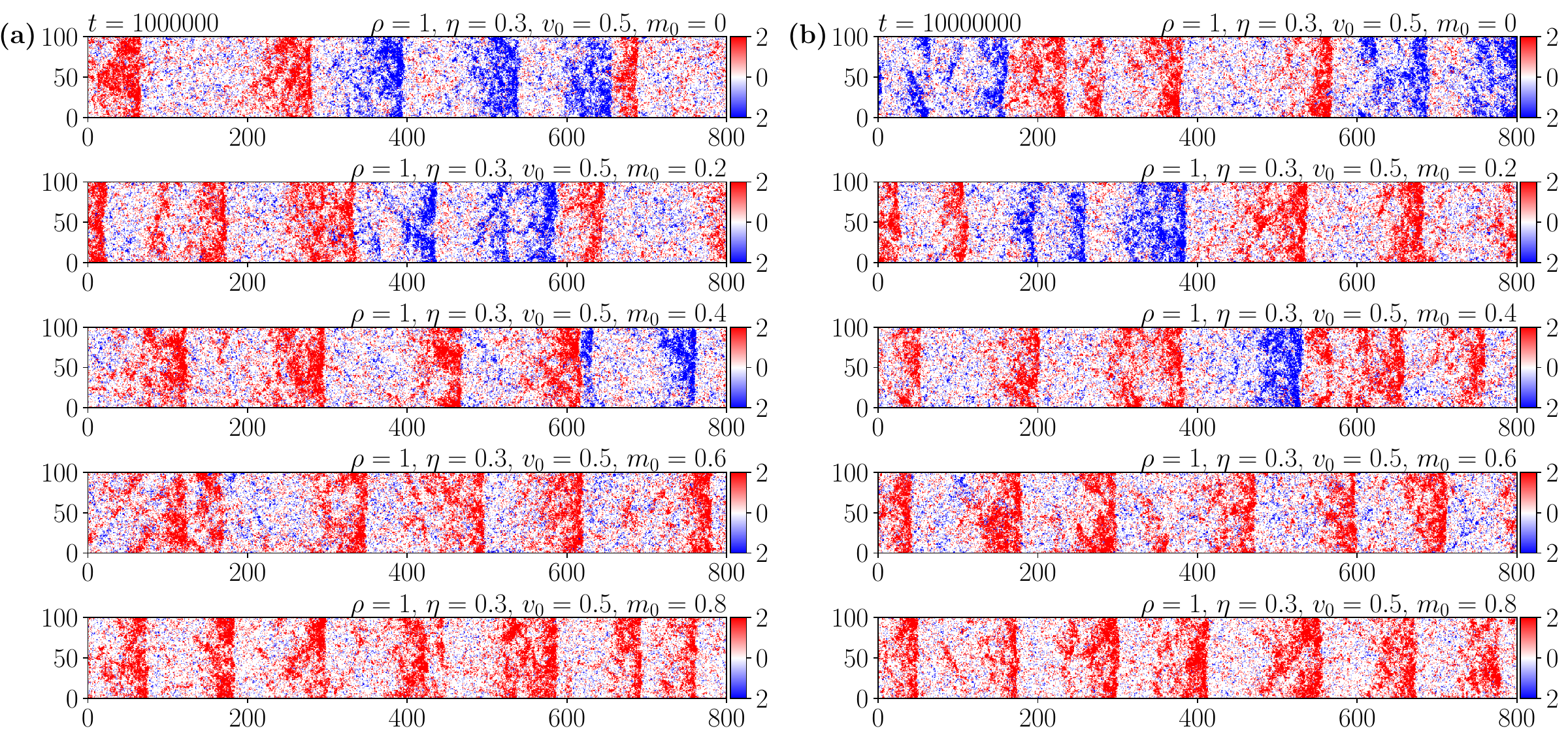}
  \caption{Comparison of band configuration at (a)~$t = 10^6$ and (b)~$t = 10^7$ showing stable band number over long times. Parameters: $\rho =1,\eta =0.3, v_0 = 0.5$.} 
  \label{fig:band_stability}
\end{figure*}

Here, we discuss the stability of the micro-phase separated bands and examine whether they exhibit any translational order, focusing on the band structures obtained in the presence of population heterogeneity (Sec.~\ref{PH-TSVM}). We find that the micro-phase separated coexistence phase corresponds to a stable steady state, and that band coarsening is limited: the number and size of bands remain finite even at long times. Fig.~\ref{fig:band_stability} shows steady-state band configurations for different values of $m_0$ at $t = 10^6$ and $t = 10^7$, and confirms that once the system reaches the phase-separated state, the average number of bands remains largely unchanged over time. It is important to note that in both the VM and TSVM, fluctuations—arising from noise and finite system size—can affect band number and spacing by inducing mergers, splits, or positional shifts, leading to temporal variations in the band structure even after apparent phase separation. In a two-species system with reciprocal interactions, as in the present model, the alignment dynamics between species introduces further complexity by coupling their spatial organization.

\begin{figure*}[htbp]
  \centering
  \includegraphics[width=\textwidth]{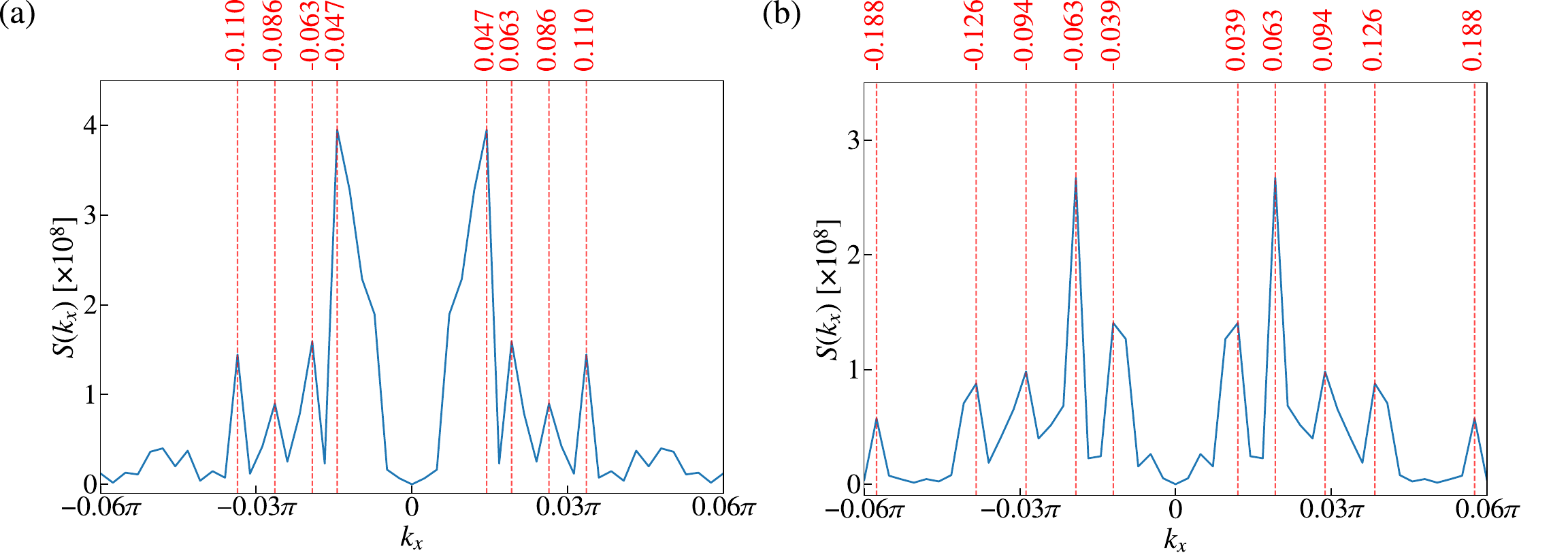}
  \caption{Structure factors (non-normalized) constructed from the spatial Fourier transform of the one-dimensional projected density profile along $x$ axis: (a) $m_0 =0.6$. (b) $m_0 =0.8$. All prominent Bragg peaks have been marked with red dashed lines. Parameters: $\rho =1,\eta =0.3, v_0 = 0.5$.} 
  \label{fig:density_fft}
\end{figure*}

To investigate the presence of long-range translational order similar to a flying smectic, we compute the static structure factor corresponding to the longitudinal particle density profile. This analysis is limited to cases with $m_0 \geq 0.6$, where the minority species no longer forms distinct bands, and the overall spatial organization becomes more regular. The system is divided along the longitudinal ($x$) direction into $L_y$ stripes, and the one-dimensional density profile $\rho(i)$ is computed. From this, we determine the density fluctuations $\delta\rho(i) = \rho(i) - \langle \rho \rangle$, and apply a discrete Fourier transform to obtain the spectral density $\tilde{\rho}(k_x)$. The non-normalized structure factor is then calculated as $S(k_x) = |\tilde{\rho}(k_x)|^2$. Fig.~\ref{fig:density_fft} presents the resulting structure factors for $m_0 = 0.6$ and $m_0 = 0.8$. In both cases, a prominent Bragg peak is observed and aligns with the expected wavevector $k_1 = 2\pi/a$, where $a = L_x/n_b$ denotes the average distance between bands and $n_b$ is the number of observed bands. However, no higher-order harmonics at integer multiples of $k_1$ are seen, and the primary peak shows considerable broadening, especially at $m_0 = 0.6$. These observations indicate notable fluctuations in both band spacing and width, inconsistent with true long-range translational order. Therefore, although the bands may appear quasi-periodic over limited regions and timescales, the system remains in a fluctuating, dynamic banded state rather than forming a true smectic.

Therefore, our analysis reveals that the micro-phase separated state in the heterogeneous TSVM exhibits finite band structures that persist over long times without coarsening into a macroscopic phase-separated state. However, the presence of broadened, irregular Bragg peaks in the structure factor signals the absence of long-range translational symmetry, distinguishing these banded states from flying smectics.

\section*{Acknowledgements}
AD sincerely acknowledges the Indian Association for the Cultivation of Science (IACS), Kolkata, India, for providing the fellowship and computational facilities. RP thanks IACS for its computational facilities and resources. SC, MM, and HR are financially supported by the German Research Foundation (DFG) within the Collaborative Research Center SFB 1027-A3. AD acknowledges many helpful discussions with Dr.~Mintu Karmakar.

%\clearpage

%\bibliographystyle{}
\bibliography{Bibmanuscript_TSVM_het}

%merlin.mbs apsrev4-1.bst 2010-07-25 4.21a (PWD, AO, DPC) hacked
%Control: key (0)
%Control: author (0) dotless jnrlst
%Control: editor formatted (1) identically to author
%Control: production of article title (0) allowed
%Control: page (1) range
%Control: year (0) verbatim
%Control: production of eprint (0) enabled
\begin{thebibliography}{70}%
\makeatletter
\providecommand \@ifxundefined [1]{%
 \@ifx{#1\undefined}
}%
\providecommand \@ifnum [1]{%
 \ifnum #1\expandafter \@firstoftwo
 \else \expandafter \@secondoftwo
 \fi
}%
\providecommand \@ifx [1]{%
 \ifx #1\expandafter \@firstoftwo
 \else \expandafter \@secondoftwo
 \fi
}%
\providecommand \natexlab [1]{#1}%
\providecommand \enquote  [1]{``#1''}%
\providecommand \bibnamefont  [1]{#1}%
\providecommand \bibfnamefont [1]{#1}%
\providecommand \citenamefont [1]{#1}%
\providecommand \href@noop [0]{\@secondoftwo}%
\providecommand \href [0]{\begingroup \@sanitize@url \@href}%
\providecommand \@href[1]{\@@startlink{#1}\@@href}%
\providecommand \@@href[1]{\endgroup#1\@@endlink}%
\providecommand \@sanitize@url [0]{\catcode `\\12\catcode `\$12\catcode
  `\&12\catcode `\#12\catcode `\^12\catcode `\_12\catcode `\%12\relax}%
\providecommand \@@startlink[1]{}%
\providecommand \@@endlink[0]{}%
\providecommand \url  [0]{\begingroup\@sanitize@url \@url }%
\providecommand \@url [1]{\endgroup\@href {#1}{\urlprefix }}%
\providecommand \urlprefix  [0]{URL }%
\providecommand \Eprint [0]{\href }%
\providecommand \doibase [0]{http://dx.doi.org/}%
\providecommand \selectlanguage [0]{\@gobble}%
\providecommand \bibinfo  [0]{\@secondoftwo}%
\providecommand \bibfield  [0]{\@secondoftwo}%
\providecommand \translation [1]{[#1]}%
\providecommand \BibitemOpen [0]{}%
\providecommand \bibitemStop [0]{}%
\providecommand \bibitemNoStop [0]{.\EOS\space}%
\providecommand \EOS [0]{\spacefactor3000\relax}%
\providecommand \BibitemShut  [1]{\csname bibitem#1\endcsname}%
\let\auto@bib@innerbib\@empty
%</preamble>
\bibitem [{\citenamefont {Vicsek}\ and\ \citenamefont
  {Zafeiris}(2012)}]{vicsek2012collective}%
  \BibitemOpen
  \bibfield  {author} {\bibinfo {author} {\bibfnamefont {T.}~\bibnamefont
  {Vicsek}}\ and\ \bibinfo {author} {\bibfnamefont {A.}~\bibnamefont
  {Zafeiris}},\ }\bibfield  {title} {\enquote {\bibinfo {title} {Collective
  motion},}\ }\href@noop {} {\bibfield  {journal} {\bibinfo  {journal} {Physics
  reports}\ }\textbf {\bibinfo {volume} {517}},\ \bibinfo {pages} {71}
  (\bibinfo {year} {2012})}\BibitemShut {NoStop}%
\bibitem [{\citenamefont {Bottinelli}\ \emph {et~al.}(2016)\citenamefont
  {Bottinelli}, \citenamefont {Sumpter},\ and\ \citenamefont
  {Silverberg}}]{bottinelli2016emergent}%
  \BibitemOpen
  \bibfield  {author} {\bibinfo {author} {\bibfnamefont {A.}~\bibnamefont
  {Bottinelli}}, \bibinfo {author} {\bibfnamefont {D.~T.~J.}\ \bibnamefont
  {Sumpter}}, \ and\ \bibinfo {author} {\bibfnamefont {J.~L.}\ \bibnamefont
  {Silverberg}},\ }\bibfield  {title} {\enquote {\bibinfo {title} {Emergent
  structural mechanisms for high-density collective motion inspired by human
  crowds},}\ }\href@noop {} {\bibfield  {journal} {\bibinfo  {journal} {Phys.
  Rev. Lett.}\ }\textbf {\bibinfo {volume} {117}},\ \bibinfo {pages} {228301}
  (\bibinfo {year} {2016})}\BibitemShut {NoStop}%
\bibitem [{\citenamefont {G{\'o}mez-Nava}\ \emph {et~al.}(2022)\citenamefont
  {G{\'o}mez-Nava}, \citenamefont {Bon},\ and\ \citenamefont
  {Peruani}}]{gomez2022intermittent}%
  \BibitemOpen
  \bibfield  {author} {\bibinfo {author} {\bibfnamefont {L.}~\bibnamefont
  {G{\'o}mez-Nava}}, \bibinfo {author} {\bibfnamefont {R.}~\bibnamefont {Bon}},
  \ and\ \bibinfo {author} {\bibfnamefont {F.}~\bibnamefont {Peruani}},\
  }\bibfield  {title} {\enquote {\bibinfo {title} {Intermittent collective
  motion in sheep results from alternating the role of leader and follower},}\
  }\href@noop {} {\bibfield  {journal} {\bibinfo  {journal} {Nat. Phys.}\
  }\textbf {\bibinfo {volume} {18}},\ \bibinfo {pages} {1494} (\bibinfo {year}
  {2022})}\BibitemShut {NoStop}%
\bibitem [{\citenamefont {Ballerini}\ \emph {et~al.}(2008)\citenamefont
  {Ballerini}, \citenamefont {Cabibbo}, \citenamefont {Candelier},
  \citenamefont {Cavagna}, \citenamefont {Cisbani}, \citenamefont {Giardina},
  \citenamefont {Lecomte}, \citenamefont {Orlandi}, \citenamefont {Parisi},
  \citenamefont {Procaccini}, \citenamefont {Viale},\ and\ \citenamefont
  {Zdravkovic}}]{ballerini2008interaction}%
  \BibitemOpen
  \bibfield  {author} {\bibinfo {author} {\bibfnamefont {M.}~\bibnamefont
  {Ballerini}}, \bibinfo {author} {\bibfnamefont {N.}~\bibnamefont {Cabibbo}},
  \bibinfo {author} {\bibfnamefont {R.}~\bibnamefont {Candelier}}, \bibinfo
  {author} {\bibfnamefont {A.}~\bibnamefont {Cavagna}}, \bibinfo {author}
  {\bibfnamefont {E.}~\bibnamefont {Cisbani}}, \bibinfo {author} {\bibfnamefont
  {I.}~\bibnamefont {Giardina}}, \bibinfo {author} {\bibfnamefont
  {V.}~\bibnamefont {Lecomte}}, \bibinfo {author} {\bibfnamefont
  {A.}~\bibnamefont {Orlandi}}, \bibinfo {author} {\bibfnamefont
  {G.}~\bibnamefont {Parisi}}, \bibinfo {author} {\bibfnamefont
  {A.}~\bibnamefont {Procaccini}}, \bibinfo {author} {\bibfnamefont
  {M.}~\bibnamefont {Viale}}, \ and\ \bibinfo {author} {\bibfnamefont
  {V.}~\bibnamefont {Zdravkovic}},\ }\bibfield  {title} {\enquote {\bibinfo
  {title} {Interaction ruling animal collective behavior depends on topological
  rather than metric distance: Evidence from a field study},}\ }\href@noop {}
  {\bibfield  {journal} {\bibinfo  {journal} {Proc. Natl. Acad. Sci.}\ }\textbf
  {\bibinfo {volume} {105}},\ \bibinfo {pages} {1232} (\bibinfo {year}
  {2008})}\BibitemShut {NoStop}%
\bibitem [{\citenamefont {Becco}\ \emph {et~al.}(2006)\citenamefont {Becco},
  \citenamefont {Vandewalle}, \citenamefont {Delcourt},\ and\ \citenamefont
  {Poncin}}]{becco2006experimental}%
  \BibitemOpen
  \bibfield  {author} {\bibinfo {author} {\bibfnamefont {C.}~\bibnamefont
  {Becco}}, \bibinfo {author} {\bibfnamefont {N.}~\bibnamefont {Vandewalle}},
  \bibinfo {author} {\bibfnamefont {J.}~\bibnamefont {Delcourt}}, \ and\
  \bibinfo {author} {\bibfnamefont {P.}~\bibnamefont {Poncin}},\ }\bibfield
  {title} {\enquote {\bibinfo {title} {Experimental evidences of a structural
  and dynamical transition in fish school},}\ }\href@noop {} {\bibfield
  {journal} {\bibinfo  {journal} {Phys. A: Stat. Mech. Appl.}\ }\textbf
  {\bibinfo {volume} {367}},\ \bibinfo {pages} {487} (\bibinfo {year}
  {2006})}\BibitemShut {NoStop}%
\bibitem [{\citenamefont {Peruani}\ \emph {et~al.}(2012)\citenamefont
  {Peruani}, \citenamefont {Starru\ss{}}, \citenamefont {Jakovljevic},
  \citenamefont {S\o{}gaard-Andersen}, \citenamefont {Deutsch},\ and\
  \citenamefont {B\"ar}}]{peruani2012collective}%
  \BibitemOpen
  \bibfield  {author} {\bibinfo {author} {\bibfnamefont {F.}~\bibnamefont
  {Peruani}}, \bibinfo {author} {\bibfnamefont {J.}~\bibnamefont
  {Starru\ss{}}}, \bibinfo {author} {\bibfnamefont {V.}~\bibnamefont
  {Jakovljevic}}, \bibinfo {author} {\bibfnamefont {L.}~\bibnamefont
  {S\o{}gaard-Andersen}}, \bibinfo {author} {\bibfnamefont {A.}~\bibnamefont
  {Deutsch}}, \ and\ \bibinfo {author} {\bibfnamefont {M.}~\bibnamefont
  {B\"ar}},\ }\bibfield  {title} {\enquote {\bibinfo {title} {Collective motion
  and nonequilibrium cluster formation in colonies of gliding bacteria},}\
  }\href@noop {} {\bibfield  {journal} {\bibinfo  {journal} {Phys. Rev. Lett.}\
  }\textbf {\bibinfo {volume} {108}},\ \bibinfo {pages} {098102} (\bibinfo
  {year} {2012})}\BibitemShut {NoStop}%
\bibitem [{\citenamefont {Giavazzi}\ \emph {et~al.}(2018)\citenamefont
  {Giavazzi}, \citenamefont {Paoluzzi}, \citenamefont {Macchi}, \citenamefont
  {Bi}, \citenamefont {Scita}, \citenamefont {Manning}, \citenamefont
  {Cerbino},\ and\ \citenamefont {Marchetti}}]{giavazzi2018flocking}%
  \BibitemOpen
  \bibfield  {author} {\bibinfo {author} {\bibfnamefont {F.}~\bibnamefont
  {Giavazzi}}, \bibinfo {author} {\bibfnamefont {M.}~\bibnamefont {Paoluzzi}},
  \bibinfo {author} {\bibfnamefont {M.}~\bibnamefont {Macchi}}, \bibinfo
  {author} {\bibfnamefont {D.}~\bibnamefont {Bi}}, \bibinfo {author}
  {\bibfnamefont {G.}~\bibnamefont {Scita}}, \bibinfo {author} {\bibfnamefont
  {M.~L.}\ \bibnamefont {Manning}}, \bibinfo {author} {\bibfnamefont
  {R.}~\bibnamefont {Cerbino}}, \ and\ \bibinfo {author} {\bibfnamefont
  {M.~C.}\ \bibnamefont {Marchetti}},\ }\bibfield  {title} {\enquote {\bibinfo
  {title} {Flocking transitions in confluent tissues},}\ }\href@noop {}
  {\bibfield  {journal} {\bibinfo  {journal} {Soft matter}\ }\textbf {\bibinfo
  {volume} {14}},\ \bibinfo {pages} {3471} (\bibinfo {year}
  {2018})}\BibitemShut {NoStop}%
\bibitem [{\citenamefont {Schaller}\ \emph {et~al.}(2010)\citenamefont
  {Schaller}, \citenamefont {Weber}, \citenamefont {Semmrich}, \citenamefont
  {Frey},\ and\ \citenamefont {Bausch}}]{schaller2010polar}%
  \BibitemOpen
  \bibfield  {author} {\bibinfo {author} {\bibfnamefont {V.}~\bibnamefont
  {Schaller}}, \bibinfo {author} {\bibfnamefont {C.}~\bibnamefont {Weber}},
  \bibinfo {author} {\bibfnamefont {C.}~\bibnamefont {Semmrich}}, \bibinfo
  {author} {\bibfnamefont {E.}~\bibnamefont {Frey}}, \ and\ \bibinfo {author}
  {\bibfnamefont {A.}~\bibnamefont {Bausch}},\ }\bibfield  {title} {\enquote
  {\bibinfo {title} {Polar patterns of driven filaments},}\ }\href@noop {}
  {\bibfield  {journal} {\bibinfo  {journal} {Nature}\ }\textbf {\bibinfo
  {volume} {467}},\ \bibinfo {pages} {73} (\bibinfo {year} {2010})}\BibitemShut
  {NoStop}%
\bibitem [{\citenamefont {Bricard}\ \emph {et~al.}(2013)\citenamefont
  {Bricard}, \citenamefont {Caussin}, \citenamefont {Desreumaux}, \citenamefont
  {Dauchot},\ and\ \citenamefont {Bartolo}}]{bricard2013emergence}%
  \BibitemOpen
  \bibfield  {author} {\bibinfo {author} {\bibfnamefont {A.}~\bibnamefont
  {Bricard}}, \bibinfo {author} {\bibfnamefont {J-B}\ \bibnamefont {Caussin}},
  \bibinfo {author} {\bibfnamefont {N.}~\bibnamefont {Desreumaux}}, \bibinfo
  {author} {\bibfnamefont {O.}~\bibnamefont {Dauchot}}, \ and\ \bibinfo
  {author} {\bibfnamefont {D.}~\bibnamefont {Bartolo}},\ }\bibfield  {title}
  {\enquote {\bibinfo {title} {Emergence of macroscopic directed motion in
  populations of motile colloids},}\ }\href@noop {} {\bibfield  {journal}
  {\bibinfo  {journal} {Nature}\ }\textbf {\bibinfo {volume} {503}},\ \bibinfo
  {pages} {95} (\bibinfo {year} {2013})}\BibitemShut {NoStop}%
\bibitem [{\citenamefont {Kaiser}\ \emph {et~al.}(2017)\citenamefont {Kaiser},
  \citenamefont {Snezhko},\ and\ \citenamefont {Aranson}}]{kaiser2017flocking}%
  \BibitemOpen
  \bibfield  {author} {\bibinfo {author} {\bibfnamefont {A.}~\bibnamefont
  {Kaiser}}, \bibinfo {author} {\bibfnamefont {A.}~\bibnamefont {Snezhko}}, \
  and\ \bibinfo {author} {\bibfnamefont {I.~S.}\ \bibnamefont {Aranson}},\
  }\bibfield  {title} {\enquote {\bibinfo {title} {Flocking ferromagnetic
  colloids},}\ }\href@noop {} {\bibfield  {journal} {\bibinfo  {journal}
  {Science advances}\ }\textbf {\bibinfo {volume} {3}},\ \bibinfo {pages}
  {e1601469} (\bibinfo {year} {2017})}\BibitemShut {NoStop}%
\bibitem [{\citenamefont {Deseigne}\ \emph {et~al.}(2010)\citenamefont
  {Deseigne}, \citenamefont {Dauchot},\ and\ \citenamefont
  {Chat{\'e}}}]{deseigne2010collective}%
  \BibitemOpen
  \bibfield  {author} {\bibinfo {author} {\bibfnamefont {J.}~\bibnamefont
  {Deseigne}}, \bibinfo {author} {\bibfnamefont {O.}~\bibnamefont {Dauchot}}, \
  and\ \bibinfo {author} {\bibfnamefont {H.}~\bibnamefont {Chat{\'e}}},\
  }\bibfield  {title} {\enquote {\bibinfo {title} {Collective motion of
  vibrated polar disks},}\ }\href@noop {} {\bibfield  {journal} {\bibinfo
  {journal} {Phys. Rev. Lett.}\ }\textbf {\bibinfo {volume} {105}},\ \bibinfo
  {pages} {098001} (\bibinfo {year} {2010})}\BibitemShut {NoStop}%
\bibitem [{\citenamefont {Vicsek}\ \emph {et~al.}(1995)\citenamefont {Vicsek},
  \citenamefont {Czir\'ok}, \citenamefont {Ben-Jacob}, \citenamefont {Cohen},\
  and\ \citenamefont {Shochet}}]{Vicsek}%
  \BibitemOpen
  \bibfield  {author} {\bibinfo {author} {\bibfnamefont {T.}~\bibnamefont
  {Vicsek}}, \bibinfo {author} {\bibfnamefont {A.}~\bibnamefont {Czir\'ok}},
  \bibinfo {author} {\bibfnamefont {E.}~\bibnamefont {Ben-Jacob}}, \bibinfo
  {author} {\bibfnamefont {I.}~\bibnamefont {Cohen}}, \ and\ \bibinfo {author}
  {\bibfnamefont {O.}~\bibnamefont {Shochet}},\ }\bibfield  {title} {\enquote
  {\bibinfo {title} {Novel type of phase transition in a system of self-driven
  particles},}\ }\href@noop {} {\bibfield  {journal} {\bibinfo  {journal}
  {Phys. Rev. Lett.}\ }\textbf {\bibinfo {volume} {75}},\ \bibinfo {pages}
  {1226} (\bibinfo {year} {1995})}\BibitemShut {NoStop}%
\bibitem [{\citenamefont {Toner}\ and\ \citenamefont
  {Tu}(1995)}]{Toner1995LROXY}%
  \BibitemOpen
  \bibfield  {author} {\bibinfo {author} {\bibfnamefont {J.}~\bibnamefont
  {Toner}}\ and\ \bibinfo {author} {\bibfnamefont {Y.}~\bibnamefont {Tu}},\
  }\bibfield  {title} {\enquote {\bibinfo {title} {Long-range order in a
  two-dimensional dynamical $\mathrm{XY}$ model: How birds fly together},}\
  }\href@noop {} {\bibfield  {journal} {\bibinfo  {journal} {Phys. Rev. Lett.}\
  }\textbf {\bibinfo {volume} {75}},\ \bibinfo {pages} {4326} (\bibinfo {year}
  {1995})}\BibitemShut {NoStop}%
\bibitem [{\citenamefont {Toner}\ and\ \citenamefont
  {Tu}(1998)}]{Toner1998flocks}%
  \BibitemOpen
  \bibfield  {author} {\bibinfo {author} {\bibfnamefont {J.}~\bibnamefont
  {Toner}}\ and\ \bibinfo {author} {\bibfnamefont {Y.}~\bibnamefont {Tu}},\
  }\bibfield  {title} {\enquote {\bibinfo {title} {Flocks, herds, and schools:
  A quantitative theory of flocking},}\ }\href@noop {} {\bibfield  {journal}
  {\bibinfo  {journal} {Phys. Rev. E}\ }\textbf {\bibinfo {volume} {58}},\
  \bibinfo {pages} {4828} (\bibinfo {year} {1998})}\BibitemShut {NoStop}%
\bibitem [{\citenamefont {Solon}\ \emph {et~al.}(2015)\citenamefont {Solon},
  \citenamefont {Chat\'e},\ and\ \citenamefont {Tailleur}}]{Solon2015phase}%
  \BibitemOpen
  \bibfield  {author} {\bibinfo {author} {\bibfnamefont {A.~P.}\ \bibnamefont
  {Solon}}, \bibinfo {author} {\bibfnamefont {H.}~\bibnamefont {Chat\'e}}, \
  and\ \bibinfo {author} {\bibfnamefont {J.}~\bibnamefont {Tailleur}},\
  }\bibfield  {title} {\enquote {\bibinfo {title} {From phase to microphase
  separation in flocking models: The essential role of nonequilibrium
  fluctuations},}\ }\href@noop {} {\bibfield  {journal} {\bibinfo  {journal}
  {Phys. Rev. Lett.}\ }\textbf {\bibinfo {volume} {114}},\ \bibinfo {pages}
  {068101} (\bibinfo {year} {2015})}\BibitemShut {NoStop}%
\bibitem [{\citenamefont {Menzel}(2012)}]{menzel2012collective}%
  \BibitemOpen
  \bibfield  {author} {\bibinfo {author} {\bibfnamefont {A.~M.}\ \bibnamefont
  {Menzel}},\ }\bibfield  {title} {\enquote {\bibinfo {title} {Collective
  motion of binary self-propelled particle mixtures},}\ }\href@noop {}
  {\bibfield  {journal} {\bibinfo  {journal} {Phys. Rev. E}\ }\textbf {\bibinfo
  {volume} {85}},\ \bibinfo {pages} {021912} (\bibinfo {year}
  {2012})}\BibitemShut {NoStop}%
\bibitem [{\citenamefont {Fruchart}\ \emph {et~al.}(2021)\citenamefont
  {Fruchart}, \citenamefont {Hanai}, \citenamefont {Littlewood},\ and\
  \citenamefont {Vitelli}}]{NRVM}%
  \BibitemOpen
  \bibfield  {author} {\bibinfo {author} {\bibfnamefont {M.}~\bibnamefont
  {Fruchart}}, \bibinfo {author} {\bibfnamefont {R.}~\bibnamefont {Hanai}},
  \bibinfo {author} {\bibfnamefont {P.~B.}\ \bibnamefont {Littlewood}}, \ and\
  \bibinfo {author} {\bibfnamefont {V.}~\bibnamefont {Vitelli}},\ }\bibfield
  {title} {\enquote {\bibinfo {title} {Non-reciprocal phase transitions},}\
  }\href@noop {} {\bibfield  {journal} {\bibinfo  {journal} {Nature}\ }\textbf
  {\bibinfo {volume} {592}},\ \bibinfo {pages} {363} (\bibinfo {year}
  {2021})}\BibitemShut {NoStop}%
\bibitem [{\citenamefont {Kreienkamp}\ and\ \citenamefont
  {Klapp}(2022)}]{kreienkamp2022clustering}%
  \BibitemOpen
  \bibfield  {author} {\bibinfo {author} {\bibfnamefont {K.~L.}\ \bibnamefont
  {Kreienkamp}}\ and\ \bibinfo {author} {\bibfnamefont {S.H.L.}\ \bibnamefont
  {Klapp}},\ }\bibfield  {title} {\enquote {\bibinfo {title} {Clustering and
  flocking of repulsive chiral active particles with non-reciprocal
  couplings},}\ }\href@noop {} {\bibfield  {journal} {\bibinfo  {journal} {New
  J. Phys.}\ }\textbf {\bibinfo {volume} {24}},\ \bibinfo {pages} {123009}
  (\bibinfo {year} {2022})}\BibitemShut {NoStop}%
\bibitem [{\citenamefont {Chatterjee}\ \emph {et~al.}(2023)\citenamefont
  {Chatterjee}, \citenamefont {Mangeat}, \citenamefont {Woo}, \citenamefont
  {Rieger},\ and\ \citenamefont {Noh}}]{SwarnajitTSVM}%
  \BibitemOpen
  \bibfield  {author} {\bibinfo {author} {\bibfnamefont {S.}~\bibnamefont
  {Chatterjee}}, \bibinfo {author} {\bibfnamefont {M.}~\bibnamefont {Mangeat}},
  \bibinfo {author} {\bibfnamefont {C-U}\ \bibnamefont {Woo}}, \bibinfo
  {author} {\bibfnamefont {H.}~\bibnamefont {Rieger}}, \ and\ \bibinfo {author}
  {\bibfnamefont {J.~D.}\ \bibnamefont {Noh}},\ }\bibfield  {title} {\enquote
  {\bibinfo {title} {Flocking of two unfriendly species: The two-species vicsek
  model},}\ }\href@noop {} {\bibfield  {journal} {\bibinfo  {journal} {Phys.
  Rev. E}\ }\textbf {\bibinfo {volume} {107}},\ \bibinfo {pages} {024607}
  (\bibinfo {year} {2023})}\BibitemShut {NoStop}%
\bibitem [{\citenamefont {Martin}\ \emph {et~al.}(2023)\citenamefont {Martin},
  \citenamefont {Seara}, \citenamefont {Avni}, \citenamefont {Fruchart},\ and\
  \citenamefont {Vitelli}}]{NRASM}%
  \BibitemOpen
  \bibfield  {author} {\bibinfo {author} {\bibfnamefont {D.}~\bibnamefont
  {Martin}}, \bibinfo {author} {\bibfnamefont {D.}~\bibnamefont {Seara}},
  \bibinfo {author} {\bibfnamefont {Y.}~\bibnamefont {Avni}}, \bibinfo {author}
  {\bibfnamefont {M.}~\bibnamefont {Fruchart}}, \ and\ \bibinfo {author}
  {\bibfnamefont {V.}~\bibnamefont {Vitelli}},\ }\bibfield  {title} {\enquote
  {\bibinfo {title} {The transition to collective motion in nonreciprocal
  active matter: coarse graining agent-based models into fluctuating
  hydrodynamics},}\ }\href@noop {} {\bibfield  {journal} {\bibinfo  {journal}
  {arXiv preprint arXiv:2307.08251}\ } (\bibinfo {year} {2023})}\BibitemShut
  {NoStop}%
\bibitem [{\citenamefont {Mangeat}\ \emph {et~al.}(2025)\citenamefont
  {Mangeat}, \citenamefont {Chatterjee}, \citenamefont {Noh},\ and\
  \citenamefont {Rieger}}]{mangeat2024emergent}%
  \BibitemOpen
  \bibfield  {author} {\bibinfo {author} {\bibfnamefont {M.}~\bibnamefont
  {Mangeat}}, \bibinfo {author} {\bibfnamefont {S.}~\bibnamefont {Chatterjee}},
  \bibinfo {author} {\bibfnamefont {J.~D.}\ \bibnamefont {Noh}}, \ and\
  \bibinfo {author} {\bibfnamefont {H.}~\bibnamefont {Rieger}},\ }\bibfield
  {title} {\enquote {\bibinfo {title} {Emergent complex phases in a discrete
  flocking model with reciprocal and non-reciprocal interactions},}\
  }\href@noop {} {\bibfield  {journal} {\bibinfo  {journal} {Commun. Phys.}\
  }\textbf {\bibinfo {volume} {8}},\ \bibinfo {pages} {186} (\bibinfo {year}
  {2025})}\BibitemShut {NoStop}%
\bibitem [{\citenamefont {Grauer}\ \emph {et~al.}(2020)\citenamefont {Grauer},
  \citenamefont {L{\"o}wen}, \citenamefont {Be’er},\ and\ \citenamefont
  {Liebchen}}]{grauer2020swarm}%
  \BibitemOpen
  \bibfield  {author} {\bibinfo {author} {\bibfnamefont {J.}~\bibnamefont
  {Grauer}}, \bibinfo {author} {\bibfnamefont {H.}~\bibnamefont {L{\"o}wen}},
  \bibinfo {author} {\bibfnamefont {A.}~\bibnamefont {Be’er}}, \ and\
  \bibinfo {author} {\bibfnamefont {B.}~\bibnamefont {Liebchen}},\ }\bibfield
  {title} {\enquote {\bibinfo {title} {Swarm hunting and cluster ejections in
  chemically communicating active mixtures},}\ }\href@noop {} {\bibfield
  {journal} {\bibinfo  {journal} {Scientific Reports}\ }\textbf {\bibinfo
  {volume} {10}},\ \bibinfo {pages} {5594} (\bibinfo {year}
  {2020})}\BibitemShut {NoStop}%
\bibitem [{\citenamefont {Tucci}\ \emph {et~al.}(2024)\citenamefont {Tucci},
  \citenamefont {Golestanian},\ and\ \citenamefont
  {Saha}}]{tucci2024nonreciprocal}%
  \BibitemOpen
  \bibfield  {author} {\bibinfo {author} {\bibfnamefont {G.}~\bibnamefont
  {Tucci}}, \bibinfo {author} {\bibfnamefont {R.}~\bibnamefont {Golestanian}},
  \ and\ \bibinfo {author} {\bibfnamefont {S.}~\bibnamefont {Saha}},\
  }\bibfield  {title} {\enquote {\bibinfo {title} {Nonreciprocal collective
  dynamics in a mixture of phoretic janus colloids},}\ }\href@noop {}
  {\bibfield  {journal} {\bibinfo  {journal} {New J. Phys.}\ }\textbf {\bibinfo
  {volume} {26}},\ \bibinfo {pages} {073006} (\bibinfo {year}
  {2024})}\BibitemShut {NoStop}%
\bibitem [{\citenamefont {Tucci}\ \emph {et~al.}(2025)\citenamefont {Tucci},
  \citenamefont {Pisegna}, \citenamefont {Golestanian},\ and\ \citenamefont
  {Saha}}]{tucci2025hydrodynamic}%
  \BibitemOpen
  \bibfield  {author} {\bibinfo {author} {\bibfnamefont {G.}~\bibnamefont
  {Tucci}}, \bibinfo {author} {\bibfnamefont {G.}~\bibnamefont {Pisegna}},
  \bibinfo {author} {\bibfnamefont {R.}~\bibnamefont {Golestanian}}, \ and\
  \bibinfo {author} {\bibfnamefont {S.}~\bibnamefont {Saha}},\ }\bibfield
  {title} {\enquote {\bibinfo {title} {Hydrodynamic stresses in a multi-species
  suspension of active janus colloids},}\ }\href@noop {} {\bibfield  {journal}
  {\bibinfo  {journal} {arXiv preprint arXiv:2502.07744}\ } (\bibinfo {year}
  {2025})}\BibitemShut {NoStop}%
\bibitem [{\citenamefont {Ben-Jacob}\ \emph {et~al.}(2016)\citenamefont
  {Ben-Jacob}, \citenamefont {Finkelshtein}, \citenamefont {Ariel},\ and\
  \citenamefont {Ingham}}]{ben2016multispecies}%
  \BibitemOpen
  \bibfield  {author} {\bibinfo {author} {\bibfnamefont {E.}~\bibnamefont
  {Ben-Jacob}}, \bibinfo {author} {\bibfnamefont {A.}~\bibnamefont
  {Finkelshtein}}, \bibinfo {author} {\bibfnamefont {G.}~\bibnamefont {Ariel}},
  \ and\ \bibinfo {author} {\bibfnamefont {C.}~\bibnamefont {Ingham}},\
  }\bibfield  {title} {\enquote {\bibinfo {title} {Multispecies swarms of
  social microorganisms as moving ecosystems},}\ }\href@noop {} {\bibfield
  {journal} {\bibinfo  {journal} {Trends in Microbiology}\ }\textbf {\bibinfo
  {volume} {24}},\ \bibinfo {pages} {257} (\bibinfo {year} {2016})}\BibitemShut
  {NoStop}%
\bibitem [{\citenamefont {Herbert-Read}\ \emph {et~al.}(2013)\citenamefont
  {Herbert-Read}, \citenamefont {Krause}, \citenamefont {Morrell},
  \citenamefont {Schaerf}, \citenamefont {Krause},\ and\ \citenamefont
  {Ward}}]{Herbert2013colmove}%
  \BibitemOpen
  \bibfield  {author} {\bibinfo {author} {\bibfnamefont {J.}~\bibnamefont
  {Herbert-Read}}, \bibinfo {author} {\bibfnamefont {S.}~\bibnamefont
  {Krause}}, \bibinfo {author} {\bibfnamefont {L.}~\bibnamefont {Morrell}},
  \bibinfo {author} {\bibfnamefont {T.}~\bibnamefont {Schaerf}}, \bibinfo
  {author} {\bibfnamefont {J.}~\bibnamefont {Krause}}, \ and\ \bibinfo {author}
  {\bibfnamefont {A.}~\bibnamefont {Ward}},\ }\bibfield  {title} {\enquote
  {\bibinfo {title} {The role of individuality in collective group movement},}\
  }\href@noop {} {\bibfield  {journal} {\bibinfo  {journal} {Proc. R. Soc. B}\
  }\textbf {\bibinfo {volume} {280}},\ \bibinfo {pages} {20122564} (\bibinfo
  {year} {2013})}\BibitemShut {NoStop}%
\bibitem [{\citenamefont {Ilkanaiv}\ \emph {et~al.}(2017)\citenamefont
  {Ilkanaiv}, \citenamefont {Kearns}, \citenamefont {Ariel},\ and\
  \citenamefont {Be’er}}]{ilkanaiv2017effect}%
  \BibitemOpen
  \bibfield  {author} {\bibinfo {author} {\bibfnamefont {B.}~\bibnamefont
  {Ilkanaiv}}, \bibinfo {author} {\bibfnamefont {D.~B.}\ \bibnamefont
  {Kearns}}, \bibinfo {author} {\bibfnamefont {G.}~\bibnamefont {Ariel}}, \
  and\ \bibinfo {author} {\bibfnamefont {A.}~\bibnamefont {Be’er}},\
  }\bibfield  {title} {\enquote {\bibinfo {title} {Effect of cell aspect ratio
  on swarming bacteria},}\ }\href@noop {} {\bibfield  {journal} {\bibinfo
  {journal} {Phys. Rev. Lett.}\ }\textbf {\bibinfo {volume} {118}},\ \bibinfo
  {pages} {158002} (\bibinfo {year} {2017})}\BibitemShut {NoStop}%
\bibitem [{\citenamefont {Peled}\ \emph {et~al.}(2021)\citenamefont {Peled},
  \citenamefont {Ryan}, \citenamefont {Heidenreich}, \citenamefont {B\"ar},
  \citenamefont {Ariel},\ and\ \citenamefont {Be'er}}]{Peled2021bacteria}%
  \BibitemOpen
  \bibfield  {author} {\bibinfo {author} {\bibfnamefont {S.}~\bibnamefont
  {Peled}}, \bibinfo {author} {\bibfnamefont {S.~D.}\ \bibnamefont {Ryan}},
  \bibinfo {author} {\bibfnamefont {S.}~\bibnamefont {Heidenreich}}, \bibinfo
  {author} {\bibfnamefont {M.}~\bibnamefont {B\"ar}}, \bibinfo {author}
  {\bibfnamefont {G.}~\bibnamefont {Ariel}}, \ and\ \bibinfo {author}
  {\bibfnamefont {A.}~\bibnamefont {Be'er}},\ }\bibfield  {title} {\enquote
  {\bibinfo {title} {Heterogeneous bacterial swarms with mixed lengths},}\
  }\href@noop {} {\bibfield  {journal} {\bibinfo  {journal} {Phys. Rev. E}\
  }\textbf {\bibinfo {volume} {103}},\ \bibinfo {pages} {032413} (\bibinfo
  {year} {2021})}\BibitemShut {NoStop}%
\bibitem [{\citenamefont {Jose}\ \emph {et~al.}(2022)\citenamefont {Jose},
  \citenamefont {Ariel},\ and\ \citenamefont {Be'er}}]{Jose2022mixedspecies}%
  \BibitemOpen
  \bibfield  {author} {\bibinfo {author} {\bibfnamefont {A.}~\bibnamefont
  {Jose}}, \bibinfo {author} {\bibfnamefont {G.}~\bibnamefont {Ariel}}, \ and\
  \bibinfo {author} {\bibfnamefont {A.}~\bibnamefont {Be'er}},\ }\bibfield
  {title} {\enquote {\bibinfo {title} {Physical characteristics of
  mixed-species swarming colonies},}\ }\href@noop {} {\bibfield  {journal}
  {\bibinfo  {journal} {Phys. Rev. E}\ }\textbf {\bibinfo {volume} {105}},\
  \bibinfo {pages} {064404} (\bibinfo {year} {2022})}\BibitemShut {NoStop}%
\bibitem [{\citenamefont {Natan}\ \emph {et~al.}(2022)\citenamefont {Natan},
  \citenamefont {Worlitzer}, \citenamefont {Ariel},\ and\ \citenamefont
  {Be’er}}]{Natan2022bacteria}%
  \BibitemOpen
  \bibfield  {author} {\bibinfo {author} {\bibfnamefont {G.}~\bibnamefont
  {Natan}}, \bibinfo {author} {\bibfnamefont {V.}~\bibnamefont {Worlitzer}},
  \bibinfo {author} {\bibfnamefont {G.}~\bibnamefont {Ariel}}, \ and\ \bibinfo
  {author} {\bibfnamefont {A.}~\bibnamefont {Be’er}},\ }\bibfield  {title}
  {\enquote {\bibinfo {title} {Mixed-species bacterial swarms show an interplay
  of mixing and segregation across scales},}\ }\href@noop {} {\bibfield
  {journal} {\bibinfo  {journal} {Sci. Rep.}\ }\textbf {\bibinfo {volume}
  {12}},\ \bibinfo {pages} {16500} (\bibinfo {year} {2022})}\BibitemShut
  {NoStop}%
\bibitem [{\citenamefont {Zuo}\ and\ \citenamefont
  {Wu}(2020)}]{zuo2020dynamic}%
  \BibitemOpen
  \bibfield  {author} {\bibinfo {author} {\bibfnamefont {W.}~\bibnamefont
  {Zuo}}\ and\ \bibinfo {author} {\bibfnamefont {Y.}~\bibnamefont {Wu}},\
  }\bibfield  {title} {\enquote {\bibinfo {title} {Dynamic motility selection
  drives population segregation in a bacterial swarm},}\ }\href@noop {}
  {\bibfield  {journal} {\bibinfo  {journal} {Proc. Natl. Acad. Sci.}\ }\textbf
  {\bibinfo {volume} {117}},\ \bibinfo {pages} {4693} (\bibinfo {year}
  {2020})}\BibitemShut {NoStop}%
\bibitem [{\citenamefont {Kolb}\ and\ \citenamefont
  {Klotsa}(2020)}]{kolb2020active}%
  \BibitemOpen
  \bibfield  {author} {\bibinfo {author} {\bibfnamefont {T.}~\bibnamefont
  {Kolb}}\ and\ \bibinfo {author} {\bibfnamefont {D.}~\bibnamefont {Klotsa}},\
  }\bibfield  {title} {\enquote {\bibinfo {title} {Active binary mixtures of
  fast and slow hard spheres},}\ }\href@noop {} {\bibfield  {journal} {\bibinfo
   {journal} {Soft Matter}\ }\textbf {\bibinfo {volume} {16}},\ \bibinfo
  {pages} {1967} (\bibinfo {year} {2020})}\BibitemShut {NoStop}%
\bibitem [{\citenamefont {Pattanayak}\ \emph {et~al.}(2020)\citenamefont
  {Pattanayak}, \citenamefont {Singh}, \citenamefont {Kumar},\ and\
  \citenamefont {Mishra}}]{pattanayak2020speed}%
  \BibitemOpen
  \bibfield  {author} {\bibinfo {author} {\bibfnamefont {S.}~\bibnamefont
  {Pattanayak}}, \bibinfo {author} {\bibfnamefont {J.~P.}\ \bibnamefont
  {Singh}}, \bibinfo {author} {\bibfnamefont {M.}~\bibnamefont {Kumar}}, \ and\
  \bibinfo {author} {\bibfnamefont {S.}~\bibnamefont {Mishra}},\ }\bibfield
  {title} {\enquote {\bibinfo {title} {Speed inhomogeneity accelerates
  information transfer in polar flock},}\ }\href@noop {} {\bibfield  {journal}
  {\bibinfo  {journal} {Phys. Rev. E}\ }\textbf {\bibinfo {volume} {101}},\
  \bibinfo {pages} {052602} (\bibinfo {year} {2020})}\BibitemShut {NoStop}%
\bibitem [{\citenamefont {Forget}\ \emph {et~al.}(2022)\citenamefont {Forget},
  \citenamefont {Adiba}, \citenamefont {Brunnet},\ and\ \citenamefont
  {De~Monte}}]{Forget2022hetmol}%
  \BibitemOpen
  \bibfield  {author} {\bibinfo {author} {\bibfnamefont {M.}~\bibnamefont
  {Forget}}, \bibinfo {author} {\bibfnamefont {S.}~\bibnamefont {Adiba}},
  \bibinfo {author} {\bibfnamefont {L.~G.}\ \bibnamefont {Brunnet}}, \ and\
  \bibinfo {author} {\bibfnamefont {S.}~\bibnamefont {De~Monte}},\ }\bibfield
  {title} {\enquote {\bibinfo {title} {Heterogeneous individual motility biases
  group composition in a model of aggregating cells},}\ }\href@noop {}
  {\bibfield  {journal} {\bibinfo  {journal} {Front. Ecol. Evol.}\ }\textbf
  {\bibinfo {volume} {10}} (\bibinfo {year} {2022})}\BibitemShut {NoStop}%
\bibitem [{\citenamefont {Maity}\ and\ \citenamefont
  {Morin}(2023)}]{maity2023spontaneous}%
  \BibitemOpen
  \bibfield  {author} {\bibinfo {author} {\bibfnamefont {S.}~\bibnamefont
  {Maity}}\ and\ \bibinfo {author} {\bibfnamefont {A.}~\bibnamefont {Morin}},\
  }\bibfield  {title} {\enquote {\bibinfo {title} {Spontaneous demixing of
  binary colloidal flocks},}\ }\href@noop {} {\bibfield  {journal} {\bibinfo
  {journal} {Phys. Rev. Lett.}\ }\textbf {\bibinfo {volume} {131}},\ \bibinfo
  {pages} {178304} (\bibinfo {year} {2023})}\BibitemShut {NoStop}%
\bibitem [{\citenamefont {Pigolotti}\ and\ \citenamefont
  {Benzi}(2014)}]{pigolotti2014selective}%
  \BibitemOpen
  \bibfield  {author} {\bibinfo {author} {\bibfnamefont {S.}~\bibnamefont
  {Pigolotti}}\ and\ \bibinfo {author} {\bibfnamefont {R.}~\bibnamefont
  {Benzi}},\ }\bibfield  {title} {\enquote {\bibinfo {title} {Selective
  advantage of diffusing faster},}\ }\href@noop {} {\bibfield  {journal}
  {\bibinfo  {journal} {Phys. Rev. Lett.}\ }\textbf {\bibinfo {volume} {112}},\
  \bibinfo {pages} {188102} (\bibinfo {year} {2014})}\BibitemShut {NoStop}%
\bibitem [{\citenamefont {Book}\ \emph {et~al.}(2017)\citenamefont {Book},
  \citenamefont {Ingham},\ and\ \citenamefont {Ariel}}]{book2017modeling}%
  \BibitemOpen
  \bibfield  {author} {\bibinfo {author} {\bibfnamefont {G.}~\bibnamefont
  {Book}}, \bibinfo {author} {\bibfnamefont {C.}~\bibnamefont {Ingham}}, \ and\
  \bibinfo {author} {\bibfnamefont {G.}~\bibnamefont {Ariel}},\ }\bibfield
  {title} {\enquote {\bibinfo {title} {Modeling cooperating micro-organisms in
  antibiotic environment},}\ }\href@noop {} {\bibfield  {journal} {\bibinfo
  {journal} {PLoS One}\ }\textbf {\bibinfo {volume} {12}},\ \bibinfo {pages}
  {e0190037} (\bibinfo {year} {2017})}\BibitemShut {NoStop}%
\bibitem [{\citenamefont {Khodygo}\ \emph {et~al.}(2019)\citenamefont
  {Khodygo}, \citenamefont {Swain},\ and\ \citenamefont
  {Mughal}}]{khodygo2019homogeneous}%
  \BibitemOpen
  \bibfield  {author} {\bibinfo {author} {\bibfnamefont {V.}~\bibnamefont
  {Khodygo}}, \bibinfo {author} {\bibfnamefont {M.~T.}\ \bibnamefont {Swain}},
  \ and\ \bibinfo {author} {\bibfnamefont {A.}~\bibnamefont {Mughal}},\
  }\bibfield  {title} {\enquote {\bibinfo {title} {Homogeneous and
  heterogeneous populations of active rods in two-dimensional channels},}\
  }\href@noop {} {\bibfield  {journal} {\bibinfo  {journal} {Phys. Rev. E}\
  }\textbf {\bibinfo {volume} {99}},\ \bibinfo {pages} {022602} (\bibinfo
  {year} {2019})}\BibitemShut {NoStop}%
\bibitem [{\citenamefont {Lardet}\ \emph {et~al.}(2025)\citenamefont {Lardet},
  \citenamefont {Chen},\ and\ \citenamefont {Bertrand}}]{lardet2025flocking}%
  \BibitemOpen
  \bibfield  {author} {\bibinfo {author} {\bibfnamefont {E.}~\bibnamefont
  {Lardet}}, \bibinfo {author} {\bibfnamefont {L.}~\bibnamefont {Chen}}, \ and\
  \bibinfo {author} {\bibfnamefont {T.}~\bibnamefont {Bertrand}},\ }\bibfield
  {title} {\enquote {\bibinfo {title} {Flocking beyond one species: Novel phase
  coexistence in a generalized two-species vicsek model},}\ }\href@noop {}
  {\bibfield  {journal} {\bibinfo  {journal} {arXiv preprint arXiv:2503.17617}\
  } (\bibinfo {year} {2025})}\BibitemShut {NoStop}%
\bibitem [{\citenamefont {Khelfa}\ \emph {et~al.}(2022)\citenamefont {Khelfa},
  \citenamefont {Korbmacher}, \citenamefont {Schadschneider},\ and\
  \citenamefont {Tordeux}}]{khelfa2022heterogeneity}%
  \BibitemOpen
  \bibfield  {author} {\bibinfo {author} {\bibfnamefont {B.}~\bibnamefont
  {Khelfa}}, \bibinfo {author} {\bibfnamefont {R.}~\bibnamefont {Korbmacher}},
  \bibinfo {author} {\bibfnamefont {A.}~\bibnamefont {Schadschneider}}, \ and\
  \bibinfo {author} {\bibfnamefont {A.}~\bibnamefont {Tordeux}},\ }\bibfield
  {title} {\enquote {\bibinfo {title} {Heterogeneity-induced lane and band
  formation in self-driven particle systems},}\ }\href@noop {} {\bibfield
  {journal} {\bibinfo  {journal} {Scientific reports}\ }\textbf {\bibinfo
  {volume} {12}},\ \bibinfo {pages} {4768} (\bibinfo {year}
  {2022})}\BibitemShut {NoStop}%
\bibitem [{\citenamefont {Ariel}\ \emph {et~al.}(2015)\citenamefont {Ariel},
  \citenamefont {Rimer},\ and\ \citenamefont {Ben-Jacob}}]{Ariel2014HetSPP}%
  \BibitemOpen
  \bibfield  {author} {\bibinfo {author} {\bibfnamefont {G.}~\bibnamefont
  {Ariel}}, \bibinfo {author} {\bibfnamefont {O.}~\bibnamefont {Rimer}}, \ and\
  \bibinfo {author} {\bibfnamefont {E.}~\bibnamefont {Ben-Jacob}},\ }\bibfield
  {title} {\enquote {\bibinfo {title} {Order--disorder phase transition in
  heterogeneous populations of self-propelled particles},}\ }\href@noop {}
  {\bibfield  {journal} {\bibinfo  {journal} {J. Stat. Phys.}\ }\textbf
  {\bibinfo {volume} {158}},\ \bibinfo {pages} {579} (\bibinfo {year}
  {2015})}\BibitemShut {NoStop}%
\bibitem [{\citenamefont {Netzer}\ \emph {et~al.}(2019)\citenamefont {Netzer},
  \citenamefont {Yarom},\ and\ \citenamefont {Ariel}}]{Netzer20191hetpop}%
  \BibitemOpen
  \bibfield  {author} {\bibinfo {author} {\bibfnamefont {G.}~\bibnamefont
  {Netzer}}, \bibinfo {author} {\bibfnamefont {Y.}~\bibnamefont {Yarom}}, \
  and\ \bibinfo {author} {\bibfnamefont {G.}~\bibnamefont {Ariel}},\ }\bibfield
   {title} {\enquote {\bibinfo {title} {Heterogeneous populations in a network
  model of collective motion},}\ }\href@noop {} {\bibfield  {journal} {\bibinfo
   {journal} {Phys. A: Stat. Mech. Appl.}\ }\textbf {\bibinfo {volume} {530}},\
  \bibinfo {pages} {121550} (\bibinfo {year} {2019})}\BibitemShut {NoStop}%
\bibitem [{\citenamefont {Ilker}\ and\ \citenamefont
  {Joanny}(2020)}]{ilker2020phase}%
  \BibitemOpen
  \bibfield  {author} {\bibinfo {author} {\bibfnamefont {E.}~\bibnamefont
  {Ilker}}\ and\ \bibinfo {author} {\bibfnamefont {J.~F.}\ \bibnamefont
  {Joanny}},\ }\bibfield  {title} {\enquote {\bibinfo {title} {Phase separation
  and nucleation in mixtures of particles with different temperatures},}\
  }\href@noop {} {\bibfield  {journal} {\bibinfo  {journal} {Phys. Rev. Res.}\
  }\textbf {\bibinfo {volume} {2}},\ \bibinfo {pages} {023200} (\bibinfo {year}
  {2020})}\BibitemShut {NoStop}%
\bibitem [{\citenamefont {Wilde}\ and\ \citenamefont
  {Mullineaux}(2017)}]{Wilde2017prokaryotes}%
  \BibitemOpen
  \bibfield  {author} {\bibinfo {author} {\bibfnamefont {A.}~\bibnamefont
  {Wilde}}\ and\ \bibinfo {author} {\bibfnamefont {C.~W.}\ \bibnamefont
  {Mullineaux}},\ }\bibfield  {title} {\enquote {\bibinfo {title}
  {Light-controlled motility in prokaryotes and the problem of directional
  light perception},}\ }\href@noop {} {\bibfield  {journal} {\bibinfo
  {journal} {FEMS Microbiol. Rev.}\ }\textbf {\bibinfo {volume} {41}},\
  \bibinfo {pages} {900} (\bibinfo {year} {2017})}\BibitemShut {NoStop}%
\bibitem [{\citenamefont {Frangipane}\ \emph {et~al.}(2018)\citenamefont
  {Frangipane}, \citenamefont {Dell'Arciprete}, \citenamefont {Petracchini},
  \citenamefont {Maggi}, \citenamefont {Saglimbeni}, \citenamefont {Bianchi},
  \citenamefont {Vizsnyiczai}, \citenamefont {Bernardini},\ and\ \citenamefont
  {Di~Leonardo}}]{FrangipaneEcoli}%
  \BibitemOpen
  \bibfield  {author} {\bibinfo {author} {\bibfnamefont {G.}~\bibnamefont
  {Frangipane}}, \bibinfo {author} {\bibfnamefont {D.}~\bibnamefont
  {Dell'Arciprete}}, \bibinfo {author} {\bibfnamefont {S.}~\bibnamefont
  {Petracchini}}, \bibinfo {author} {\bibfnamefont {C.}~\bibnamefont {Maggi}},
  \bibinfo {author} {\bibfnamefont {F.}~\bibnamefont {Saglimbeni}}, \bibinfo
  {author} {\bibfnamefont {S.}~\bibnamefont {Bianchi}}, \bibinfo {author}
  {\bibfnamefont {G.}~\bibnamefont {Vizsnyiczai}}, \bibinfo {author}
  {\bibfnamefont {M.~L.}\ \bibnamefont {Bernardini}}, \ and\ \bibinfo {author}
  {\bibfnamefont {R.}~\bibnamefont {Di~Leonardo}},\ }\bibfield  {title}
  {\enquote {\bibinfo {title} {Dynamic density shaping of photokinetic
  \textit{E. coli}},}\ }\href@noop {} {\bibfield  {journal} {\bibinfo
  {journal} {eLife}\ }\textbf {\bibinfo {volume} {7}},\ \bibinfo {pages}
  {e36608} (\bibinfo {year} {2018})}\BibitemShut {NoStop}%
\bibitem [{\citenamefont {S\"oker}\ \emph {et~al.}(2021)\citenamefont
  {S\"oker}, \citenamefont {Auschra}, \citenamefont {Holubec}, \citenamefont
  {Kroy},\ and\ \citenamefont {Cichos}}]{andreas2021microswimmer}%
  \BibitemOpen
  \bibfield  {author} {\bibinfo {author} {\bibfnamefont {N.~A.}\ \bibnamefont
  {S\"oker}}, \bibinfo {author} {\bibfnamefont {S.}~\bibnamefont {Auschra}},
  \bibinfo {author} {\bibfnamefont {V.}~\bibnamefont {Holubec}}, \bibinfo
  {author} {\bibfnamefont {K.}~\bibnamefont {Kroy}}, \ and\ \bibinfo {author}
  {\bibfnamefont {F.}~\bibnamefont {Cichos}},\ }\bibfield  {title} {\enquote
  {\bibinfo {title} {How activity landscapes polarize microswimmers without
  alignment forces},}\ }\href@noop {} {\bibfield  {journal} {\bibinfo
  {journal} {Phys. Rev. Lett.}\ }\textbf {\bibinfo {volume} {126}},\ \bibinfo
  {pages} {228001} (\bibinfo {year} {2021})}\BibitemShut {NoStop}%
\bibitem [{\citenamefont {Auschra}\ \emph {et~al.}(2021)\citenamefont
  {Auschra}, \citenamefont {Holubec}, \citenamefont {S{\"o}ker}, \citenamefont
  {Cichos},\ and\ \citenamefont {Kroy}}]{auschra2021polarization}%
  \BibitemOpen
  \bibfield  {author} {\bibinfo {author} {\bibfnamefont {S.}~\bibnamefont
  {Auschra}}, \bibinfo {author} {\bibfnamefont {V.}~\bibnamefont {Holubec}},
  \bibinfo {author} {\bibfnamefont {N.~A.}\ \bibnamefont {S{\"o}ker}}, \bibinfo
  {author} {\bibfnamefont {F.}~\bibnamefont {Cichos}}, \ and\ \bibinfo {author}
  {\bibfnamefont {K.}~\bibnamefont {Kroy}},\ }\bibfield  {title} {\enquote
  {\bibinfo {title} {Polarization-density patterns of active particles in
  motility gradients},}\ }\href@noop {} {\bibfield  {journal} {\bibinfo
  {journal} {Phys. Rev. E}\ }\textbf {\bibinfo {volume} {103}},\ \bibinfo
  {pages} {062601} (\bibinfo {year} {2021})}\BibitemShut {NoStop}%
\bibitem [{\citenamefont {Wysocki}\ \emph {et~al.}(2022)\citenamefont
  {Wysocki}, \citenamefont {Dasanna},\ and\ \citenamefont
  {Rieger}}]{Wysocki_2022}%
  \BibitemOpen
  \bibfield  {author} {\bibinfo {author} {\bibfnamefont {A.}~\bibnamefont
  {Wysocki}}, \bibinfo {author} {\bibfnamefont {A.~K.}\ \bibnamefont
  {Dasanna}}, \ and\ \bibinfo {author} {\bibfnamefont {H.}~\bibnamefont
  {Rieger}},\ }\bibfield  {title} {\enquote {\bibinfo {title} {Interacting
  particles in an activity landscape},}\ }\href@noop {} {\bibfield  {journal}
  {\bibinfo  {journal} {New J. Phys.}\ }\textbf {\bibinfo {volume} {24}},\
  \bibinfo {pages} {093013} (\bibinfo {year} {2022})}\BibitemShut {NoStop}%
\bibitem [{\citenamefont {Huang}\ and\ \citenamefont
  {Shao}(2024)}]{BCVM_Chirality_Huang2024}%
  \BibitemOpen
  \bibfield  {author} {\bibinfo {author} {\bibfnamefont {J.}~\bibnamefont
  {Huang}}\ and\ \bibinfo {author} {\bibfnamefont {Z.}~\bibnamefont {Shao}},\
  }\bibfield  {title} {\enquote {\bibinfo {title} {Collective motion of binary
  chiral particle mixtures with environmental complex noise},}\ }\href@noop {}
  {\bibfield  {journal} {\bibinfo  {journal} {Phys. Rev. E}\ }\textbf {\bibinfo
  {volume} {110}},\ \bibinfo {pages} {034135} (\bibinfo {year}
  {2024})}\BibitemShut {NoStop}%
\bibitem [{\citenamefont {Reichhardt}\ and\ \citenamefont
  {Reichhardt}(2018)}]{reichhardt2018avalanche}%
  \BibitemOpen
  \bibfield  {author} {\bibinfo {author} {\bibfnamefont {C.~J.~O.}\
  \bibnamefont {Reichhardt}}\ and\ \bibinfo {author} {\bibfnamefont
  {C.}~\bibnamefont {Reichhardt}},\ }\bibfield  {title} {\enquote {\bibinfo
  {title} {Avalanche dynamics for active matter in heterogeneous media},}\
  }\href@noop {} {\bibfield  {journal} {\bibinfo  {journal} {New J. Phys.}\
  }\textbf {\bibinfo {volume} {20}},\ \bibinfo {pages} {025002} (\bibinfo
  {year} {2018})}\BibitemShut {NoStop}%
\bibitem [{\citenamefont {Saavedra}\ and\ \citenamefont
  {Peruani}(2024)}]{saavedra2024self}%
  \BibitemOpen
  \bibfield  {author} {\bibinfo {author} {\bibfnamefont {R.}~\bibnamefont
  {Saavedra}}\ and\ \bibinfo {author} {\bibfnamefont {F.}~\bibnamefont
  {Peruani}},\ }\bibfield  {title} {\enquote {\bibinfo {title} {Self-trapping
  of active particles with nonreciprocal interactions in disordered media},}\
  }\href@noop {} {\bibfield  {journal} {\bibinfo  {journal} {Phys. Rev. E}\
  }\textbf {\bibinfo {volume} {110}},\ \bibinfo {pages} {064602} (\bibinfo
  {year} {2024})}\BibitemShut {NoStop}%
\bibitem [{\citenamefont {Rahmani}\ \emph {et~al.}(2021)\citenamefont
  {Rahmani}, \citenamefont {Peruani},\ and\ \citenamefont
  {Romanczuk}}]{rahmani2021topological}%
  \BibitemOpen
  \bibfield  {author} {\bibinfo {author} {\bibfnamefont {P.}~\bibnamefont
  {Rahmani}}, \bibinfo {author} {\bibfnamefont {F.}~\bibnamefont {Peruani}}, \
  and\ \bibinfo {author} {\bibfnamefont {P.}~\bibnamefont {Romanczuk}},\
  }\bibfield  {title} {\enquote {\bibinfo {title} {Topological flocking models
  in spatially heterogeneous environments},}\ }\href@noop {} {\bibfield
  {journal} {\bibinfo  {journal} {Comm. Phys.}\ }\textbf {\bibinfo {volume}
  {4}},\ \bibinfo {pages} {206} (\bibinfo {year} {2021})}\BibitemShut {NoStop}%
\bibitem [{\citenamefont {Dutta}\ \emph {et~al.}(2025)\citenamefont {Dutta},
  \citenamefont {Mangeat}, \citenamefont {Rieger}, \citenamefont {Paul},\ and\
  \citenamefont {Chatterjee}}]{zenodo}%
  \BibitemOpen
  \bibfield  {author} {\bibinfo {author} {\bibfnamefont {A.~K.}\ \bibnamefont
  {Dutta}}, \bibinfo {author} {\bibfnamefont {M.}~\bibnamefont {Mangeat}},
  \bibinfo {author} {\bibfnamefont {H.}~\bibnamefont {Rieger}}, \bibinfo
  {author} {\bibfnamefont {R.}~\bibnamefont {Paul}}, \ and\ \bibinfo {author}
  {\bibfnamefont {S.}~\bibnamefont {Chatterjee}},\ }\bibfield  {title}
  {\enquote {\bibinfo {title} {Supplementary movies for \textit{{Stability} of
  flocking in the reciprocal two-species Vicsek model: {Effects} of relative
  population, motility, and noise}},}\ }\href {\doibase
  10.5281/zenodo.15241518} {\bibfield  {journal} {\bibinfo  {journal} {Zenodo}\
  } (\bibinfo {year} {2025}),\ 10.5281/zenodo.15241518}\BibitemShut {NoStop}%
\bibitem [{\citenamefont {Chen}\ and\ \citenamefont
  {Toner}(2013)}]{chen2013universality}%
  \BibitemOpen
  \bibfield  {author} {\bibinfo {author} {\bibfnamefont {L.}~\bibnamefont
  {Chen}}\ and\ \bibinfo {author} {\bibfnamefont {J.}~\bibnamefont {Toner}},\
  }\bibfield  {title} {\enquote {\bibinfo {title} {Universality for moving
  stripes: A hydrodynamic theory of polar active smectics},}\ }\href@noop {}
  {\bibfield  {journal} {\bibinfo  {journal} {Phys. Rev. Lett.}\ }\textbf
  {\bibinfo {volume} {111}},\ \bibinfo {pages} {088701} (\bibinfo {year}
  {2013})}\BibitemShut {NoStop}%
\bibitem [{\citenamefont {Adhyapak}\ \emph {et~al.}(2013)\citenamefont
  {Adhyapak}, \citenamefont {Ramaswamy},\ and\ \citenamefont
  {Toner}}]{adhyapak2013live}%
  \BibitemOpen
  \bibfield  {author} {\bibinfo {author} {\bibfnamefont {T.~C.}\ \bibnamefont
  {Adhyapak}}, \bibinfo {author} {\bibfnamefont {S.}~\bibnamefont {Ramaswamy}},
  \ and\ \bibinfo {author} {\bibfnamefont {J.}~\bibnamefont {Toner}},\
  }\bibfield  {title} {\enquote {\bibinfo {title} {Live soap: stability, order,
  and fluctuations in apolar active smectics},}\ }\href@noop {} {\bibfield
  {journal} {\bibinfo  {journal} {Phys. Rev. Lett.}\ }\textbf {\bibinfo
  {volume} {110}},\ \bibinfo {pages} {118102} (\bibinfo {year}
  {2013})}\BibitemShut {NoStop}%
\bibitem [{SM()}]{SM}%
  \BibitemOpen
  \href@noop {} {}\bibinfo {note} {See Supplemental Material}\BibitemShut
  {NoStop}%
\bibitem [{\citenamefont {Kumar}\ \emph {et~al.}(2014)\citenamefont {Kumar},
  \citenamefont {Soni}, \citenamefont {Ramaswamy},\ and\ \citenamefont
  {Sood}}]{kumar2014flocking}%
  \BibitemOpen
  \bibfield  {author} {\bibinfo {author} {\bibfnamefont {N.}~\bibnamefont
  {Kumar}}, \bibinfo {author} {\bibfnamefont {H.}~\bibnamefont {Soni}},
  \bibinfo {author} {\bibfnamefont {S.}~\bibnamefont {Ramaswamy}}, \ and\
  \bibinfo {author} {\bibfnamefont {A.~K.}\ \bibnamefont {Sood}},\ }\bibfield
  {title} {\enquote {\bibinfo {title} {Flocking at a distance in active
  granular matter},}\ }\href@noop {} {\bibfield  {journal} {\bibinfo  {journal}
  {Nat. Comm.}\ }\textbf {\bibinfo {volume} {5}},\ \bibinfo {pages} {4688}
  (\bibinfo {year} {2014})}\BibitemShut {NoStop}%
\bibitem [{\citenamefont {Koumakis}\ \emph {et~al.}(2016)\citenamefont
  {Koumakis}, \citenamefont {Gnoli}, \citenamefont {Maggi}, \citenamefont
  {Puglisi},\ and\ \citenamefont {Di~Leonardo}}]{koumakis2016mechanism}%
  \BibitemOpen
  \bibfield  {author} {\bibinfo {author} {\bibfnamefont {N.}~\bibnamefont
  {Koumakis}}, \bibinfo {author} {\bibfnamefont {A.}~\bibnamefont {Gnoli}},
  \bibinfo {author} {\bibfnamefont {C.}~\bibnamefont {Maggi}}, \bibinfo
  {author} {\bibfnamefont {A.}~\bibnamefont {Puglisi}}, \ and\ \bibinfo
  {author} {\bibfnamefont {R.}~\bibnamefont {Di~Leonardo}},\ }\bibfield
  {title} {\enquote {\bibinfo {title} {Mechanism of self-propulsion in
  3d-printed active granular particles},}\ }\href@noop {} {\bibfield  {journal}
  {\bibinfo  {journal} {New J. Phys.}\ }\textbf {\bibinfo {volume} {18}},\
  \bibinfo {pages} {113046} (\bibinfo {year} {2016})}\BibitemShut {NoStop}%
\bibitem [{\citenamefont {Chen}\ \emph
  {et~al.}(2024{\natexlab{a}})\citenamefont {Chen}, \citenamefont {Lei},
  \citenamefont {Xiang}, \citenamefont {Duan}, \citenamefont {Peng},\ and\
  \citenamefont {Zhang}}]{chen2024emergent}%
  \BibitemOpen
  \bibfield  {author} {\bibinfo {author} {\bibfnamefont {J.}~\bibnamefont
  {Chen}}, \bibinfo {author} {\bibfnamefont {X.}~\bibnamefont {Lei}}, \bibinfo
  {author} {\bibfnamefont {Y.}~\bibnamefont {Xiang}}, \bibinfo {author}
  {\bibfnamefont {M.}~\bibnamefont {Duan}}, \bibinfo {author} {\bibfnamefont
  {X.}~\bibnamefont {Peng}}, \ and\ \bibinfo {author} {\bibfnamefont {H.~P.}\
  \bibnamefont {Zhang}},\ }\bibfield  {title} {\enquote {\bibinfo {title}
  {Emergent chirality and hyperuniformity in an active mixture with
  nonreciprocal interactions},}\ }\href@noop {} {\bibfield  {journal} {\bibinfo
   {journal} {Phys. Rev. Lett.}\ }\textbf {\bibinfo {volume} {132}},\ \bibinfo
  {pages} {118301} (\bibinfo {year} {2024}{\natexlab{a}})}\BibitemShut
  {NoStop}%
\bibitem [{\citenamefont {Bera}\ and\ \citenamefont
  {Sood}(2020)}]{bera2020motile}%
  \BibitemOpen
  \bibfield  {author} {\bibinfo {author} {\bibfnamefont {P.~K.}\ \bibnamefont
  {Bera}}\ and\ \bibinfo {author} {\bibfnamefont {A.~K.}\ \bibnamefont
  {Sood}},\ }\bibfield  {title} {\enquote {\bibinfo {title} {Motile dissenters
  disrupt the flocking of active granular matter},}\ }\href@noop {} {\bibfield
  {journal} {\bibinfo  {journal} {Phys. Rev. E}\ }\textbf {\bibinfo {volume}
  {101}},\ \bibinfo {pages} {052615} (\bibinfo {year} {2020})}\BibitemShut
  {NoStop}%
\bibitem [{\citenamefont {Chen}\ \emph
  {et~al.}(2024{\natexlab{b}})\citenamefont {Chen}, \citenamefont {Yang},
  \citenamefont {Zhang}, \citenamefont {Li}, \citenamefont {Pan},\ and\
  \citenamefont {Wang}}]{chen2024scale}%
  \BibitemOpen
  \bibfield  {author} {\bibinfo {author} {\bibfnamefont {T.}~\bibnamefont
  {Chen}}, \bibinfo {author} {\bibfnamefont {X.}~\bibnamefont {Yang}}, \bibinfo
  {author} {\bibfnamefont {B.}~\bibnamefont {Zhang}}, \bibinfo {author}
  {\bibfnamefont {J.}~\bibnamefont {Li}}, \bibinfo {author} {\bibfnamefont
  {J.}~\bibnamefont {Pan}}, \ and\ \bibinfo {author} {\bibfnamefont
  {Y.}~\bibnamefont {Wang}},\ }\bibfield  {title} {\enquote {\bibinfo {title}
  {Scale-inspired programmable robotic structures with concurrent shape
  morphing and stiffness variation},}\ }\href@noop {} {\bibfield  {journal}
  {\bibinfo  {journal} {Sci. Robot.}\ }\textbf {\bibinfo {volume} {9}},\
  \bibinfo {pages} {eadl0307} (\bibinfo {year}
  {2024}{\natexlab{b}})}\BibitemShut {NoStop}%
\bibitem [{\citenamefont {Storms}\ \emph {et~al.}(2019)\citenamefont {Storms},
  \citenamefont {Carere}, \citenamefont {Zoratto},\ and\ \citenamefont
  {Hemelrijk}}]{storms2019complex}%
  \BibitemOpen
  \bibfield  {author} {\bibinfo {author} {\bibfnamefont {R.~F.}\ \bibnamefont
  {Storms}}, \bibinfo {author} {\bibfnamefont {C.}~\bibnamefont {Carere}},
  \bibinfo {author} {\bibfnamefont {F.}~\bibnamefont {Zoratto}}, \ and\
  \bibinfo {author} {\bibfnamefont {C.~K.}\ \bibnamefont {Hemelrijk}},\
  }\bibfield  {title} {\enquote {\bibinfo {title} {Complex patterns of
  collective escape in starling flocks under predation},}\ }\href@noop {}
  {\bibfield  {journal} {\bibinfo  {journal} {Behav. Ecol. Sociobio.}\ }\textbf
  {\bibinfo {volume} {73}},\ \bibinfo {pages} {1} (\bibinfo {year}
  {2019})}\BibitemShut {NoStop}%
\bibitem [{\citenamefont {Papadopoulou}\ \emph {et~al.}(2022)\citenamefont
  {Papadopoulou}, \citenamefont {Hildenbrandt}, \citenamefont {Sankey},
  \citenamefont {Portugal},\ and\ \citenamefont
  {Hemelrijk}}]{papadopoulou2022emergence}%
  \BibitemOpen
  \bibfield  {author} {\bibinfo {author} {\bibfnamefont {M.}~\bibnamefont
  {Papadopoulou}}, \bibinfo {author} {\bibfnamefont {H.}~\bibnamefont
  {Hildenbrandt}}, \bibinfo {author} {\bibfnamefont {D.~W.~E.}\ \bibnamefont
  {Sankey}}, \bibinfo {author} {\bibfnamefont {S.~J.}\ \bibnamefont
  {Portugal}}, \ and\ \bibinfo {author} {\bibfnamefont {C.~K.}\ \bibnamefont
  {Hemelrijk}},\ }\bibfield  {title} {\enquote {\bibinfo {title} {Emergence of
  splits and collective turns in pigeon flocks under predation},}\ }\href@noop
  {} {\bibfield  {journal} {\bibinfo  {journal} {Royal Society Open Science}\
  }\textbf {\bibinfo {volume} {9}},\ \bibinfo {pages} {211898} (\bibinfo {year}
  {2022})}\BibitemShut {NoStop}%
\bibitem [{\citenamefont {Herbert-Read}\ \emph {et~al.}(2017)\citenamefont
  {Herbert-Read}, \citenamefont {Ros{\'e}n}, \citenamefont {Szorkovszky},
  \citenamefont {Ioannou}, \citenamefont {Rogell}, \citenamefont {Perna},
  \citenamefont {Ramnarine}, \citenamefont {Kotrschal}, \citenamefont {Kolm},
  \citenamefont {Krause},\ and\ \citenamefont
  {Sumpter}}]{herbert2017predation}%
  \BibitemOpen
  \bibfield  {author} {\bibinfo {author} {\bibfnamefont {J.~E.}\ \bibnamefont
  {Herbert-Read}}, \bibinfo {author} {\bibfnamefont {E.}~\bibnamefont
  {Ros{\'e}n}}, \bibinfo {author} {\bibfnamefont {A.}~\bibnamefont
  {Szorkovszky}}, \bibinfo {author} {\bibfnamefont {C.~C.}\ \bibnamefont
  {Ioannou}}, \bibinfo {author} {\bibfnamefont {B.}~\bibnamefont {Rogell}},
  \bibinfo {author} {\bibfnamefont {A.}~\bibnamefont {Perna}}, \bibinfo
  {author} {\bibfnamefont {I.~W.}\ \bibnamefont {Ramnarine}}, \bibinfo {author}
  {\bibfnamefont {A.}~\bibnamefont {Kotrschal}}, \bibinfo {author}
  {\bibfnamefont {N.}~\bibnamefont {Kolm}}, \bibinfo {author} {\bibfnamefont
  {J.}~\bibnamefont {Krause}}, \ and\ \bibinfo {author} {\bibfnamefont
  {D.~J.~T.}\ \bibnamefont {Sumpter}},\ }\bibfield  {title} {\enquote {\bibinfo
  {title} {How predation shapes the social interaction rules of shoaling
  fish},}\ }\href@noop {} {\bibfield  {journal} {\bibinfo  {journal} {Proc.
  Royal Soc. B: Bio. Sci.}\ }\textbf {\bibinfo {volume} {284}},\ \bibinfo
  {pages} {20171126} (\bibinfo {year} {2017})}\BibitemShut {NoStop}%
\bibitem [{\citenamefont {Momeni}\ \emph {et~al.}(2013)\citenamefont {Momeni},
  \citenamefont {Brileya}, \citenamefont {Fields},\ and\ \citenamefont
  {Shou}}]{momeni2013strong}%
  \BibitemOpen
  \bibfield  {author} {\bibinfo {author} {\bibfnamefont {B.}~\bibnamefont
  {Momeni}}, \bibinfo {author} {\bibfnamefont {K.~A.}\ \bibnamefont {Brileya}},
  \bibinfo {author} {\bibfnamefont {M.~W.}\ \bibnamefont {Fields}}, \ and\
  \bibinfo {author} {\bibfnamefont {W.}~\bibnamefont {Shou}},\ }\bibfield
  {title} {\enquote {\bibinfo {title} {Strong inter-population cooperation
  leads to partner intermixing in microbial communities},}\ }\href@noop {}
  {\bibfield  {journal} {\bibinfo  {journal} {elife}\ }\textbf {\bibinfo
  {volume} {2}},\ \bibinfo {pages} {e00230} (\bibinfo {year}
  {2013})}\BibitemShut {NoStop}%
\bibitem [{\citenamefont {Farine}\ \emph {et~al.}(2014)\citenamefont {Farine},
  \citenamefont {Aplin}, \citenamefont {Garroway}, \citenamefont {Mann},\ and\
  \citenamefont {Sheldon}}]{farine2014collective}%
  \BibitemOpen
  \bibfield  {author} {\bibinfo {author} {\bibfnamefont {D.~R.}\ \bibnamefont
  {Farine}}, \bibinfo {author} {\bibfnamefont {L.~M.}\ \bibnamefont {Aplin}},
  \bibinfo {author} {\bibfnamefont {C.~J.}\ \bibnamefont {Garroway}}, \bibinfo
  {author} {\bibfnamefont {R.~P.}\ \bibnamefont {Mann}}, \ and\ \bibinfo
  {author} {\bibfnamefont {B.~C.}\ \bibnamefont {Sheldon}},\ }\bibfield
  {title} {\enquote {\bibinfo {title} {Collective decision making and social
  interaction rules in mixed-species flocks of songbirds},}\ }\href@noop {}
  {\bibfield  {journal} {\bibinfo  {journal} {Animal behaviour}\ }\textbf
  {\bibinfo {volume} {95}},\ \bibinfo {pages} {173} (\bibinfo {year}
  {2014})}\BibitemShut {NoStop}%
\bibitem [{\citenamefont {Papageorgiou}\ \emph {et~al.}(2024)\citenamefont
  {Papageorgiou}, \citenamefont {Nyaguthii},\ and\ \citenamefont
  {Farine}}]{papageorgiou2024compromise}%
  \BibitemOpen
  \bibfield  {author} {\bibinfo {author} {\bibfnamefont {D.}~\bibnamefont
  {Papageorgiou}}, \bibinfo {author} {\bibfnamefont {B.}~\bibnamefont
  {Nyaguthii}}, \ and\ \bibinfo {author} {\bibfnamefont {D.~R.}\ \bibnamefont
  {Farine}},\ }\bibfield  {title} {\enquote {\bibinfo {title} {Compromise or
  choose: shared movement decisions in wild vulturine guineafowl},}\
  }\href@noop {} {\bibfield  {journal} {\bibinfo  {journal} {Comm. Biol.}\
  }\textbf {\bibinfo {volume} {7}},\ \bibinfo {pages} {95} (\bibinfo {year}
  {2024})}\BibitemShut {NoStop}%
\bibitem [{\citenamefont {Ginelli}\ \emph {et~al.}(2015)\citenamefont
  {Ginelli}, \citenamefont {Peruani}, \citenamefont {Pillot}, \citenamefont
  {Chat{\'e}}, \citenamefont {Theraulaz},\ and\ \citenamefont
  {Bon}}]{ginelli2015intermittent}%
  \BibitemOpen
  \bibfield  {author} {\bibinfo {author} {\bibfnamefont {F.}~\bibnamefont
  {Ginelli}}, \bibinfo {author} {\bibfnamefont {F.}~\bibnamefont {Peruani}},
  \bibinfo {author} {\bibfnamefont {M-H.}\ \bibnamefont {Pillot}}, \bibinfo
  {author} {\bibfnamefont {H.}~\bibnamefont {Chat{\'e}}}, \bibinfo {author}
  {\bibfnamefont {G.}~\bibnamefont {Theraulaz}}, \ and\ \bibinfo {author}
  {\bibfnamefont {R.}~\bibnamefont {Bon}},\ }\bibfield  {title} {\enquote
  {\bibinfo {title} {Intermittent collective dynamics emerge from conflicting
  imperatives in sheep herds},}\ }\href@noop {} {\bibfield  {journal} {\bibinfo
   {journal} {Proc. Natl. Acad. Sci.}\ }\textbf {\bibinfo {volume} {112}},\
  \bibinfo {pages} {12729} (\bibinfo {year} {2015})}\BibitemShut {NoStop}%
\bibitem [{\citenamefont {Martinez}\ \emph {et~al.}(2018)\citenamefont
  {Martinez}, \citenamefont {Alarcon}, \citenamefont {Rodriguez}, \citenamefont
  {Aragones},\ and\ \citenamefont {Valeriani}}]{martinez2018collective}%
  \BibitemOpen
  \bibfield  {author} {\bibinfo {author} {\bibfnamefont {R.}~\bibnamefont
  {Martinez}}, \bibinfo {author} {\bibfnamefont {F.}~\bibnamefont {Alarcon}},
  \bibinfo {author} {\bibfnamefont {D.~R.}\ \bibnamefont {Rodriguez}}, \bibinfo
  {author} {\bibfnamefont {J.~L.}\ \bibnamefont {Aragones}}, \ and\ \bibinfo
  {author} {\bibfnamefont {C.}~\bibnamefont {Valeriani}},\ }\bibfield  {title}
  {\enquote {\bibinfo {title} {Collective behavior of vicsek particles without
  and with obstacles},}\ }\href@noop {} {\bibfield  {journal} {\bibinfo
  {journal} {Eur. Phys. J. E}\ }\textbf {\bibinfo {volume} {41}},\ \bibinfo
  {pages} {1} (\bibinfo {year} {2018})}\BibitemShut {NoStop}%
\bibitem [{\citenamefont {Karmakar}\ \emph {et~al.}(2023)\citenamefont
  {Karmakar}, \citenamefont {Chatterjee}, \citenamefont {Mangeat},
  \citenamefont {Rieger},\ and\ \citenamefont {Paul}}]{karmakar2023jamming}%
  \BibitemOpen
  \bibfield  {author} {\bibinfo {author} {\bibfnamefont {M.}~\bibnamefont
  {Karmakar}}, \bibinfo {author} {\bibfnamefont {S.}~\bibnamefont
  {Chatterjee}}, \bibinfo {author} {\bibfnamefont {M.}~\bibnamefont {Mangeat}},
  \bibinfo {author} {\bibfnamefont {H.}~\bibnamefont {Rieger}}, \ and\ \bibinfo
  {author} {\bibfnamefont {R.}~\bibnamefont {Paul}},\ }\bibfield  {title}
  {\enquote {\bibinfo {title} {Jamming and flocking in the restricted active
  potts model},}\ }\href@noop {} {\bibfield  {journal} {\bibinfo  {journal}
  {Phys. Rev. E}\ }\textbf {\bibinfo {volume} {108}},\ \bibinfo {pages}
  {014604} (\bibinfo {year} {2023})}\BibitemShut {NoStop}%
\end{thebibliography}%

\end{document}